\documentclass[aoas]{imsart}
\usepackage{fix-cm}
\usepackage{colortbl}
\usepackage{booktabs}
\usepackage{xcolor}
\usepackage{newtxtext}
\usepackage{enumitem}
\RequirePackage{amsthm,amsmath,amsfonts,amssymb,hyperref,graphicx,comment,bm}
\RequirePackage[authoryear]{natbib}
\RequirePackage{xcolor}
\usepackage[flushleft]{threeparttable}

\usepackage{scalerel}
\makeatletter
\newcommand*{\rom}[1]{\expandafter\@slowromancap\romannumeral #1@}
\makeatother

\startlocaldefs

\endlocaldefs

\begin{document}

\begin{frontmatter}
\title{A joint model for individual mean and
within-subject variability of a longitudinal outcome with
competing-risk time-to-event outcomes}
\runtitle{Joint Models for Mean, Within-Subject Variability, and Competing Risks}
\runauthor{S. LI ET AL.}


\begin{aug}
\author[A,G]{\fnms{Shanpeng}~\snm{Li}\ead[label=e1]{lishanpeng0913@ucla.edu}},
\author[D,F]{\fnms{Daniel}~\snm{S.}~\snm{Nuyujukian}},
\author[E]{\fnms{Robyn}~\snm{L.}~\snm{McClelland}},
\author[D]{\fnms{Peter}~\snm{D.}~\snm{Reaven}},
\author[A,B]{\fnms{Jin}~\snm{Zhou}},
\author[A,C]{\fnms{Hua}~\snm{Zhou}}
\and
\author[A,C]{\fnms{Gang}~\snm{Li}\ead[label=e2]{vli@ucla.edu}}
\address[A]{Department of Biostatistics,
University of California, Los Angeles, USA\printead[presep={,\ }]{e1,e2}}

\address[B]{Department of Medicine,
University of California, Los Angeles, USA}

\address[C]{Department of Computational Medicine,
University of California, Los Angeles, USA}
\address[D]{Phoenix VA Health Care System, Phoenix, USA}

\address[E]{Department of Biostatistics, University of Washington, Seattle, USA}

\address[F]{Department of Epidemiology and Biostatistics, University of Arizona, Tucson, USA}

\address[G]{Department of Computational and Quantitative Medicine, City of Hope, Duarte, USA}

\end{aug}

\begin{abstract} 
Motivated by a growing body of research emphasizing the importance of modeling within-subject (WS) variability in longitudinal biomarkers and its association with health outcomes, this paper proposes a semiparametric joint model for both the mean and WS variability of a longitudinal biomarker, jointly with competing-risk time-to-event outcomes.  We derive an expectation-maximization algorithm for parameter estimation and a profile-likelihood method for standard error estimation and inference, which allows time-dependent covariates and general forms of the latent association structure. Furthermore, we optimize the implementation of our joint model when the survival submodel includes only time-independent baseline covariates and shared random effects, allowing it to scale effectively to biobank-scale data involving tens of thousands of subjects. Our method demonstrates satisfactory performance in simulations, whereas classical joint models that assume homogeneous WS variability may suffer from substantial estimation bias, invalid inference, and inferior prediction when confronted with heterogeneous WS variability. We illustrate the utility of our method using the Multi-Ethnic Study of Atherosclerosis (MESA) cohort. Our analysis demonstrates that associations between WS blood pressure variability and cardiovascular outcomes, previously observed in clinical trials involving relatively homogeneous populations, extend to a more ethnically diverse and generally healthier cohort, and that explicitly modeling heterogeneous WS variability substantially enhances risk discrimination. A user-friendly R package, \textbf{JMH}, has been developed for the proposed shared random effects model with efficient implementation and is publicly available on the Comprehensive R Archive Network \url{https://CRAN.R-project.org/package=JMH}.
\end{abstract}


\end{frontmatter}

\section{Introduction}\label{sec1}
In recent years, there has been increasing interest in modeling the within-subject (WS) variability of a longitudinal biomarker and studying its effects on health outcomes. In many studies, WS variability itself is often of scientific interest, and researchers aim to identify covariates such as risk factors, genetic variants, and environmental factors that influence WS variances \citep[among others]{hertzog2003assessing, fleeson2004moving, martin2004intraindividual,hedeker2008application, german2022wiser}. For instance, \cite{hedeker2008application} used a mixed effects location-scale model to characterize mood variation in an adolescent smoking study. In other studies, researchers are interested in modeling the WS variability, along with the level, of a longitudinal biomarker, and studying its effects on health outcomes
\citep[among others]{rothwell2010prognostic,zhou2018glycemic,ceriello2019glycaemic,barrett2019estimating,gao2011joint,martins2022flexible,courcoul2025location}.
Research based on clinical trials  such as the UK Prospective Diabetes Study (UKPDS)~\citep{uk1998intensive}, Action to Control Cardiovascular Risk in Diabetes (ACCORD)~\citep{action2008effects,zhou2018glycemic, ismail2010effect}, and Veterans Affairs Diabetes Trial (VADT)~\citep{duckworth2009glucose, reaven2019intensive} has suggested that individual variability in glycemic and blood pressure levels is associated with cardiovascular disease risks \citep{zhou2018glycemic,nuyujukian2021refining}, heart failure \citep{Nuyujukian20BPVADT}, nephropathy \citep{zhou2020long, zhou2020fasting}, and retinopathy \citep{zhou2020fasting}, independent of traditional glycemic and blood pressure control. 
These findings underscore the importance of modeling the WS variability of a biomarker and their impacts on health outcomes. 

This paper considers joint modeling of the mean and WS variability of a longitudinal biomarker, together with competing-risk time-to-event outcomes. Our work was originally motivated by a collaborative research to study the links between blood pressure variation (BPV) and heart failure (HF) and death using the Multi-Ethnic Study of Atherosclerosis (MESA) data \citep{bild2002multi}. Note that although there is growing recognition of BPV as a significant risk factor for cardiovascular diseases \citep{stevens2016blood, wang2017visit, chiriaco2019association, muntner2015visit, Nuyujukian20BPVADT, kaze2021long},  
 most previous studies investigating this relationship have been conducted with cohorts predominantly consisting of type 2 diabetes patients of European descent and in clinical trials. Therefore, it is crucial to assess whether these findings are applicable to more diverse and healthier populations.  The MESA, which enrolled 6,814 men and women aged 45 to 84 from four  racial/ethnic groups without overt clinical cardiovascular disease  across six U.S.\ field centers from 2000 to 2002, offers a unique opportunity to further explore this issue.
 However, current approaches to modeling the WS variability of a biomarker and studying its effects on clinical outcomes present significant statistical and computational challenges, especially when analyzing large-scale data like MESA, as discussed below.

To date, there are two main approaches to studying the WS variability of a biomarker, and its effects on an event outcome. A commonly used method is the ad hoc two-stage approach. This method initially estimates the time-dependent WS variability of a longitudinal biomarker using descriptive
sample variability measures such as coefficient of variation (CV) and average real variability (ARV)
\citep{mena2005reliable}
in stage 1 and subsequently correlates these estimates with a time-to-event outcome in the stage 2 analysis \citep{rothwell2010prognostic,zhou2018glycemic,ceriello2019glycaemic}. However, this approach is well known to have several practical and theoretical shortcomings, such as unstable variability measures due to insufficient number of repeated measurements \citep{mena2014many}, unaddressed biases and variances from the first stage that may compromise the second stage analysis, and potential biases arising from neglecting the correlation between the random effects associated with the mean trajectory and WS variability \citep{barrett2019estimating}. 

An alternative approach is the joint modeling of longitudinal and time-to-event data. A joint model typically includes a mixed-effects submodel for the longitudinal outcome and a survival submodel for the time-to-event outcome, where the two submodels are linked by the random effects \citep{huang2011general,tsiatis2004joint,wu2012analysis,Rizopoulos12JMBook,elashoff2016joint,hickey2018comparison,papageorgiou2019overview,alsefri2020bayesian,sudell2016joint,mccrink2013advances}. 
While much of the joint model literature has traditionally concentrated on modeling the level of a longitudinal biomarker trajectory and its association with time-to-event outcomes, assuming homogeneous (constant) WS variance across all subjects and time, there have been several recent extensions of the joint model to residually model WS variability and study its effects on event outcomes. One extension, proposed by \citet{wang2023modeling}, quantifies subject-specific fluctuations in the biomarker trajectory through a non-linear mixed-effects submodel for the longitudinal process, though it assumes homogeneous WS variance. 
The fluctuation summary measure is then incorporated into a Cox proportional hazards submodel for the time-to-event data. However, this method is inadequate for analyzing the MESA data because of its inability to incorporate covariates and lack of a framework for incorporating additional factors that may affect subject-specific WS variability above and beyond the estimated mean trajectory of the subject. Furthermore, it requires a large number of longitudinal measurements per subject to work well, while the MESA data is sparse with at most six measurements per subject.

Another joint model extension adapts the mixed-effects location-scale submodel to model both the mean trajectory and WS variance of a biomarker, linking it to a survival outcome via a proportional hazards submodel with a latent association structure that includes the random effects from both the mean trajectory and WS variability components \citep{gao2011joint,barrett2019estimating, martins2022flexible, courcoul2025location}. 
We note that these joint models  all impose a parametric baseline hazard in the proportional hazards submodel, using Weibull, piecewise constants, or splines. These fully parametric joint models allow for the direct application of standard Bayesian MCMC methods or maximum likelihood estimation methods using readily available software. On the other hand, 
using a parametric baseline hazard can be unnecessarily restrictive and vulnerable to model misspecification.

In this paper, we propose a more flexible joint model that simultaneously models the level and  WS variability of a longitudinal biomarker together with a competing risks event outcome. Specifically, as detailed in Section~ \ref{prelim}, our proposed joint model includes a linear mixed-effects multiple location-scale submodel for the level and WS variability of a longitudinal biomarker \citep{dzubur2020mixwild,german2022wiser,koGWASLongitudinalTrajectories2022}, coupled with a semi-parametric cause-specific Cox proportional hazards submodel for the competing risks survival outcomes. These submodels are intricately linked through a latent association structure involving the random effects in the longitudinal submodel. Our proposed joint model extends previous models by \citet{gao2011joint, barrett2019estimating, martins2022flexible, courcoul2025location} in one or more aspects, and the primary contributions of this paper are fourfold.
\begin{enumerate}
\item Model flexibility - Unlike other developed joint models for the level and WS variability of a longitudinal outcome together with an event outcome, our joint model does not require a parametric baseline hazard in the competing risks time-to-event submodel.
\item Estimation algorithms - Unlike existing parametric joint models with heterogeneous WS variance, standard maximum likelihood or Bayesian estimation methods and available software are not directly applicable to our proposed semi-parametric joint model. In Section \ref{Sec:2}, we develop tailored estimation and inference procedures for the proposed semiparametric joint model within a general framework that accommodates time-dependent covariates and flexible latent association structures. These procedures build on the methodology in \citet{li2022efficient} but differ in two important aspects. First, additional formulations are required to accommodate the inclusion of the WS variability submodel. Second, we adopt a different numerical integration strategy for the E-step of our EM algorithm because the approach used in \citet{li2022efficient} is not practically feasible for the proposed model, as detailed in Section \ref{sec:integration}.

\item Efficient implementation and software - The use of large-scale datasets, such as MESA, derived from electronic health records or biobanks has become ubiquitous. 
 In Section \ref{sec:2.4}, we discuss strategies to optimize the implementation of our joint model when the survival submodel includes only time-independent baseline covariates and shared random effects, enabling it to scale efficiently to large datasets involving tens of thousands of subjects. We have developed an R package, \textbf{JMH}, which is publicly available on the Comprehensive R Archive Network at \url{https://CRAN.R-project.org/package=JMH}. These scalable linear-scan algorithms open the door to computationally efficient extensions incorporating time-dependent covariates and general association structures via landmarking, as further discussed in the last paragraph of the Discussion section. 

\item MESA Analysis – As discussed earlier, an open question in the blood pressure variability literature is whether associations between WS blood pressure variability and cardiovascular outcomes—largely established in clinical trials involving relatively homogeneous populations—generalize to a more ethnically diverse and generally healthier population. Moreover, commonly used approaches, such as ad hoc two-stage methods, have important limitations. The proposed method enables a principled analysis of the MESA cohort, which is more ethnically diverse and generally healthier (Section \ref{Sec::MESA}), with the aim of assessing whether previously reported associations can be externally validated, while also addressing key statistical and computational challenges of existing approaches. Our MESA analysis (Section \ref{Sec::MESA}) demonstrates strong associations between WS blood pressure variability and cardiovascular outcomes by formally estimating and testing WS variability effects, thereby supporting their generalizability and relevance for cardiovascular disease risk in the general population. In addition, we find that incorporating heterogeneous WS variability can lead to substantial improvements in clinically relevant discrimination.
\end{enumerate}

The rest of the paper is organized as follows. Section \ref{Sec:2} describes the mathematical formulation of our proposed joint model, an EM algorithm for semi-parametric maximum likelihood estimation, a profile likelihood method for standard error estimation, efficient implementation when the survival submodel includes only time-independent covariates and shared random effects, dynamic prediction, and performance metrics for prediction. Section \ref{simulation} assesses the empirical performance of our method through simulation studies, comparing them to classical models and demonstrating scalability across different sample sizes. Section \ref{Sec::MESA} applies our method to the MESA study. Concluding remarks and further discussions are presented in Section \ref{discussion}.

\section{Methods}
\label{Sec:2}
\subsection{Model and data specifications}
\label{prelim}
Assume that there are $n$ subjects in the study.  For subject $i$, one observes a longitudinal outcome $Y_i(o)$ at multiple time points $o_{ij}$, $j = 1,  \ldots, n_i$, $i=1,\ldots, n$. In addition, each subject may experience one of $K$ distinct failure types or be right censored during the follow-up.  Let $\Tilde{T}_i$ denote the failure time of interest, $\Tilde{D}_i$ the failure type taking values in $\{1, \ldots, K\}$, and $C_i$ be an non-informative, independent censoring time for subject $i$. Then the observed right-censored competing risks time-to-event data for subject $i$ has the form  $(T_i, D_i)\equiv \left\{\text{min}(\Tilde{T}_i, C_i),\Tilde{D}_i I(\Tilde{T}_i \le C_i)\right\}$, $i=1,\ldots, n$. 

Assume that the longitudinal outcome $Y_i(o)$ is characterized by the following \textit{mixed-effects multiple location-scale submodel}:
\begin{eqnarray}
\label{eq1.1}
Y_{i}(o) & = & X_{i}^{(1)\top}(o) \beta + Z_{i}^{\top}(o) b_i + \sigma_{i}(o)\epsilon_{i}(o),\\ 
\label{eq1.2}
\sigma_{i}^2(o) & = & \exp\left\{W_{i}^{\top}(o) \tau + V_{i}^{\top}(o)\omega_i\right\},
\end{eqnarray}
where  $X_i^{(1)}(o)$,  $Z_i(o)$, $W_{i}(o)$, and $V_{i}(o)$ are vectors of possibly time-varying covariates, $\beta$ and $b_i$ represent the fixed effects and random effects, respectively, associated with the location component $m_{i}(o)$ for the mean trajectory, and $\tau$ and $\omega_i$ represent the fixed effects and random effects, respectively, associated with the scale component $\sigma_{i}(o)$ for the WS variability. 
Assume that the measurement error $\epsilon_{i}(o)\sim N(0,1)$ is independent of $b_i$ and $\omega_i$, and mutually independent across all time points and subjects, and the random effects follows a multivariate normal distribution:
\begin{eqnarray*}
\label{eqMVN}
\theta_i \equiv (b_i^{\top}, \omega_i^{\top})^{\top} \sim MVN(0, \Sigma_{\theta}), \quad
\Sigma_{\theta} =  \left( {\begin{array}{cc}
   \Sigma_{bb} & \Sigma_{b\omega} \\
   \Sigma_{b\omega}^{\top} & \Sigma_{\omega\omega} \\
  \end{array} } \right),
\end{eqnarray*}
where $\Sigma_{b\omega}=cov(b_i,\omega_i)$, $\Sigma_{bb}=cov(b_i,b_i)$, and $\Sigma_{\omega\omega}=cov(\omega_i,\omega_i)$. 

Assume further that the competing risks time-to-event outcome follows the cause-specific Cox proportional hazards submodel:
\begin{eqnarray}
\label{eq2}
\lambda_{ik}(t\mid X_i^{(2)}(t), M_{i}(\theta_i, t)) & = & \lim_{h \to 0} \frac{P(t \le \Tilde{T}_i < t + h, \Tilde{D}_i = k \mid  T_i \ge t, X_i^{(2)}(t), M_{i}(\theta_i, t))}{h} \cr
& = & \lambda_{0k}(t) \exp\{X_i^{(2)\top}(t) \gamma_k + M_{i}^{\top}(\theta_i, t)\alpha_k\},\quad k=1,\ldots, K,
\end{eqnarray}
where  $\lambda_{0k}(t)$ is a completely unspecified baseline hazard function, $X_i^{(2)}(t)$ is a vector of possibly time-varying covariates for the competing risks time-to-event outcome, $\gamma_k$ is a vector of fixed effects of $X_i^{(2)}(t)$, $M_{i}(\theta_i, t)$ is a vector of pre-specified functions of $\theta_i$ and $t$, and $\alpha_k$ is a vector of association parameters between the longitudinal and time-to-event outcomes. 

Note that the three submodels (\ref{eq1.1})-(\ref{eq2}) are linked together via the latent association structure $M_{i}^{\top}(\theta_i, t)\alpha_k$. Some useful examples of $M_{i}^{\top}(\theta_i, t)\alpha_k$ include 
\begin{enumerate}
    \item ``present value'' parameterization: $M_{i}^{\top}(\theta_i, t)\alpha_k=\alpha_{bk} m_i(t) + \alpha_{\omega k} \log\left\{\sigma_{i}^2(t)\right\}$,
    \item ``present value of latent process'' parameterization: $M_{i}^{\top}(\theta_i, t)\alpha_k=\alpha_{bk} Z_i^{\top}(t)b_i \\ + \alpha_{\omega k}  V_i^{\top}(t)\omega_i$,
    \item ``time-dependent slopes'' parameterization: $M_{i}^{\top}(\theta_i, t)\alpha_k= \left[m_i(t), \frac{d}{dt} m_i(t)\right]\alpha_{bk} \\+ \left[\log\left\{\sigma_{i}^2(t)\right\}, \frac{d}{dt} \log\left\{\sigma_{i}^2(t)\right\}\right]\alpha_{\omega k}$,
    \item ``shared random effects'' parameterization: $M_{i}^{\top}(\theta_i, t)\alpha_k = \alpha_{bk}^{\top} b_i + \alpha_{\omega k}^{\top} \omega_i$,
\end{enumerate}
The first three parameterizations incorporate the trajectory functions as time-dependent covariates in the survival submodel (\ref{eq2}), which are straightforward to interpret but can make the estimation process computationally intensive unless additional parametric assumptions are imposed on the baseline hazard functions. The last parameterization with shared random effects  uses the random effects as features extracted from the subject-specific mean and variance components to influence the survival outcome. It might lead to less interpretable association parameters, specifically when the spline functions are considered as random effects covariates \citep{rizopoulos2012joint,lawrence2015joint}. On the other hand, this time-independent association structure is useful for dynamic prediction of the survival outcome, and it opens a new path to facilitate efficient implementation of the cumulative baseline hazard function from a computational perspective \citep{lawrence2015joint, li2022efficient}. In Section \ref{sec:2.4}, we will discuss efficient fitting of the shared random effects model for large data.
 Finally, it is evident that our joint model 
(\ref{eq1.1})-(\ref{eq2}) reduces to the classical joint models with homogeneous WS variance such as that of \citet{li2022efficient}
if the submodel (\ref{eq1.2}) is replaced by $\sigma_i^2(t)\equiv \sigma^2$.

{\bf Remark 1:} (Interpretation of association parameters) 
The random effects $b_i$ and $\omega_i$ from submodels (\ref{eq1.1}) and (\ref{eq1.2}) are often highly correlated, which can complicate the interpretation of the association parameters $\alpha_k = (\alpha_{bk}^{\top}, \alpha_{\omega k}^{\top})^{\top}$ in the survival submodel (\ref{eq2}).
Below, we illustrate how this challenge can be handled in the context of a shared random effects joint model, in which the survival submodel (\ref{eq2}) is specified as:
\begin{eqnarray}
\label{eq4remark}
\lambda_{ik}(t \mid X_i^{(2)}(t), \theta_i)
= \lambda_{0k}(t) \exp\{X_i^{(2)\top}(t)\gamma_k
+ \alpha_{bk}^{\top} b_i + \alpha_{\omega k}^{\top} \omega_i\},
\quad k = 1, \ldots, K,
\end{eqnarray}
where $\alpha_{\omega k}$ cannot be simply interpreted as the effect of $\omega_i$ conditional on $b_i$ in the presence of strong collinearity between $b_i$ and $\omega_i$. 

To address this issue, and for simplicity of exposition, we consider the special case where $\omega_i$ is scalar. The multivariate normality assumption for $\theta_i = (b_i^{\top}, \omega_i)^{\top}$
allows the following decomposition:
\begin{eqnarray}
\label{decomposition}
\omega_i = \nu^{\top} b_i + e_i, 
\end{eqnarray}
where $\nu = \Sigma_{bb}^{-1}\Sigma_{b\omega}$ and $e_i \sim N(0, \sigma_e^2)$ is independent of $b_i$, and can be interpreted as residual WS variability after accounting for $b_i$.
The decomposition \eqref{decomposition} implies that submodel (\ref{eq4remark}) can be rewritten as
\begin{eqnarray}
\label{eq5remark}
\lambda_{ik}(t \mid X_i^{(2)}(t), \theta_i)
= \lambda_{0k}(t) \exp\{X_i^{(2)\top}(t)\gamma_k
+ \alpha_{bk}^{\top *} b_i + \alpha_{\omega k} e_i\},
\quad k = 1, \ldots, K,
\end{eqnarray}
where $\alpha_{bk}^{\top *} = \alpha_{bk}^{\top} + \alpha_{\omega k}\nu$, and the coefficient $\alpha_{\omega k}$ of $e_i$ is the same as that of $\omega_i$ in \eqref{eq4remark}.

Since $e_i$ is independent of $b_i$, $\alpha_{\omega k}$ can be interpreted as the effect of $e_i$—the residual WS variability after accounting for $b_i$—on the $k$th cause-specific hazard. Therefore, the decomposition \eqref{decomposition} helps disentangle the correlation between $b_i$ and $\omega_i$ and leads to a clearer interpretation of the association parameters. Finally, after fitting the original joint model \eqref{eq1.1}-\eqref{eq2}, an estimate of 
$\alpha_{bk}^{\top *} = \alpha_{bk}^{\top} + \alpha_{\omega k}\nu$ can be obtained by plugging in the estimates of $\alpha_{bk}^{\top}$, $\alpha_{\omega k}$, and $\nu = \Sigma_{bb}^{-1}\Sigma_{b\omega}$. Its variance can be estimated using the delta method.

Throughout the paper, we assume that for each subject $i$, the longitudinal measurements $Y_i$ are independent of the competing risks outcome $(\Tilde{T}_i, \Tilde{D}_i)$ conditional on the observed covariates and the unobserved random effects. We further assume that the censoring time $C_i$ is independent of $(\Tilde{T}_i, \Tilde{D}_i)$ and $Y_i$ conditional on the observed covariates and the unobserved random effects $\theta_i$. These assumptions imply that the longitudinal outcome $Y_i$ and the observed right-censored competing risks time-to-event outcome $(T_i,D_i)$ are independent conditional on the covariates and the random effects, which are crucial for deriving the observed data likelihood as detailed in the following section. Similar assumptions are also commonly used in the joint models literature (see, e.g.,  
\citet{wulfsohn1997joint, henderson2000joint, 
song2002semiparametric, hsieh2006joint, elashoff2016joint}, among others).
 
\subsection{Likelihood and EM estimation}
\label{Sec:EM}
Denote by $\Psi$ = ($\beta$, $\tau$, $\gamma$, $\alpha$, $\Sigma_{\theta}$, $\lambda_{01}(\cdot)$, ..., $\lambda_{0K}(\cdot)$) the collection of all unknown parameters and functions from the submodels (\ref{eq1.1})-(\ref{eq2}), where $\gamma = (\gamma_1^{\top}, \ldots, \gamma_K^{\top})^{\top}$ and $\alpha = (\alpha_{b1}^{\top}, \ldots, \alpha_{bK}^{\top}, \alpha_{\omega 1}^{\top}, \ldots, \alpha_{\omega K}^{\top})^{\top}$. Denote by $Y_i = (Y_{i1}, ..., Y_{in_i})^{\top}$, where $Y_{ij}=Y_i(o_{ij})$. Omitting the covariates for the sake of brevity, the observed-data likelihood is given by
\begin{eqnarray*}
     L(\Psi; Y, T, D) &\propto& \prod_{i=1}^n f(Y_i, T_i, D_i\mid \Psi)\\
    & = &\prod_{i=1}^n \int f(Y_i\mid \theta_i, \Psi)f(T_i, D_i\mid \theta_i, \Psi)f(\theta_i\mid \Psi)d\theta_i\\
    & = &\prod_{i=1}^n \int \prod_{j=1}^{n_i}\frac{1}{\sqrt{2\pi \sigma_{i}^2(o_{ij})}}\exp\left[-\frac{\left\{Y_{i}(o_{ij})-m_i(o_{ij})\right\}^2}{2\sigma_{i}^2(o_{ij})}\right]\\
    && \times \prod_{k=1}^K\lambda_{ik}\left\{T_i\mid X_i^{(2)}(T_i), M_{i}(\theta_i, T_i)\right\}^{I(D_i=k)} \cr
    &&\times \exp\left[-\sum_{k=1}^K \int_0^{T_i} \lambda_{ik}\left\{t\mid X_i^{(2)}(t), M_{i}(\theta_i, t)\right\}dt\right] \\
    && \times \frac{1}{\sqrt{(2\pi)^{q}\mid \Sigma_{\theta}\mid }}\exp\left(-\frac{1}{2}\theta_i^{\top} \Sigma_{\theta}^{-1} \theta_i\right)d\theta_i,
\end{eqnarray*}
where the first equality follows from the assumption that $Y_i$ and $(T_i, D_i)$ are independent conditional on the covariates and the random effects. 

Because $\Psi$ contains $K$ unknown cause-specific baseline hazard functions and the likelihood function involves integrals,  directly maximizing the above observed-data likelihood is difficult. To tackle this issue, we derive an EM algorithm to compute the semi-parametric maximum likelihood estimate (SMLE) of $\Psi$ by regarding the latent random effects $\theta_i$ as missing data \citep{dempster1977maximum, elashoff2008joint}.The complete-data likelihood based on $(Y, T, D, \theta)$ is given by 
\begin{eqnarray*}
     L(\Psi; Y, T, D, \theta) &\propto& \prod_{i=1}^n \prod_{j=1}^{n_i}\frac{1}{\sqrt{2\pi\exp\left\{W_{i}^{\top}(o_{ij}) \tau + V_{i}^{\top}(o_{ij})\omega_i\right\}}}\cr
     &&\times \exp\left[-\frac{\left\{Y_i(o_{ij})-X_{i}^{(1)\top}(o_{ij})\beta - Z_{i}^{\top}(o_{ij}) b_i\right\}^2}{2\exp\left\{W_{i}^{\top}(o_{ij}) \tau + V_{i}^{\top}(o_{ij})\omega_i\right\}}\right] \cr
     && \times \prod_{k=1}^K \left[\Delta\Lambda_{0k}(T_i) \exp\left\{X_i^{(2)\top}(T_i) \gamma_k + M_{i}^{\top}(\theta_i, T_i)\alpha_k\right\}\right]^{I(D_i=k)} \cr
     &&\times  \exp\left[-\sum_{k=1}^K \int_0^{T_i} \exp\left\{X_i^{(2)\top}(t) \gamma_k + M_{i}^{\top}(\theta_i, t)\alpha_k\right\}d\Lambda_{0k}(t)\right] \cr
    && \times \frac{1}{\sqrt{(2\pi)^{q}\mid \Sigma_{\theta}\mid }}\exp\left(-\frac{1}{2}\theta_i^{\top} \Sigma_{\theta}^{-1} \theta_i\right),
\end{eqnarray*}
where $\Lambda_{0k}(.)$ is the cumulative baseline hazard function for type $k$ failure and $\Delta\Lambda_{0k}(T_i)=\Lambda_{0k}(T_i) - \Lambda_{0k}(T_i-)$.
The EM algorithm iterates between an expectation step (E-step):
\begin{eqnarray}
\label{EstepFunction}
Q(\Psi;\Psi^{(m)})\equiv E^{(m)}_{\theta \mid Y,T, D,\Psi^{(m)}}\left\{\log L(\Psi;Y, T, D, \theta)\right\},
\end{eqnarray}
 and a maximization step (M-step):
 \begin{equation} \label{Mstep}
    \Psi^{(m+1)} = \arg\max_{\Psi} Q(\Psi;\Psi^{(m)}),
\end{equation}
until the algorithm converges, where $\Psi^{(m)}$ is the estimate of $\Psi$ from the $m$-th iteration. 
Each E-step involves calculating integrals of the form
\begin{eqnarray}
\label{Epost}
 E^{(m)}\{h(\theta_i)\} 
 &=&\int h(\theta_i)f(\theta_i \mid Y_i, T_i, D_i, \Psi^{(m)})d\theta_i
\end{eqnarray}
for every subject $i$, $i=1,\ldots, n$, which are evaluated using the adaptive Gauss-Hermite quadrature approximation rule \citep{naylor1982applications} as discussed later in Section \ref{sec:2.4} and in Section 1.2 of the Supplementary Material.
As shown in Section 1.1 of the Supplementary Material, the M-step (\ref{Mstep}) has closed-form solutions for a number of parameters including the nonparametric baseline cumulative hazard functions $\Lambda_{0k}(t)$, $k=1,\ldots, K$, which is a key advantage of the EM-algorithm. Other
parameters without closed-form solutions in the M-step are updated using the one-step Newton-Raphson method. 
Details of the EM algorithm are provided in equations (S4)-(S9) of the Supplementary Material.

\subsection{Standard error estimation}
As discussed in \cite{elashoff2016joint} (Section 4.1, p.72), several approaches including 
profile-likelihood, observed information matrix, and bootstrap method
have been proposed in the literature for estimating the standard errors of the parametric components of the SMLE. Here we adopt the profile-likelihood approach because it can be readily computed from the EM algorithm and performed well in our simulation studies. 

Let $\Omega = (\beta, \tau, \gamma, \alpha, \Sigma_{\theta})$ denote the parametric component of $\Psi$ and $\hat{\Omega}$ its SMLE. We propose to estimate the variance-covariance matrix of $\hat{\Omega}$ by inverting the empirical Fisher information obtained from the profile likelihood of $\Omega$ \citep{lin2004latent, zeng2005asymptotic, zeng2005simultaneous} as follows:
\begin{equation}
\label{SEestimation}
    \sum_{i=1}^n [\nabla_{\Omega}l^{(i)}(\hat{\Omega};Y, T, D)][\nabla_{\Omega} l^{(i)}(\hat{\Omega};Y, T, D)]^{\top},
\end{equation}
 where $\nabla_{\Omega}l^{(i)}(\hat{\Omega};Y, T, D)$ is the observed score vector from the profile-likelihood $l^{(i)}({\Omega};Y, T, D)$ of $\Omega$ on the $i$th subject by profiling out the baseline hazards. Detailed formulas for calculating the observed score vector for each parametric component are provided in Section 2, equations (S10) through (S14), of the Supplementary Material. 

 We conclude this section by pointing out that there is a subtle yet important difference between our formulas for 
$\nabla_{\Omega}l^{(i)}(\hat{\Omega};Y, T, D)$
and those discussed by \citet{hsieh2006joint}. In a nutshell, when taking the derivative, the latter approach does not account for the fact that the profile likelihood also depends on $\Omega$ through $\hat{\lambda_0}(.)$, making it intuitively invalid. \citet{hsieh2006joint} has shown both theoretically and empirically that the method they discussed suffers from information loss compared to the true Hessian, thus leads to underestimated standard errors. However, our formulas for calculating the observed score vector in Section 2 of the Supplementary Material do account for the fact that the profile likelihood also depends on  $\Omega$ through $\hat{\lambda_0}(.)$, and thus do not suffer from the same issue and have demonstrated satisfactory performance in our simulation studies. A more detailed discussion can be found in the Remark in Section 2 of the Supplementary Material.

\subsection{Computational aspects}
\label{sec:2.4}

\subsubsection{Numerical integration}
\label{sec:integration}
Evaluating equation (\ref{EstepFunction}) in the E-step involves numerical integration, commonly employing the standard Gauss-Hermite quadrature rule \citep{press2007numerical}. However, this method is computationally inefficient as it often requires 20-30 quadrature points, even for 2-dimensional integration (see Section 1.2 of the Supplementary Material). For a joint model with homogeneous WS variance, \citet{li2022efficient} utilized the pseudo-adaptive Gauss-Hermite rule proposed by \citet{rizopoulos2012fast}, which requires fewer quadrature points (3 to 6) and only centers and scales the integrand once before the EM iterations, thereby avoiding the need for relocation and significantly reducing the computational load. However, this method is impractical for our proposed joint model with heterogeneous WS variance, due to challenges in efficiently fitting the required mixed-effects multiple location and scale model.

In this paper, we adopt the adaptive Gauss-Hermite quadrature approximation rule \citep{naylor1982applications} for numerical integration, with further details provided in Page 3, Paragraph 2, Section 1.2 of the Supplementary Material. Compared to the pseudo-adaptive Gauss-Hermite quadrature rule used by \citet{li2022efficient}, this adaptive version represents a compromise as it requires updates to quadrature points at each E-step rather than pre-calculating their locations before starting the EM iterations. Nevertheless, it guarantees the accuracy of integral approximation with just a few quadrature points (3-6), significantly reducing the computational burden relative to the standard Gauss-Hermite rule. 

\subsubsection{Efficient implementation of the shared random effects model with time-independent covariates}
\label{sec:efficient}
Note that our EM algorithm and standard error estimation method are developed for the joint model (\ref{eq1.1})–(\ref{eq2}), which accommodates time-dependent covariates and general forms of the latent association structure.
As detailed in Section 1.1 of the Supplementary Material, our EM algorithm for the proposed joint model requires many double summations, with each necessitating
$O(n^2)$ evaluations of exponential functions when implemented naively, which are computationally expensive. This can lead to significant computational bottlenecks, particularly when the EM algorithm is slow to converge. Moreover, the formulas for standard error estimation outlined in Section 2 of the Supplementary Material also involve double summations for each subject, requiring  $O(n^3)$ evaluations of exponential functions overall. However, when the survival submodel includes only time-independent covariates and shared random effects, we can reduce the computational complexity to $O(n)$ by applying the linear scan algorithms developed by \citet{li2022efficient}.  In Section 3, Page 5 of the Supplementary Material, we provide details on these linear scan algorithms used for both the EM algorithm and standard error estimation in our joint model with time-independent covariates and shared random effects.

\subsection{Dynamic prediction for competing risks time-to-event data}
\label{sec:dp}
The proposed joint model (\ref{eq1.1})-(\ref{eq2}) not only offers a general framework to model the individual mean and WS variability of a longitudinal outcome and study their association with competing-risk time-to-event outcomes, but also facilitates subject-level dynamic prediction of cumulative incidence probabilities of a competing risks event for a new subject $i^*$ based on his/her longitudinal biomarker history.  Specifically, given the longitudinal biomarker history 
$Y_{i^*}^{(s)}=\{Y_{i^*}(t_{{i^*}j}), t_{{i^*}j}\leq s\}$ prior to a landmark time $s>0$ and that an event has yet to happen by time $s$,  the cumulative incidence probability for type $k$ failure at a horizon time $u>s$ is 
\begin{eqnarray}
\label{competingdy}
P_{i^* k}(u, s|\Psi) &=& \text{Pr}(T_{i^*} \leq u, D_{i^*} = k |T_{i^*} > s, Y_{i^*}^{(s)}, \Psi)\cr
&=& \int \text{Pr}(T_{i^*} \leq u, D_{i^*} = k | T_{i^*} > s, Y_{i^*}^{(s)}, \theta_{i^*}, \Psi)f(\theta_{i^*}|T_{i^*} > s, Y_{i^*}^{(s)}, \Psi) d\theta_{i^*} \cr
&=& \int \frac{\text{Pr}(T_{i^*} \leq u, D_{i^*} = k, T_{i^*} > s| \theta_{i^*}, \Psi)}{\text{Pr}(T_{i^*} > s| \theta_{i^*}, \Psi)}f(\theta_{i^*}|T_{i^*} > s, Y_{i^*}^{(s)}, \Psi) d\theta_{i^*} \cr
&=& \frac{\int \frac{CIF_{i^*k}(u,s|\theta_{i^*}, \Psi)}{S_{i^*}(s|\theta_{i^*}, \Psi)} f(Y_{i^*}^{(s)}|\theta_{i^*}, \Psi)S_{i^*}(s|\theta_{i^*}, \Psi)f(\theta_{i^*} | \Psi) d\theta_{i^*}}{\int f(Y_{i^*}^{(s)}|\theta_{i^*}, \Psi)S_{i^*}(s|\theta_{i^*}, \Psi)f(\theta_{i^*} | \Psi) d\theta_{i^*}},\cr
&=& \frac{\int CIF_{i^*k}(u,s|\theta_{i^*}, \Psi) f(Y_{i^*}^{(s)}|\theta_{i^*}, \Psi)f(\theta_{i^*} | \Psi) d\theta_{i^*}}{\int f(Y_{i^*}^{(s)}|\theta_{i^*}, \Psi)S_{i^*}(s|\theta_{i^*}, \Psi)f(\theta_{i^*} | \Psi) d\theta_{i^*}},
\label{pikus}
\end{eqnarray}
where 
\[S_{i^*}(t|\theta_{i^*}, \Psi) = \exp\left[-\sum_{k=1}^K \int_0^{t} \exp\left\{X_{i^*}^{(2)\top}(l) \gamma_k + M_{i^*}^{\top}(\theta_{i^*}, l)\alpha_k\right\}d\Lambda_{0k}(l)\right],\]
is the overall survival function, 
\[CIF_{i^* k}(u,s|\theta_{i^*}, \Psi) = \int_s^u S_{i^*}(t|\theta_{i^*}, \Psi)d\left[\Lambda_{0k}(t)\exp\left\{X_{i^*}^{(2)\top}(t) \gamma_k + M_{i^*}^{\top}(\theta_{i^*}, t)\alpha_k\right\}\right]\]
is the cumulative incidence function (CIF) for type $k$ failure, 
$f(Y_{i^*}^{(s)}|\theta_{i^*}, \Psi)$ is the likelihood for $Y_{i^*}^{(s)}$, $S_{i^*}(s|\theta_{i^*}, \Psi)$ is the overall survival function evaluated at $s$, and $f(\theta_{i^*} | \Psi)$ is the prior distribution of $\theta_{i^*}$, and their explicit formulas are provided in equations (S18)-(S22), Section 4.1 of the Supplementary Material. The integrals in (\ref{pikus}) can be evaluated using a Gauss-Hermite quadrature rule. An estimate of $P_{i^* k}(u, s|\Psi)$ is then obtained by replacing $\Psi$, $S_{i^*}(.)$, and $CIF_{i^* k}(.)$ with their sample estimates $\hat{\Psi}$, $\hat{S}_{i^*}(.)$, and $\widehat{CIF}_{i^* k}(.)$, and respectively, as detailed in Section 4.1 of the Supplementary Material. 

The prediction performance of a joint model for competing risks outcomes can be evaluated using cross-validated calibration and discrimination measures, such as the Brier score \citep{wu2018quantifying} and the concordance index (C-index) \citep{wolbers2014concordance}. In Section \ref{sec:sim2}, we will also consider another calibration measure: the mean absolute prediction error.

\section{Simulations}
We present simulations to evaluate the performance of the proposed joint model. It is also compared to a classical joint model for longitudinal and time-to-event data that assumes homogeneous WS variance, illustrating that ignoring heterogeneous WS variability may lead to biased estimation, invalid inference, and inferior prediction performance. Additionally, it is compared to a multiple mixed-effects location-scale model for longitudinal data \citep{german2022wiser} to demonstrate that failing to account for nonignorable missing data due to informative dropout may result in biased estimation and inferences, highlighting that the proposed model offers a useful approach to address these issues. Lastly, we include a simulation to demonstrate the scalability of our model relative to other joint models.

\label{simulation}
\subsection{(Estimation and Inference)}
\label{sec:sim1}
This section studies the finite sample performance of parameter estimation, standard error estimation, and confidence intervals for the proposed joint model (\ref{eq1.1})-(\ref{eq2}), as well as some other related methods. 

{\bf Simulation 1: Generative joint model with heterogeneous WS variance and non-linear time evolution.}
We consider a generative joint model with heterogeneous WS variance, where the longitudinal measurements $Y_i(o_{ij})$ were generated from the following mixed-effects multiple location-scale model: 
\begin{eqnarray}
\label{sim:hetereq1.1}
Y_{i}(o_{ij}) & = & \beta_0 + \beta_1 X_{1i} + \sum_{\kappa=1}^2 \beta_{\kappa+1} B_{\kappa}(o_{ij}) + b_{i0} + \sum_{\kappa=1}^2 b_{i\kappa} B_{\kappa}(o_{ij}) + \sigma_{i}(o_{ij})\epsilon_{i}(o_{ij}),\\
\label{sim:hetereq1.2}
 \;\;\;\;   \sigma_{i}^2(o_{ij}) &=& \exp\left\{\tau_0 + \tau_1 X_{1i} + \sum_{\kappa=1}^2 \tau_{\kappa+1} B_{\kappa}(o_{ij}) + \omega_i\right\},
\end{eqnarray}
and the competing risks event data were generated from the  proportional cause-specific hazards models: 
\begin{eqnarray}
\label{sim:hetereq2.1}
\lambda_{i1}(t) &=& \lambda_{01}(t) \exp\{\gamma_{11}X_{1i} + \gamma_{12}X_{2i} + \gamma_{13}X_{3i} + \sum_{\kappa=0}^2 \alpha_{b\kappa1} b_{i\kappa} + \alpha_{\omega1}\omega_i\}, \\
\label{sim:hetereq2.2}
\lambda_{i2}(t) &=& \lambda_{02}(t) \exp\{\gamma_{21}X_{1i} + \gamma_{22}X_{2i} + \gamma_{23}X_{3i} + \sum_{\kappa=0}^2 \alpha_{b\kappa2} b_{i\kappa} + \alpha_{\omega2}\omega_i\},
\end{eqnarray}
where $\theta_i = (b_{i0}, b_{i1}, b_{i2}, \omega_i)^{\top} \sim N(0, \Sigma_{\theta})$ 
with
\begin{eqnarray*}
\Sigma_{\theta} =  \left( {\begin{array}{cccc}
   10 & 3 & 2 & 0.5\\
   3 & 4 & 2 &  0.5\\
   2 & 2 & 4 &  0.5\\
   0.5 & 0.5& 0.5 & 1 
  \end{array} } \right).
\end{eqnarray*}
Here, $B_{\kappa}(o_{ij}), \kappa =1, 2,$ denotes the 2-degree B-spline basis at the scheduled visiting times $o_{ij}$ for subject $i$ with increments of 0.25, $X_{1i} \sim Bernoulli(0.5)$, $X_{2i} \sim Uni(-1, 1)$, and $X_{3i} \sim N(1,4)$. The true parameter values are $\beta = (5, 1.5, 2, 1)$, $\tau = (2, 1, 0.05)$, $\gamma_1 = (1, 0.5, 0.5)$, $\gamma_2 = (-0.5, 0.5, 0.25)$, $\alpha_{b1} = (0.05, 0.01, 0.02)$, $\alpha_{b2} = (-0.05, 0.02, 0.03)$, $\alpha_{\omega1} = 0.7$, and $\alpha_{\omega2} = 0.8$. Here, the specifications of $\Sigma_{\theta}$, $\alpha_b$, and $\alpha_{\omega}$ are designed to simulate patterns of the biomarker profile and their relationship to the competing risks outcome that are similar in nature 
to those observed in the MESA analysis in the next section. The baseline hazards $\lambda_{01}(t), \lambda_{02}(t)$ are set to constants 0.05 and 0.1, respectively. For each subject, we generated a non-informative censoring times $C_i \sim Uni(4,8)$ and let $T_i = min\{T_{i1}^*, T_{i2}^*, C_i\}$ be the observed survival time (possibly censored), where $T_{i1}^*$ and $T_{i2}^*$ are independent conditional on the covariates $X_i$, $b_i$, and $\omega_i$, from models (\ref{sim:hetereq2.1}) and (\ref{sim:hetereq2.2}), respectively, $i=1,\ldots, n$. 
The longitudinal measurements for subject $i$ are assumed missing when $o_{ij}>T_i$.  The median censoring rate is 30.7\%, and the median event rates are 39.6\% and 29.7\% for type 1 and type 2 failures. The average number of longitudinal measurements per subject is 13.

We evaluated the performance of the proposed joint model (\ref{sim:hetereq1.1}) - (\ref{sim:hetereq2.2}), referred to as Model 1, using our developed method and R package ``JMH''  described in Section \ref{Sec:2}, and compared it to a 
classical joint model, referred to as Model 2, which is identical to Model 1 except that equation (\ref{sim:hetereq1.2}) is replaced with homogeneous WS variance ($\sigma_i^2(o_{ij}) \equiv \sigma^2$), using the R-package ``FastJM'' \citep{li2022efficient}.
Table \ref{tab:simNL} summarizes the simulation results, including bias, sample standard deviations of the parameter estimates (SE), average estimated standard errors (Est. SE), and coverage probabilities of the 95\% confidence intervals (CP), based on 300 Monte Carlo replicates with a sample size of 
$n=3000$.  

It is observed from Table \ref{tab:simNL} that the proposed joint model (Model 1) performs well, exhibiting small biases in all parameter and standard error estimates, with CP close to the nominal 95\% level. In contrast, the joint model assuming homogeneous WS variability (Model 2) exhibits substantial biases in most parameters—including $\beta_2$, $\beta_3$, all fixed effects $\gamma_{ij}$, and all association parameters $\alpha_{bij}$ and $\alpha_{\omega}$—as well as in the covariance matrix of the random effects. These biases result in markedly low coverage probabilities for the corresponding confidence intervals, ranging from 5\% to 85.9\%. 

\begin{table}
\begin{center}
\def~{\hphantom{0}}
\caption{Comparison of the bias, standard error (SE),
estimated standard error (Est. SE), and coverage probability (CP) 
between the proposed joint model with heterogeneous WS variability (Model 1) and a classical joint model 
with homogeneous WS variability (Model 2) 
for the longitudinal outcome ($n=3000$) }
\resizebox{14cm}{!}{\begin{tabular}{lrrcccrccc}
\hline \hline \multicolumn{2}{c}{} &\multicolumn{4}{c}{Model 1 (heterogeneous WS variability)}
&\multicolumn{4}{c}{Model 2 (homogeneous WS variability)} \\ \cline{3-10}
Parameter &True &Bias &SE &Est. SE &CP (\%) &Bias &SE &Est. SE &CP (\%) \\
\hline
$Longitudinal$ &  &    & &  &  &    & & & \\
\  Fixed effects &  &   & &   &  &    & & & \\
\ \ Mean trajectory &  &   & &   &  &    & & & \\
\ \ \ $\beta_{0}$ &5 &-0.008 &0.101 &0.098 &94.9 &-0.019 &0.114 &0.119 &95.6 \\
\ \ \ $\beta_{1}$ &1.5 &0.017 &0.152 &0.145 &94.3 &0.026 &0.163 &0.154 &93.3 \\
\ \ \ $\beta_{2}$ &2 &0.001 &0.138 &0.136 &94.3 &\textbf{0.028} &\textbf{0.217} &\textbf{0.136} &\textbf{77.4} \\
\ \ \ $\beta_{3}$ &1 &$<$0.001 &0.104 &0.105 &94.3 &\textbf{0.030} &\textbf{0.144} &\textbf{0.098} &\textbf{80.8} \\
\ \ WS variability &  &   & &   &  &    & & & \\
\ \ \ $\tau_{0}$ &2 &-0.002 &0.034 &0.033 &94.6 &- &- &- & - \\
\ \ \ $\tau_{1}$ &1 &$<$0.001 &0.042 &0.043 &94.9 &- &- &- & -  \\
\ \ \ $\tau_{2}$ &1 &0.003 &0.045 &0.045 &96.0 &- &- &- & -  \\
\ \ \ $\tau_{3}$ &0.05 &$<$0.001 &0.029 &0.029 &95.6 &- &- &- & -  \\
$Competing \ risks $&  &  & &    &  &    & & & \\
\  Fixed effects &  &    & &  &  &    & & & \\
\ \ $\gamma_{11}$ &1 &0.015 &0.069 &0.069 &93.6 &\textbf{-0.049} &0.067 &0.065 &\textbf{85.9} \\
\ \ $\gamma_{12}$ &0.5 &0.005 &0.053 &0.053 &95.6 &\textbf{-0.066} &0.060 &0.052 &\textbf{72.4} \\
\ \ $\gamma_{13}$ &0.5 &0.003 &0.019 &0.019 &94.9 &\textbf{-0.055} &0.028 &0.019 &\textbf{27.3} \\
\ \ $\gamma_{21}$ &-0.5 &-0.003 &0.078 &0.081 &96.0 &\textbf{-0.065} &0.083 &0.078 &\textbf{86.5} \\
\ \ $\gamma_{22}$ &0.5 &-0.001 &0.061 &0.063 &97.3 &\textbf{-0.069} &0.073 &0.064 &\textbf{76.1} \\
\ \ $\gamma_{23}$ &0.25 &0.002 &0.021 &0.021 &96.6 &\textbf{-0.050} &0.029 &0.021 &\textbf{33.0} \\
\ Association &  &      &  &    & & & & &\\
\ \  $\alpha_{b01}$ &0.05 &0.001 &0.025 &0.025 &94.9 &\textbf{-0.031} &\textbf{0.126} &\textbf{0.042} &\textbf{49.5} \\
\ \  $\alpha_{b11}$ &0.01 &0.002 &0.076 &0.078 &95.3 &\textbf{0.091} &\textbf{0.240} &\textbf{0.079} &\textbf{45.5} \\
\ \  $\alpha_{b21}$ &0.02 &-0.004 &0.055 &0.056 &92.6 &\textbf{0.035} &\textbf{0.066} &\textbf{0.034} &\textbf{72.4} \\
\ \  $\alpha_{b02}$ &-0.05 &-0.002 &0.029 &0.029 &96.3 &\textbf{-0.069} &\textbf{0.143} &\textbf{0.059} &\textbf{60.9} \\
\ \  $\alpha_{b12}$ &0.02 &0.002 &0.088 &0.09 &95.3 &\textbf{0.162} &\textbf{0.273} &\textbf{0.111} &\textbf{49.2} \\
\ \  $\alpha_{b22}$ &0.03 &0.004 &0.058 &0.063 &99.0 &\textbf{0.076} &\textbf{0.094} &\textbf{0.044} &\textbf{54.2} \\
\ \  $\alpha_{\omega 1}$ &0.7 &0.01 &0.052 &0.053 &96.3 &- &- &- & -  \\
\ \  $\alpha_{\omega 2}$ &0.8 &0.006 &0.06 &0.061 &95.6 &- &- &- & -   \\
\ Covariance matrix &  &      &  &    & & & & & \\
\ of random effects &  &      &  &    & & & & & \\
\ \  $\Sigma_{11}$ &10  &-0.047 &0.383 &0.38 &93.3 &\textbf{0.726} &\textbf{0.638} &\textbf{0.409} &\textbf{52.9} \\
\ \  $\Sigma_{22}$ &4  &0.041 &0.444 &0.471 &96.0 &\textbf{1.591} &\textbf{1.658} &\textbf{0.759} &\textbf{55.9} \\
\ \  $\Sigma_{33}$ &4  &0.008 &0.344 &0.334 &94.9 &\textbf{2.667} &\textbf{1.12} &\textbf{0.484} &\textbf{5.1} \\
\ \  $\Sigma_{44}$ &1  &-0.005 &0.037 &0.036 &93.6 &- &- &- & - \\
\ \  $\Sigma_{12}$ &3  &-0.006 &0.324 &0.32 &95.6 &\textbf{1.651} &\textbf{0.722} &\textbf{0.446} &\textbf{14.8} \\
\ \  $\Sigma_{23}$ &2  &0.010 &0.284 &0.283 &95.6 &\textbf{-0.715} &\textbf{1.028} &\textbf{0.472} &\textbf{51.5} \\
\ \  $\Sigma_{34}$ &0.5  &-0.002 &0.103 &0.093 &91.9 &- &- &- & - \\
\ \  $\Sigma_{13}$ &2  &-0.014 &0.27 &0.265 &93.3 &\textbf{0.517} &\textbf{0.491} &\textbf{0.367} &\textbf{64.0} \\
\ \  $\Sigma_{24}$ &0.5  &-0.001 &0.118 &0.112 &94.3 &- &- &- & - \\
\ \  $\Sigma_{14}$ &0.5  &-0.010 &0.091 &0.087 &93.9 &- &- &- & - \\
\hline \hline
\end{tabular}
\label{tab:simNL}}
\end{center}
\noindent Note: Large error in confidence interval coverage probability (CP) compared to the 95\% nominal level are highlighted in boldface. Each entry is based on 300 Monte Carlo samples.
\end{table}

{\bf Simulation 2: Impact of random effects correlation and sample size.}
We have conducted additional simulations to examine the impact of random effects correlations and sample size on the performance of our proposed model, compared to the classical joint model that assumes homogeneous WS variance.
To save computation time, we consider linear time evolutions with fewer random effects in the generative model,
where the longitudinal measurements $Y_i(o_{ij})$ were generated from the  mixed-effects multiple location-scale model
\begin{eqnarray}
\label{longmean}
Y_{i}(o_{ij}) & = & \beta_0 + \beta_1 X_{1i} + \beta_2 X_{2i} + \beta_3 X_{3i} + \beta_4 o_{ij} + b_{i} + \sigma_{i}(o_{ij})\epsilon_{i}(o_{ij}),\\
\label{longvar}
\sigma_{i}^2(o_{ij}) &=& \exp(\tau_0 + \tau_1 X_{1i} + \tau_2 X_{2i} + \tau_3 X_{3i} + \tau_4 o_{ij} +\omega_i),
\end{eqnarray}
and the event data were generated from the  proportional cause-specific hazards models:
\begin{eqnarray}
\label{h01}
\lambda_{i1}(t) &=& \lambda_{01}(t) \exp\{\gamma_{11}X_{1i} + \gamma_{12}X_{2i} + \gamma_{13}X_{3i} + \alpha_{b1} b_i + \alpha_{\omega 1} \omega_i\}, \\
\label{h02}
\lambda_{i2}(t) &=& \lambda_{02}(t) \exp\{\gamma_{21}X_{1i} + \gamma_{22}X_{2i} + \gamma_{23}X_{3i} + \alpha_{b2} b_i + \alpha_{\omega 2} \omega_i\},
\end{eqnarray}
where $\theta_i \sim N_{2}(0, \Sigma_{\theta})$ with $\sigma_{b}^2 = 0.5, \sigma_{\omega}^2 = 0.5, \text{ and } \sigma_{b\omega} = \text{cov}(b, \omega)=0.5\rho_{b\omega}$.  Details of the model specifications are provided in Section 5.1 of the Supplementary Material. 

Similar to Simulation 1, results on the performance of the proposed model (Model 1) versus the classical joint model with homogeneous variance (Model 2) are reported in Tables S1 through S12 of the Supplementary Material, for different combinations of correlations $\rho_{b\omega} = {0.75, 0.5, 0.25, 0}$ (high, medium, low, zero) and sample sizes $n={800, 2000, 10000}$. The conclusions are 
consistent with those from Simulation 1, with Model 1 demonstrating negligible bias for all parameters and standard error estimates and CPs close to 95\%, while Model 2 exhibiting substantial bias in several parameters and standard error estimates, leading to significant under-coverage of the associated confidence intervals across all scenarios (see Tables S1 through S12). 

It is worth pointing out that for Model 2, when the correlation $\rho_{b\omega}$ is reduced to zero, the estimation bias in the association parameters $\alpha_{b1}$ and $\alpha_{b2}$ diminishes, resulting in CPs approaching the nominal 95\% level. However, estimates for some other fixed effects parameters, such as $\gamma_{11}$ and $\gamma_{13}$, and the random effects variance $\sigma_b^2$, remain biased (see Table S8 of the Supplementary Material).

{\bf Simulation 3: Impact of non-ignorable monotone missing data on the mixed-effects multiple location-scale model.}
When WS variability is itself of scientific interest, the mixed-effects multiple location-scale model is commonly used for longitudinal data analysis. \citep[among others]{german2022wiser, hedeker2008application}. 
Under the Simulation 2 setting, we have performed an additional simulation to
evaluate the performance of the mixed-effects multiple location-scale model (\ref{longmean}) - (\ref{longvar}) 
for the longitudinal outcome only  using the WiSER method \citep{german2022wiser}, which disregards the survival submodel (\ref{h01}) - (\ref{h02}) and thus does not account for non-ignorable missing data due to terminal events. The results are summarized in Table S13, Section 5.2 of the Supplementary Material, which show that non-ignorable missing data due to informative dropout can induce substantial bias and invalid inferences when fitting the mixed-effects multiple location-scale model alone, and that the proposed joint model offers an effective method to address this issue.

{\bf Simulation 4: Generative joint model with homogeneous WS variance.}
Lastly, we considered a scenario where the generative joint model has homogeneous WS variance. Note that both our proposed joint model with heterogeneous WS variance (Model 1) and the classical joint model with homogeneous variance (Model 2) are valid under this setting. The results are presented in Table S14, Section 5.3 of the Supplementary Material. As expected, both methods exhibit small biases in the estimation of parameters and standard errors. The parameter estimates for Model 1 show slightly larger standard errors compared to those from Model 2, although the differences are minimal.

\subsection{(Prediction Performance)}
\label{sec:sim2}
We conducted additional simulations to compare the prediction performance of our proposed joint model (Model 1) with the classical joint model (Model 2), which ignores heterogeneous WS variability, for dynamic event prediction based on a subject’s history prior to a specified landmark time. Prediction performance was evaluated using two 4-fold cross-validated calibration metrics—the mean absolute prediction error (MAPE) and the Brier score \citep{wu2018quantifying}—along with a cross-validated discrimination metric, the C-index \citep{wolbers2014concordance}, with details provided in Section 4 of the Supplementary Material. To reduce variability from random partitioning, the procedure was repeated using 10 random splits for each dataset, and the scores for each metric were averaged across these partitions. We then computed the overall average scores across all 10 datasets.

\begin{figure}[ht]
    \includegraphics[width=10cm]{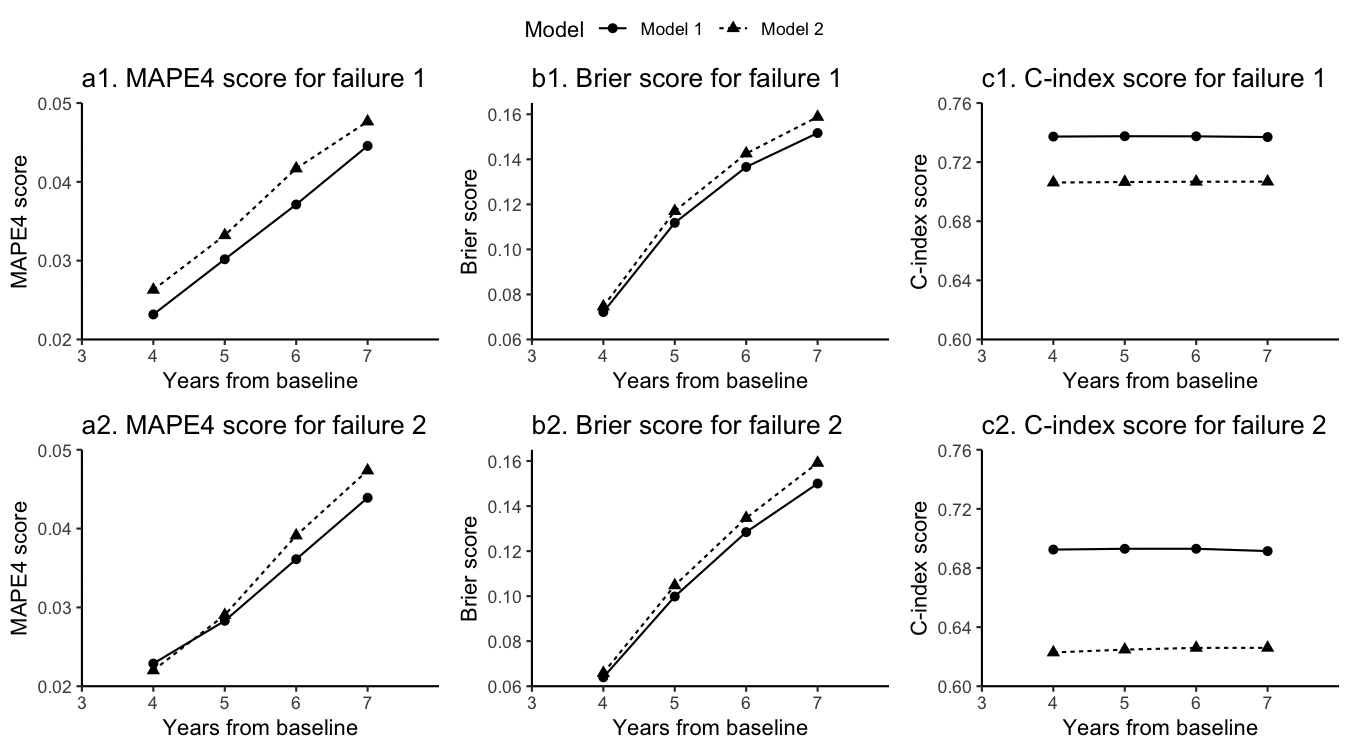}
    \caption{Prediction performance of Models 1 and 2 evaluated by MAPE4, Brier score, and C-index at horizon times $u = (4, 5, 6, 7)$ from landmark time $s = 3$, under the generative model (\ref{sim:hetereq1.1})–(\ref{sim:hetereq2.2}).}
    \label{fig:NLheter}
\end{figure}

{\bf Simulation 5: Generative joint model with non-linear mean evolution and heterogeneous WS variance.}
Under the Simulation 1 setting (\ref{sim:hetereq1.1})–(\ref{sim:hetereq2.2}), we generated 10 datasets of size $n = 3000$. Figure \ref{fig:NLheter} presents the average MAPE4 (Panels a1 and a2), Brier score (Panels b1 and b2), and C-index (Panels c1 and c2) for Models 1 and 2 across various horizon times. Model 1 consistently shows lower MAPE4 and Brier scores and higher C-index values than Model 2, indicating superior predictive performance. Notably, the Brier score appears less sensitive than the other two metrics in differentiating between the models. 

{\bf Simulation 6: Generative joint model with non-linear mean trajectory and homogeneous WS variance.} 
We also considered a simulation scenario, detailed in Section 5.4 of the Supplementary Material, where the longitudinal biomarker follows a non-linear mean trajectory with homogeneous WS variability. In this setting, we compared three joint models:
\begin{enumerate}[label=Model \arabic*:]
    \item Linear mean with heterogeneous WS variability (mis-specified model),  
\item Non-linear mean with homogeneous WS variability (correctly specified model), 
\item Model 2 + heterogeneous WS variability (nest Model 2 as a special case). 
\end{enumerate}
Their prediction performances are presented in Figure S1, Section 5.4 of the Supplementary Material. As expected,  Model 1 exhibited the highest MAPE4 and Brier's score, and the
lowest C-index for failure type 2, indicating that simply incorporating heterogeneous WS
variance is insufficient when a linear mean trajectory is used to model a non-linear time
evolution. Furthermore, Model 3, which includes Model 2 as a special case, demonstrated
nearly identical prediction performance to Model 2 across all three metrics. This is not
surprising, as Model 3 captures biomarker fluctuations along both dimensions.

\subsection{(Scalability)}
\label{sec:simcomp}
\begin{figure}[ht]
    \centering
    \includegraphics[width=10cm]{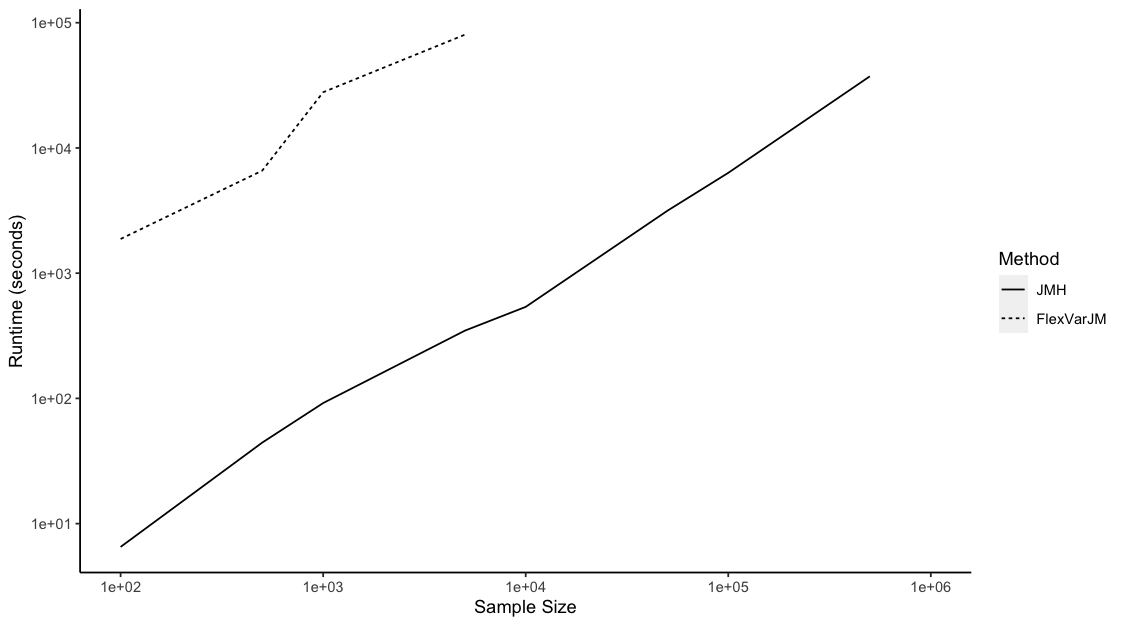}
    \caption{Runtime (seconds) comparison between our R package JMH rendered by the linear scan algorithm and the FlexVarJM package differing sample size $n$ from 100 to 500000. FlexVarJM is an established R package which uses a 2-step quasi Monte Carlo approximation for numerical integration to fit a parametric joint model with a B-spline baseline hazard in the competing risk submodel \citep{courcoul2025location}.}
    \label{fig:Sim_time}
\end{figure}

This section presents a 
simulation study to illustrate the computational efficiency of our joint model package "JMH" as $n$ grows from 100 to 500000. For comparison, we included the runtime of "FlexVarJM", another joint model R package based on \citet{courcoul2025location}. This is, to our knowledge, the only existing model similar to ours that handles both heterogeneous WS variance and competing risks in time-to-event data. All simulations were run on a MacBook Pro with M1 Pro processor and 16GB RAM running MacOS. The data generating mechanism follows the joint model (\ref{longmean}) - (\ref{h02}). Figure \ref{fig:Sim_time} shows the runtime for both packages regarding the estimation procedure as the sample size grows from 100 to 500000 per simulated dataset.  It is observed that the runtime of JMH is at least 100-fold speed-up compared to FlexVarJM. For FlexVarJM, we limited results to $n=5000$ because running the package for larger sample sizes becomes computationally prohibitive in real-time.

\section{Multi-Ethnic Study of Atherosclerosis (MESA)}
\label{Sec::MESA}
The Multi-Ethnic Study of Atherosclerosis (MESA) is a large prospective cohort study of adults aged 45–84 years from diverse racial and ethnic backgrounds who were free of clinically apparent cardiovascular disease at baseline, designed to investigate the prevalence, correlates, and progression of subclinical cardiovascular disease. It enrolled 6,814 participants from four racial/ethnic groups (non-Hispanic White, Black, Chinese, and Hispanic). The initial examination occurred at enrollment (baseline), with up to six total examinations conducted during follow-up. Baseline covariates included age (in years), sex (1 = female, 0 = male), race (1 = non-White, 0 = White), and Framingham cardiovascular risk factors \citep{d2013cardiovascular}, such as total cholesterol (TC), high-density lipoprotein cholesterol (HDL-C), smoking status, and diabetes status.
Systolic and diastolic blood pressures (SBP and DBP, in mmHg) were measured at each examination. Event surveillance, including telephone follow-ups every 9–12 months, was conducted separately to collect and adjudicate cardiovascular disease (CVD) events. After excluding 71 individuals with missing event times or covariate values, our analytic cohort consisted of 6,743 participants with an average of 4.8 SBP measurements per subject (range: 1–6), yielding a total of 32,353 SBP measurements. Among these participants, 379 (5.6\%) experienced heart failure, 1,326 (19.7\%) died, and 5,038 (74.7\%) were right-censored without experiencing either event.

As noted in the Introduction, an open question in the blood pressure variability literature is whether previously reported associations between WS blood pressure variability and cardiovascular outcomes—largely established in clinical trial populations and relatively homogeneous patient groups—extend to a more ethnically diverse and generally healthier population. The Multi-Ethnic Study of Atherosclerosis (MESA), which includes participants from four racial/ethnic groups without overt cardiovascular disease at baseline, provides a unique setting in which to examine the external validity and generalizability of these findings. This broader scientific motivation naturally leads to the following three research questions in the MESA analysis presented in this section:
\begin{enumerate}
\item Does SBP WS variability differ across individuals and/or change over time?
\item Is SBP WS variability associated with the risk of heart failure (HF) and death, after controlling for SBP level?
\item Does incorporating WS variability improve the individual-level prediction performance for HF and death?
\end{enumerate}

These questions are designed to disentangle inferential and predictive objectives while assessing the role of WS variability in both risk association and risk prediction. The first two are inferential in nature and can be answered by explicit modeling of WS variability together with formal statistical inference, including estimation of regression effects, hypothesis testing, and construction of confidence intervals. In contrast, the third question is predictive, shifting both the scientific focus and methodological requirements from parameter inference to the assessment of gains in individual-level risk prediction, encompassing evaluation of discrimination and calibration using appropriate predictive performance metrics.

It is important to note that classical joint models assume homogeneous WS variance and are therefore unable to address any of the three questions considered here. Mixed-effects multiple location–scale models can accommodate heterogeneity in WS variability and thus address the first question; however, they do not account for possible nonignorable missing data due to terminal events such as death, nor do they extend to joint modeling with competing risks, and therefore cannot address the second and third questions. In contrast, the proposed joint model enables a principled analysis that addresses all three questions within a single coherent framework, while overcoming key statistical and computational limitations of common approaches as shown below.

We applied our proposed joint model (Model 1) to the MESA data using the following mixed-effects, local-scale submodel to characterize SBP levels and WS variability:
{\small
\begin{eqnarray}
\label{realdataeq1.1}
        Y_i(o_{ij}) &=& \beta_0 + 
         \sum_{\kappa=1}^2 \beta_{\kappa} B_{\kappa}(o_{ij}) + \beta_{3} \text{Age}_{i} + \beta_{4} \text{Race}_{i} + \beta_{5} \text{Sex}_{i} + b_{i0} +  \sum_{\kappa=1}^2 b_{i\kappa} B_{\kappa}(o_{ij})\\
         &&+ \sigma_{i}(o_{ij})\epsilon_{i}(o_{ij}),\nonumber \\
        \label{realdataeq1.2}
    \sigma_{i}^2(o_{ij}) &=& \exp\left\{\tau_0 + \sum_{\kappa=1}^2 \tau_{\kappa} B_{\kappa}(o_{ij}) + \tau_{3} \text{Age}_{i} + \tau_{4} \text{Race}_{i} + \tau_{5} \text{Sex}_{i} +\omega_i\right\},
\end{eqnarray}
}and the following cause-specific Cox proportional hazards submodel for the
competing-risk events, HF and death:
{\small
\begin{eqnarray}
\label{realdataeq1.3}
    \lambda_{i1}(t) &=& \lambda_{01}(t)\exp(\gamma_{11} \text{Age}_{i} + \gamma_{12}\text{Race}_{i} + \gamma_{13}\text{Sex}_{i} + \gamma_{14}\text{TC}_i \cr
    && + \gamma_{15}\text{HDL-C}_i + \gamma_{16}\text{Smoking}_i + \gamma_{17}\text{Diabetes}_i + \alpha_{b1}^{\top} b_{i} + \alpha_{\omega 1}\omega_i), \\
    \label{realdataeq1.4}
    \lambda_{i2}(t) &=& \lambda_{02}(t)\exp(\gamma_{21} \text{Age}_{i} + \gamma_{22}\text{Race}_{i} + \gamma_{23}\text{Sex}_{i} + \gamma_{24}\text{TC}_i \cr
    && + \gamma_{25}\text{HDL-C}_i + \gamma_{26}\text{Smoking}_i + \gamma_{27}\text{Diabetes}_i + \alpha_{b2}^{\top} b_{i} + \alpha_{\omega 2}\omega_i),
\end{eqnarray}
}where $B_{\kappa}(o_{ij}), \kappa = 1, 2,$ is a 2-degree B-spline basis function to extrapolate a non-linear trajectory, and $\theta_i \equiv (b_{i0}, b_{i1}, b_{i2}, \omega_i)^{\top}$ follows a multivariate normal distribution with mean zero and 
variance-covariance matrix
\begin{eqnarray*}
\Sigma_{\theta} =  \left( {\begin{array}{cccc}
   \sigma_{b_0}^2 & \sigma_{b_{01}} &\sigma_{b_{02}} &\sigma_{b_{0}\omega} \\
   \sigma_{b_{01}} & \sigma_{b_1}^2 & \sigma_{b_{12}} &\sigma_{b_{1}\omega} \\
   \sigma_{b_{02}} & \sigma_{b_{12}} &\sigma_{b_2}^2 &\sigma_{b_{2}\omega} \\
   \sigma_{b_{0}\omega} & \sigma_{b_{1}\omega} &\sigma_{b_{2}\omega} &\sigma_{\omega}^2 \\
  \end{array} } \right).
\end{eqnarray*}
Here, the choice of quadratic spline for modeling the time trend of mean SBP in the submodel \eqref{realdataeq1.1} was guided by inspection of the time plot of mean SBP shown in Figure~S2(a) of the Supplementary Material, which suggests a non-linear temporal trend for which a quadratic spline with a single internal knot appears adequate. A linear spline would likely be insufficient unless additional internal knots were introduced; however, this may be undesirable as it can lead to overfitting and the capture of spurious variation, particularly given the sparsity of SBP measurements in the MESA data (median = 5, IQR = [4, 6]). While a cubic spline with one or two internal knots could also be considered, it introduces additional parameters and complexity, which may offer limited benefit in this setting, as the more parsimonious quadratic spline appears to adequately capture the observed non-linearity in the mean response.

Exploring the temporal trend of the WS variability in submodel~\eqref{realdataeq1.2} is more challenging, as it cannot be directly inferred from a simple time plot of SBP variance. The observed variance of SBP at a given time and covariate level reflects a combination of multiple sources, including between-subject (BS) variability characterized by $b_i$, WS variability captured by the residual component, their potential correlation, and the underlying temporal evolution of SBP. One possible approach is to examine a time plot of the logarithm of the residual variance, obtained after fitting a preliminary submodel~\eqref{realdataeq1.1} under the assumption of homogeneous WS variance. While this assumption may introduce some bias in the estimated residuals, the plot nevertheless provides a useful exploratory tool for assessing the temporal patterns of WS variability. As shown in Figure~S2(b)  of the Supplementary Material, it suggests that a quadratic spline with a single internal knot may provide an adequate specification for the time-varying WS variability in submodel~\eqref{realdataeq1.2}.

The results of our joint analysis are summarized in Table \ref{tab:real2}. For comparison, Table \ref{tab:real2} also presents results from two alternative models: (i) a joint model of SBP level, heart failure (HF), and death assuming homogeneous WS variability (Model 2), which replaces submodel (\ref{realdataeq1.2}) with $\sigma_{i}^2(o_{ij}) \equiv \sigma^2$ and omits the $\omega_i$ term from equations (\ref{realdataeq1.3}) and (\ref{realdataeq1.4}); and (ii) only the mixed-effects multiple location-scale submodel (\ref{realdataeq1.1}) and (\ref{realdataeq1.2}) (Model 3) using the WiSER method \citep{german2022wiser}. 
\begin{table}\footnotesize
\begin{center}
\def~{\hphantom{0}}
\caption{MESA Analysis Results: Comparison of Joint Models with Heterogeneous (Model 1) and Homogeneous (Model 2) WS Variability, and the Mixed-Effects Location-Scale Model (Model 3: WiSER) Abbreviations: SE=standard error; HR=hazard ratio; CI=confidence interval.)}
\label{tab:real2}
\resizebox{14cm}{!}{\begin{tabular}{llllll}
\hline \hline
\multicolumn{1}{l}{} &\multicolumn{2}{c}{Model 1} &\multicolumn{1}{c}{Model 2} &\multicolumn{2}{c}{Model 3 (WiSER)}\\
\hline
Longitudinal outcome &  Mean trajectory&  WS variability &Mean trajectory &Mean trajectory &WS variability\\
({\sl Systolic blood pressure (SBP, mmHg)}) & Estimate (SE) &Estimate (SE) & Estimate (SE) &Estimate (SE) & Estimate (SE)\\ 
\ \ \ Intercept & 128.39 (0.33)*** &4.81 (0.04)*** &128.99 (0.36)*** & 128.69 (0.38)*** &5.04 (0.05)***\\
\ \ \ $B_1(\text{Time})$ &0.78 (0.59) &0.79 (0.09)*** &-0.47 (0.69) &0.10 (0.66) &0.69 (0.12)***\\
\ \ \ $B_2(\text{Time})$ &11.75 (0.63)*** &0.53 (0.09)*** &11.30 (0.63)*** &10.99 (0.58)*** &0.47 (0.17)**\\
\ \ \ Age at baseline &6.66 (0.20)*** &0.37 (0.01)*** &6.73 (0.20)*** &6.68 (0.19)*** &0.32 (0.02)***\\
\ \ \  Race (Non-White/White) &-5.61 (0.39)*** &-0.22 (0.03)*** &-5.74 (0.40)*** &-5.71 (0.38)*** &-0.22 (0.03)***\\
\ \  \ Sex (Female/Male) &-0.41 (0.45) &-0.30 (0.05)*** &-1.06 (0.49)* &-0.81 (0.46) &-0.31 (0.06)***\\
\ \ \ Sex (Female/Male) : $B_1(\text{Time})$ &-3.46 (0.80)*** &0.40 (0.13)** &-3.09 (0.98)** &-3.28 (0.91)*** &0.39 (0.17)*\\
\ \ \ Sex (Female/Male) : $B_2(\text{Time})$ &-4.71 (0.82)*** &0.22 (0.11)* &-5.14 (0.87)*** &-4.85 (0.84)*** &0.21 (0.14)\\
\ \ \ Random effects &  &\\
\ \ \ (variance-covariance matrix) &\multicolumn{2}{l}{Estimate (SE)}  &Estimate (SE) &\multicolumn{2}{l}{Estimate (SE)}\\
\ \ \ \ \ $\sigma_{b_0}^2$ &\multicolumn{2}{l}{238.35 (6.19)***} &256.96 (6.41)*** &259.73 (3.34)***\\
\ \ \ \ \ $\sigma_{b_1}^2$ &\multicolumn{2}{l}{265.31 (18.56)***} &473.06 (23.67)*** &430.62 (21.52)***\\
\ \ \ \ \ $\sigma_{b_2}^2$ &\multicolumn{2}{l}{225.27 (17.55)***} &330.05 (15.40)*** &224.82 (41.03)***\\
\ \ \ \ \  $\sigma_{\omega}^2$ &\multicolumn{2}{l}{0.55 (0.02)***} &N/A &N/A\\
\ \ \ \ \ $\sigma_{b_{01}}$ &\multicolumn{2}{l}{-115.62 (9.10)***} &-155.26 (10.27)*** &-162.25 (7.75)***\\
\ \ \ \ \ $\sigma_{b_{12}}$  &\multicolumn{2}{l}{116.80 (11.82)***} &182.82 (14.33)*** &155.413 (6.45)***\\
\ \ \ \ \  $\sigma_{b_2\omega}$ &\multicolumn{2}{l}{-1.88 (0.45)***} &N/A &N/A\\
\ \ \ \ \ $\sigma_{b_{02}}$ &\multicolumn{2}{l}{-86.67 (7.83)***} &-113.97 (8.21)*** &-114.00 (1.77)***\\
\ \ \ \ \  $\sigma_{b_1\omega}$ &\multicolumn{2}{l}{-2.59 (0.44)***} &N/A &N/A\\
\ \ \ \ \  $\sigma_{b_0\omega}$ &\multicolumn{2}{l}{8.52 (0.28)***} &N/A &N/A\\ \hline
Cause-specific hazard &  &  \\
 ({\sl Heart failure}) & \multicolumn{2}{l}{HR (95\% CI)} & HR (95\% CI) \\
\ \ \ Age at baseline &\multicolumn{2}{l}{2.18 (1.91-2.48)***} &2.10 (1.86-2.38)*** \\
\ \ \ Race (Non-White/White) &\multicolumn{2}{l}{1.21 (0.96-1.51)} &1.22 (0.98-1.52) \\
\ \ \ Sex (Female/Male) &\multicolumn{2}{l}{1.38 (1.09-1.75)**} &1.39 (1.11-1.75)*** \\
\ \ \ TC &\multicolumn{2}{l}{1.00 (1.00-1.00)} &1.00 (1.00-1.00) \\
\ \ \ HDL-C &\multicolumn{2}{l}{0.99 (0.99-1.00)} &0.99 (0.99-1.00) \\
\ \ \ Smoking &\multicolumn{2}{l}{1.28 (1.09-1.50)**} &1.31 (1.12-1.52)*** \\
\ \ \ Diabetes &\multicolumn{2}{l}{1.38 (1.26-1.51)***} &1.38 (1.27-1.51)*** \\
\ \ \ Random effects  &  &  \\
\ \ \ \ \  \ Mean trajectory ($b_i$) &  & \\
\ \ \ \ \  \ \ $\alpha_{b_01}^*$ &\multicolumn{2}{l}{1.01 (0.96, 1.07)} &1.01 (1.01-1.02)**\\
\ \ \ \ \  \ \ $\alpha_{b_11}^*$ &\multicolumn{2}{l}{0.98 (0.84, 1.15)} &1.00 (0.99-1.01)\\
\ \ \ \ \  \ \ $\alpha_{b_21}^*$ &\multicolumn{2}{l}{0.97 (0.91, 1.04)} &0.98 (0.97-1.00)*\\
\ \ \ \ \ \  Residual WS variability ($e_i$) &  & \\
\ \ \ \ \ \ \ $\alpha_{\omega1}$ &\multicolumn{2}{l}{2.26 (1.33-3.84)**} &N/A\\

Cause-specific hazard &  &   \\
 ({\sl Death}) & \multicolumn{2}{l}{HR (95\% CI)} & HR (95\% CI) \\
\ \ \ Age at baseline &\multicolumn{2}{l}{2.98 (2.77-3.20)***} &2.88 (2.70-3.07)*** \\
\ \ \ Race (Non-White/White) &\multicolumn{2}{l}{0.94 (0.83-1.06)} &0.95 (0.85-1.07) \\
\ \ \ Sex (Female/Male) &\multicolumn{2}{l}{1.31 (1.15-1.48)***} &1.30 (1.15-1.46)*** \\
\ \ \ TC &\multicolumn{2}{l}{1.00 (1.00-1.00)*} &1.00 (1.00-1.03)* \\
\ \ \ HDL-C &\multicolumn{2}{l}{1.00 (0.99-1.00)} &1.00 (0.99-1.00) \\
\ \ \ Smoking &\multicolumn{2}{l}{1.51 (1.39-1.64)***} &1.52 (1.41-1.64)*** \\
\ \ \ Diabetes &\multicolumn{2}{l}{1.12 (1.06-1.18)***} &1.12 (1.06-1.18)*** \\
\ \ \ Random effects  &  &  \\
\ \ \ \ \  \ Mean trajectory ($b_i$) &  & \\
\ \ \ \ \  \ \ $\alpha_{b_02}^*$ &\multicolumn{2}{l}{1.02 (0.98, 1.06)} &1.01 (1.01-1.02)***\\
\ \ \ \ \  \ \ $\alpha_{b_12}^*$ &\multicolumn{2}{l}{1.00 (0.88, 1.13)} &1.00 (0.99-1.00)\\
\ \ \ \ \  \ \ $\alpha_{b_22}^*$ &\multicolumn{2}{l}{1.03 (0.97, 1.08)} &1.01 (1.00-1.03)*\\
\ \ \ \ \ \  Residual WS variability ($e_i$) &  & \\
\ \ \ \ \ \ \ $\alpha_{\omega2}$ &\multicolumn{2}{l}{1.88 (1.39-2.53)***} &N/A\\
\hline \hline
\end{tabular}
\label{tab:realnon-linear}}
\end{center}
\noindent * p-value$<$0.05; ** p-value$<$0.01; *** p-value$<$0.001. 

\noindent HR of the reparameterized association parameters for the mean trajectory ($\alpha_{b1}^*, \alpha_{b2}^*$) and residual WS variability ($\alpha_{\omega1}, \alpha_{\omega2}$) are reported. See Remark 1 for the details of reparameterization.

\vspace{2mm}
\end{table}

The first question, concerning whether SBP WS variability differs across individuals and/or changes over time, is addressed by Models 1 and 3 through examination of the estimated coefficients and their corresponding p-values in columns 3 and 6 of Table \ref{tab:real2}, respectively.
The results show that Models 1 and 3 yielded largely consistent conclusions: SBP WS variability differs significantly across individuals and over time in the MESA cohort, with all covariates showing strong associations with WS variability (p-value <0.05) under Model 1.
The only exception is the Sex (Female/Male) x $B_2$(Time) interaction term, for which the effect is statistically significant (p-value < 0.05) under Model 1 but not significant under Model 3. This difference may be attributed to the fact that Model 1 accounts for non-ignorable missing SBP data due to heart failure and death through joint modeling, whereas Model 3 does not.

The second question, concerning whether SBP WS variability is associated with the risk of heart failure (HF) and death after controlling for SBP level, is addressed only by Model 1 through examination of the estimated association parameters and their corresponding p-values. Given the high estimated correlation between $b_{i0}$ and $\omega_i$ ($\rho_{b_0\omega} = 0.74$), we adopted the reparameterization described in Section \ref{Sec:2} (Remark 1, equation \eqref{eq5remark}) and reported the effects of $b_i$, the random effects for SBP level, and $e_i$, the residual WS variability after accounting for $b_i$, on the competing-risk time-to-event outcomes—heart failure and death.
We found that the residual WS variability ($e_i$), beyond what is explained by $b_i$, is strongly associated with increased risks of heart failure (HR = 2.26, 95\% CI: 1.33–3.84) and death (HR = 1.88, 95\% CI: 1.39–2.53). These results demonstrate that the strong association between WS systolic blood pressure variability and cardiovascular disease risk—previously documented mainly in selected clinical populations—extends to a large, diverse, and initially healthy population, highlighting its relevance for cardiovascular disease risk in the general population.

\begin{figure}[ht]
    \includegraphics[width=14cm]{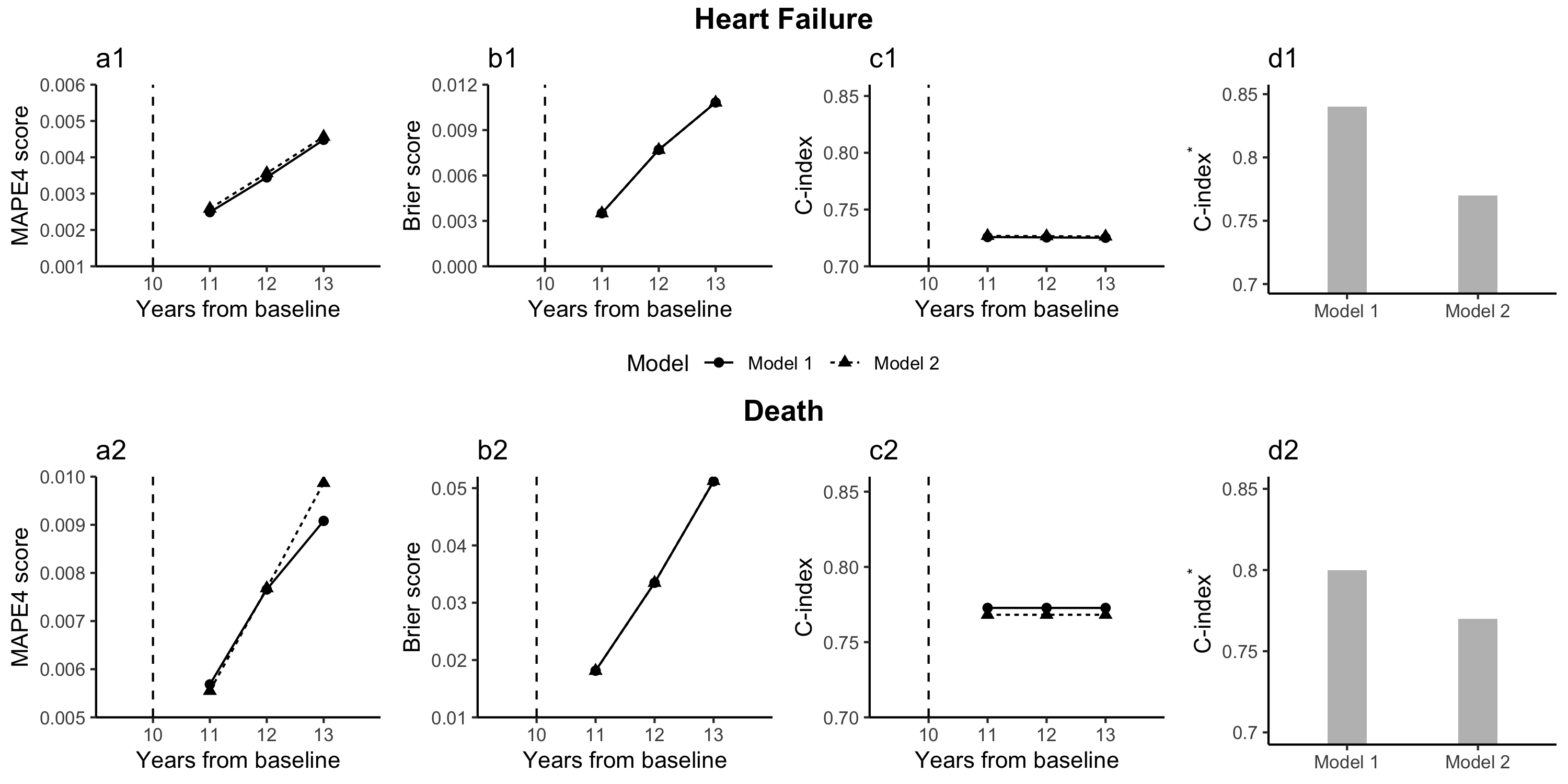}
    \caption{Prediction performance of Model 1 and Model 2 at landmark time $s = 10$ and horizon times $u = (11, 12, 13)$ years, evaluated using time-dependent MAPE4 (Panels a1, a2), Brier score (Panels b1, b2), C-index (Panels c1, c2), and the overall concordance metric $\text{C-index}^*$ (Panels d1, d2), based on 20 random splits. Model 1—the proposed joint model (Equations (\ref{realdataeq1.1})–(\ref{realdataeq1.4}))—assumes heterogeneous WS variability, while Model 2 assumes homogeneous SBP WS variability.}
    \label{fig:MESA_DP}
\end{figure}

With respect to the third question—whether incorporating WS variability improves individual-level prediction performance for heart failure and death—we compared the individual-level risk prediction performance of Models 1 and 2, evaluating both discrimination and calibration using appropriate predictive performance metrics.
Figure \ref{fig:MESA_DP} (panels a–c) presents the prediction performance of both models at the landmark time $s = 10$ and horizon times $u = 11, 12,$ and $13$ years from baseline, evaluated using the average 4-fold cross-validated time-dependent MAPE4, Brier score, and C-index based on the predicted cumulative incidence rate, as described in Section 4 of the Supplementary Material. Additionally, for each competing risk, we evaluated the $\text{C-index}^*$ shown in panel d, based on a prognostic index (PI) derived from the risk score of the corresponding cause-specific proportional hazards model (see equation (S25), Section 4.2 of the supplementary Material).
All performance metrics were computed based on 20 random splits. 

Figure \ref{fig:MESA_DP} shows that, while the first three time-dependent metrics yield similar performance across the two models, the $\text{C-index}^{*}$ results in Figure \ref{fig:MESA_DP}(d) demonstrate a meaningful improvement under the proposed model. In particular, the $\text{C-index}^{*}$ increases from 0.77 to 0.84 for heart failure and from 0.77 to 0.80 for death, indicating improved discriminative ability when WS variability is explicitly modeled. In summary, although predictive gains may not be uniform across all metrics, incorporating heterogeneous WS variability has led to substantive improvements in clinically relevant discrimination.

\begin{figure}[ht]
    \includegraphics[width=12cm]{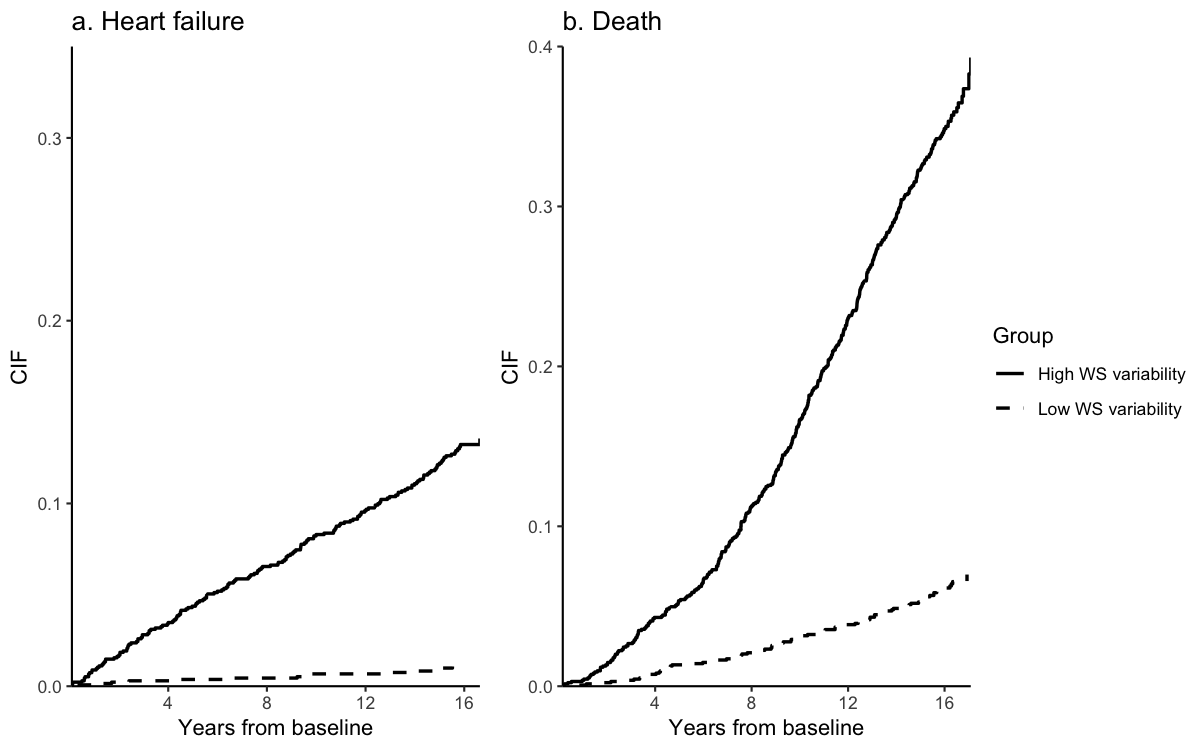}
    \caption{Empirical cause-specific cumulative incidence function (CIF) of the high (top 20\%, solid line) and low (bottom 20\%, dashed line) estimated WS variability of SBP (mmHg) groups for heart failure (left) and death (right) for the MESA cohort.}
    \label{CIF}
\end{figure}

Figure \ref{CIF} further illustrates the substantial impact of SBP WS variability on event outcomes. Subjects were ranked by the mean estimated values of $\log \sigma_i^2(o_{ij})$ across all $o_{ij}$, based on Model 1. We then plotted the cumulative incidence of \textsl{heart failure} (Figure \ref{CIF}a) and \textsl{death} (Figure \ref{CIF}b) for the top 20\% (high SBP WS variability) and bottom 20\% (low SBP WS variability) of the ranked distribution. The results show that participants in the top 20\% exhibited substantially higher cumulative incidence for both clinical events compared to those in the bottom 20\%.

\begin{figure}[ht]
    \includegraphics[width=12cm]{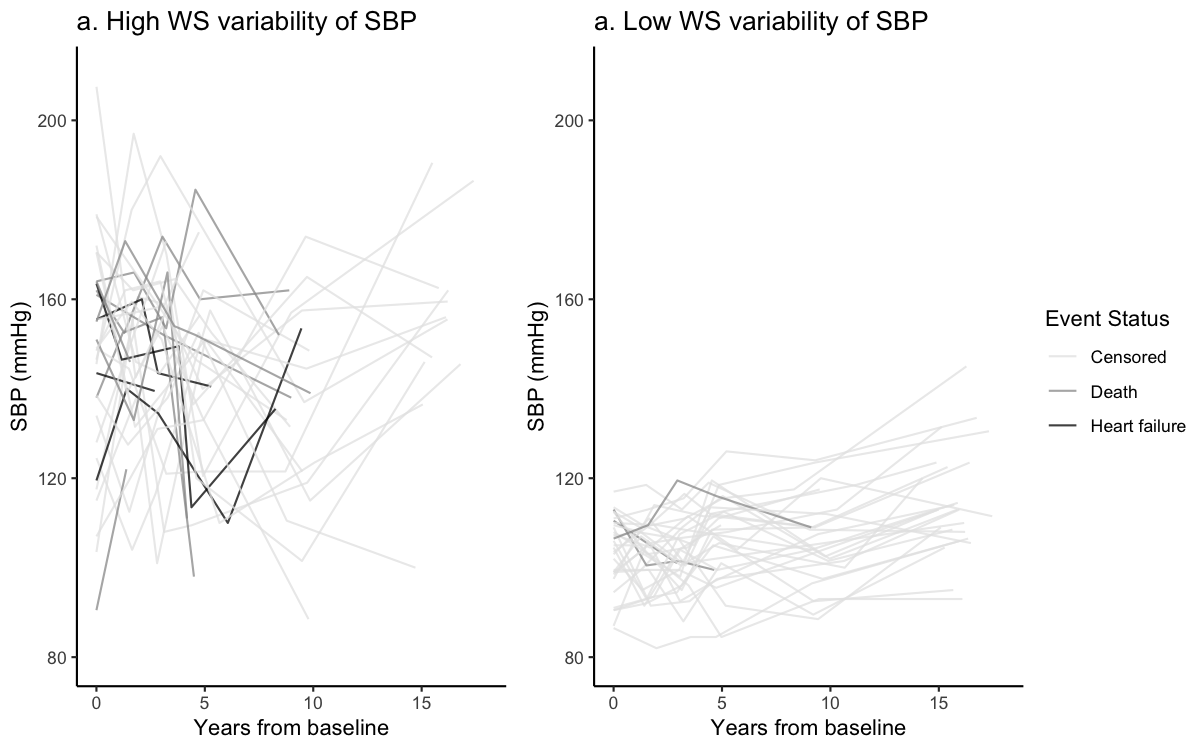}
    \caption{Spaghetti profile plot of 35 randomly selected participants from the high (top 20\%, left) and low (bottom 20\%, right) estimated SBP WS variability groups for the MESA cohort. Shades of gray indicate event status during follow-up.}
    \label{Spaghetti}
\end{figure}

A similar pattern is observed in the spaghetti plots of longitudinal SBP measurements for 35 randomly selected participants from each group over the study period, as shown in Figure \ref{Spaghetti}, where both events are more likely to happen among the high WS variability cohort than the low WS variability cohort. 

\section{Discussion}
\label{discussion}
We have developed a semi-parametric joint model for longitudinal and time-to-event data that addresses heterogeneous WS variability in the longitudinal outcome. Our model introduces enhanced flexibility through several features: it models competing risks in time-to-event data, does not specify the baseline cause-specific hazard in the survival submodel, and accommodates both baseline and time-dependent covariates. Additionally, it supports various latent association structures and incorporates both linear and non-linear temporal trends for the longitudinal biomarker. This distinguishes it from existing models, which typically depend on a parametric baseline hazard assumption and lack these comprehensive features.
We have implemented an expectation-maximization
(EM) algorithm for semi-parametric maximum likelihood estimation, which accommodates time-dependent covariates and various latent association structures, and applied a profile likelihood method for standard error estimation and inference within the proposed joint model. Furthermore, we have optimized the implementation of our joint model when the survival submodel includes only time-independent baseline covariates and shared random effects, scalable to large biobank-scale data. 

When applying our joint model framework, careful consideration of non-linear time trends for both the mean and WS variability is important. Various approaches can be used to model these trends, with splines being a flexible and widely applicable option. Exploratory plots of the mean trajectory can help guide the choice of specification, balancing flexibility and parsimony; as illustrated and detailed in Section~\ref{Sec::MESA}, a quadratic spline with a single internal knot provided a reasonable fit for the MESA data without introducing unnecessary complexity. Modeling WS variability is more challenging, as the observed variance reflects contributions from BS variability, WS residuals, their correlation, and temporal dynamics. A practical exploratory approach is to examine the logarithm of residual variance from a preliminary model assuming homogeneous WS variance, as illustrated in the MESA analysis in Section~\ref{Sec::MESA}. Although this may introduce some bias, it provides a useful tool for selecting an appropriate specification for time-varying WS variability. Overall, we recommend using such exploratory plots to guide model selection, carefully balancing flexibility to capture meaningful trends with parsimony to avoid overfitting, particularly when WS measurements are sparse.

Another important issue is the interpretation of association parameters when both BS and WS random effects are included in the survival submodel—particularly when these components are highly correlated. As illustrated in Section \ref{Sec:2} (Remark 1) and in the MESA data analysis in Section \ref{Sec::MESA}, the reparameterization allows for a clear interpretation of the residual effect of WS variability on event risk after accounting for the BS random effects.

Furthermore, leaving the baseline hazard unspecified provides robustness to misspecification but can yield unstable predictions in small samples or near follow-up boundaries, whereas parametric models enable smoother predictions and extrapolation but depend on correct hazard specification, which is challenging for complex or nonmonotone risk patterns.

Jointly modeling the level and WS variability of a longitudinal biomarker alongside a time-to-event outcome has applications beyond the scenarios considered in this paper. Future work could explore more complex data and model settings, including joint models incorporating multiple longitudinal biomarkers, multivariate time-to-event outcomes, recurrent events, and other types of time-to-event data, such as left-truncated or interval-censored data.

Finally, our scalable implementation and algorithms are developed for time-independent covariates and a shared random-effects association structure, but they open the door to extensions accommodating more general association structures as well as time-dependent covariates via the landmarking framework \citep{van2007dynamic,nicolaie2013dynamic}. As outlined in Remark 2 of the Supplementary Material, evaluating time-dependent covariates and longitudinal association components at prespecified landmark times allows efficient application of the linear-scan algorithms on landmark risk sets. This extension preserves modeling flexibility while substantially broadening the scope of scalable joint modeling, and will be further developed in a sequel paper and R package.

\section{Software}
A user-friendly R package \textbf{JMH} to fit the shared parameter joint model developed in this paper is publicly available at The Comprehensive R Archive Network \url{https://CRAN.R-project.org/package=JMH}.

\section*{Data availability}
The data and samples used in this study were obtained from MESA \url{https://www.mesa-nhlbi.org}, in accordance with their published data access policies, including a written proposal. Access requires submission and approval of a proposal in accordance with MESA data access policies. Details of the submission process are available at \url{https://tools.mesa-nhlbi.org/MESA_Files/publications/Information%20for%20New%20Investigators.pdf}.

\section*{Acknowledgments}
We are grateful to the editor, associate
editor, and three referees for their constructive and insightful feedback that significantly improved our paper. The authors also thank the investigators, staff, and participants of MESA for their valuable contributions. A full list of participating MESA investigators and institutions can be found at \url{https://www.mesa-nhlbi.org}.

\begin{funding}
This research was partially supported by National Institutes of Health (P30 CA-16042, UL1TR000124-02, and P01AT003960, GL; R35GM141798, HZ; R01HG006139, HZ and JJZ; R21HL150374, JJZ; R01DK142026, GL, HZ, and JJZ) and the National Science Foundation (DMS-2054253, HZ and JJZ; IIS-2205441, GL, HZ, and JJZ). This paper has been reviewed and approved by the MESA Publications and Presentations Committee.
\end{funding}
\begin{supplement}
    \stitle{Supplement to ``A joint model of the individual mean and
within-subject variability of a longitudinal outcome with
competing-risk time-to-event outcomes'' (DOI: 10.1214/[provided by typesetter])}
\sdescription{The Supplementary Material contains details of the implementations and algorithms (Sections 1, 2, and 3), dynamic prediction and prediction performance metrics (Section 4), additional simulation results (Section 5), and additional figures (Section 6).}
\end{supplement}

\begin{supplement}
    \stitle{Supplemental code (DOI: 10.1214/[provided by typesetter])}
    \sdescription{R code and accompanying instructions for reproducing all simulation studies and real-data analyses are provided.}
\end{supplement}

\bibliographystyle{imsart-nameyear} 
\bibliography{reference}

\end{document}


\begin{frontmatter}
\title{Supplement to ``A joint model for individual mean and
within-subject variability of a longitudinal outcome with
competing-risk time-to-event outcome''}


\begin{aug}
\author[A,G]{\fnms{Shanpeng}~\snm{Li}\ead[label=e1]{lishanpeng0913@ucla.edu}},
\author[D,F]{\fnms{Daniel}~\snm{S.}~\snm{Nuyujukian}},
\author[E]{\fnms{Robyn}~\snm{L.}~\snm{McClelland}},
\author[D]{\fnms{Peter}~\snm{D.}~\snm{Reaven}},
\author[A,B]{\fnms{Jin}~\snm{Zhou}},
\author[A,C]{\fnms{Hua}~\snm{Zhou}}
\and
\author[A,C]{\fnms{Gang}~\snm{Li}\ead[label=e2]{vli@ucla.edu}}
\address[A]{Department of Biostatistics,
University of California, Los Angeles, USA\printead[presep={,\ }]{e1,e2}}

\address[B]{Department of Medicine,
University of California, Los Angeles, USA}

\address[C]{Department of Computational Medicine,
University of California, Los Angeles, USA}
\address[D]{Phoenix VA Health Care System, Phoenix, USA}

\address[E]{Department of Biostatistics, University of Washington, Seattle, USA}

\address[F]{Department of Epidemiology and Biostatistics, University of Arizona, Tucson, USA}

\address[G]{Department of Computational and Quantitative Medicine, City of Hope, Duarte, USA}

\end{aug}

\end{frontmatter}

\section{The EM algorithm}
\label{sec:S1}
\subsection{The M-step  (equation (8))}
\label{Appen:Mstep}
 It can be shown that the parameters $\beta$, $\Sigma_{\theta}$, as well as $\Lambda_{0k}(t_{kj})$ at the $q_k$ distinct observed type $k$ event times $t_{k1} > \cdots > t_{kq_k}$, have closed form solutions
in the M-step.
Using $E^{(m)}$ to denote $E^{(m)}_{b_i,\omega_i|Y_i, T_i, D_i, \Psi^{(m)}}$ and $a^{\bigotimes 2} = aa^{\top}, \; \forall a \in \mathbb{R}^d$, we have
\begin{eqnarray}
\beta^{(m+1)} & = & \left\{ \sum_{i=1}^n \sum_{j=1}^{n_i} \frac{X_{i}^{(1)\bigotimes 2}(o_{ij})}{\exp\left\{W_{i}^{\top}(o_{ij}) \tau^{(m)}\right\}}E^{(m)}\left[\exp\left\{-V_{i}^{\top}(o_{ij})\omega_i\right\}\right] \right\}^{-1}  \cr
&& \times \left( \sum_{i=1}^n \sum_{j=1}^{n_i} \frac{E^{(m)}\left[\exp\left\{-V_{i}^{\top}(o_{ij})\omega_i\right\}\right]X_{i}^{(1)}(o_{ij})Y_i(o_{ij})}{\exp\left\{W_{i}^{\top}(o_{ij}) \tau^{(m)}\right\}} \right. \cr
&&\left. - \frac{Z_{i}^{\top}(o_{ij}) E^{(m)}\left[b_i\exp\left\{-V_{i}^{\top}(o_{ij})\omega_i\right\}\right] X_{i}^{(1)}(o_{ij})}{\exp\left\{W_{i}^{\top}(o_{ij}) \tau^{(m)}\right\}}\right)
\end{eqnarray}
\begin{eqnarray}
\Sigma_{\theta}^{(m+1)} & = & \frac{1}{n} \sum_{i=1}^n E^{(m)}(\theta_i^{\bigotimes 2}) = \frac{1}{n} \sum_{i=1}^n \left( {\begin{array}{cc}
   E^{(m)}(b_i^{\bigotimes 2}) & E^{(m)}(b_i \omega_i^{\top}) \\
   E^{(m)}(\omega_i b_i^{\top}) & E^{(m)}(\omega_i^{\bigotimes 2}) \\
  \end{array}} \right),
  \end{eqnarray}
{\small
  \begin{eqnarray}
  \label{eq:Lambda0}
\Lambda_{0k}^{(m+1)}(t) & =\!\! & \sum_{l: t_{kl} \leq t} \frac{d_{kl}}{\sum_{r \in R(t_{kl})} \exp\left\{X_r^{(2)\top}(t_{kl}) \gamma_k^{(m)}\right\}E^{(m)}\left[\exp\left\{M_{r}^{\top}(\theta_r, t_{kl})\alpha_k\right\}\right]},
\end{eqnarray}}
\noindent 
where $R(t_{kl})$ is the risk set at time $t_{kl}$, and $d_{kl}$ is the number of type $k$ failures at time $t_{kl}$, for $k=1, \ldots, K.$

The other parameters $\tau$, $\gamma$, $\alpha$, and $\nu$ do not have closed form solutions and we update them using the one-step Newton-Raphson method
\begin{eqnarray*} 
    \tau^{(m+1)} &=& \tau^{(m)} + I_{\tau}^{(m)-1} U_{\tau}^{(m)}, \\
    \gamma_k^{(m+1)} &=& \gamma_k^{(m)} + I_{\gamma_k}^{(m)-1} U_{\gamma_k}^{(m)}, \quad k=1, \ldots, K,\\
    \alpha_k^{(m+1)} &=& \alpha_k^{(m)} + I_{\alpha_k}^{(m)-1} U_{\alpha_k}^{(m)} ,  \quad k=1, \ldots, K,
\end{eqnarray*}
where
\begin{eqnarray}
\label{Appen:tau1}
I_{\tau}^{(m)}&=&\sum_{i=1}^n \sum_{j=1}^{n_i} \frac{1}{2}\exp\left\{-W_{i}^{\top}(o_{ij}) \tau^{(m)}\right\} \cr
&&\times \left(r_{i}^{(m+1)2}(o_{ij}) E^{(m)}\left[\exp\left\{-V_{i}^{\top}(o_{ij})\omega_i\right\}\right]\right. \cr
&& -\left. 2r_{i}^{(m+1)}(o_{ij})Z_{i}^{\top}(o_{ij}) E^{(m)}\left[b_i\exp\left\{-V_{i}^{\top}(o_{ij})\omega_i\right\}\right] \right.\cr 
&&+ \left. \text{tr}\left[Z_{i}^{\bigotimes 2}(o_{ij}) E^{(m)}\left\{b_i^{\bigotimes 2} \exp\left\{-V_{i}^{\top}(o_{ij})\omega_i\right\}\right\}\right]\right)W_{i}^{\bigotimes 2}(o_{ij}), \\
\label{Appen:tau2}
U_{\tau}^{(m)} &=& \sum_{i=1}^n \sum_{j=1}^{n_i}\frac{1}{2}\left[\exp\left\{-W_{i}^{\top}(o_{ij}) \tau^{(m)}\right\} \right.\cr
&&\times \left. \left\{r_{i}^{(m+1)2}(o_{ij})E^{(m)}\left[\exp\left\{-V_{i}^{\top}(o_{ij})\omega_i\right\}\right] \right.\right.\cr
&&- \left.\left. 2r_{i}^{(m+1)}(o_{ij})Z_{i}^{\top}(o_{ij}) E^{(m)}\left[b_i\exp\left\{-V_{i}^{\top}(o_{ij})\omega_i\right\}\right] \right.\right.\cr
&&+ \left.\left. \text{tr}\left(Z_i^{\bigotimes 2}(o_{ij}) E^{(m)}\left[b_i^{\bigotimes 2} \exp\left\{-V_{i}^{\top}(o_{ij})\omega_i\right\}\right]\right)\right\} - 1 \right]W_i(o_{ij}), 
\end{eqnarray}
with $r_{i}^{(m+1)}(o_{ij}) = Y_i(o_{ij}) - X_{i}^{(1)\top}(o_{ij})\beta^{(m+1)}$,
\begin{eqnarray}
\label{Appen:gamma1}
I_{\gamma_k}^{(m)} &=& \sum_{i=1}^n \sum_{t_{kj} \le T_i}\Delta \Lambda_{0k}(t_{kj})^{(m+1)} \exp\left\{X_i^{(2)\top}(t_{kj}) \gamma_k^{(m)}\right\} \cr
&& \times E^{(m)}\left[\exp\left\{M_{i}^{\top}(\theta_i, t_{kj})\alpha_k\right\}\right] X_i^{(2)\bigotimes 2}(t_{kj}), \\
U_{\gamma_k}^{(m)} &=& \sum_{i=1}^n I(D_i=k)X_i^{(2)}(T_i) \cr
\label{Appen:gamma2}
&& - \sum_{i=1}^n \sum_{t_{kj} \le T_i} \Delta \Lambda_{0k}(t_{kj})^{(m+1)} \exp\left\{X_i^{(2)\top}(t_{kj}) \gamma_k^{(m)}\right\} \cr
&&\times E^{(m)}\left[\exp\left\{M_{i}^{\top}(\theta_i, t_{kj})\alpha_k\right\}\right] X_i^{(2)}(t_{kj}), \\
\label{Appen:alpha1}
I_{\alpha_k}^{(m)} &=& \sum_{i=1}^n \sum_{t_{kj} \le T_i}\Delta \Lambda_{0k}(t_{kj})^{(m+1)} \exp\left\{X_i^{(2)\top}(t_{kj}) \gamma_k^{(m)}\right\} \cr
&&\times E^{(m)}\left[M_{i}^{\bigotimes 2}(\theta_i, t_{kj}) \exp\left\{M_{i}^{\top}(\theta_i, t_{kj})\alpha_k\right\}\right], \\
\label{Appen:alpha2}
U_{\alpha_k}^{(m)} &=& \sum_{i=1}^n I(D_i = k) E(\theta_i) \cr
&& - \sum_{i=1}^n \sum_{t_{kj} \le T_i}\Delta \Lambda_{0k}(t_{kj})^{(m+1)}\exp\left\{X_i^{(2)\top}(t_{kj}) \gamma_k^{(m)}\right\} \cr
&&\times E^{(m)}\left[M_{i}(\theta_i, t_{kj}) \exp\left\{M_{i}^{\top}(\theta_i, t_{kj})\alpha_k\right\}\right].
\end{eqnarray}

\subsection{The E-step}
\label{Appen:Estep}
A common method for approximating the integral in (9) is the standard Gauss-Hermite quadrature rule \citep{press2007numerical}, as given below:
{\small
\begin{eqnarray*}
 E^{(m)}\{h(\theta_i)\} &=&\int h(\theta_i)f(\theta_i | Y_i, T_i, D_i, \Psi^{(m)})d\theta_i \cr
 & = &\frac{\int h(\theta_i)f( Y_i, T_i, D_i, \theta_i| \Psi^{(m)})d\theta_i}{f(Y_i, T_i, D_i|\Psi^{(m)})} \cr
&=& \frac{\int h(\theta_i)f(Y_i|\theta_i,\Psi^{(m)})f(T_i, D_i|\theta_i,\Psi^{(m)})f(\theta_i|\Psi^{(m)})d\theta_i}{\int f(Y_i|\theta_i,\Psi^{(m)})f(T_i, D_i|\theta_i,\Psi^{(m)})f(\theta_i|\Psi^{(m)})d\theta_i}\\
& \approx & \frac{\sum_{t_1, t_2, ..., t_q}\pi_t h(\Tilde{\theta}_t^{(m)})f(Y_i \mid \Tilde{\theta}_t^{(m)},\Psi^{(m)})f(T_i, D_i\mid \Tilde{\theta}_t^{(m)},\Psi^{(m)}) f(\Tilde{\theta}_t^{(m)} \mid \Psi^{(m)})\exp(||\theta_t||^2)}{\sum_{t_1, t_2, ..., t_q} \pi_t f(Y_i \mid \Tilde{\theta}_t^{(m)},\Psi^{(m)}) f(T_i, D_i\mid \Tilde{\theta}_t^{(m)},\Psi^{(m)})f(\Tilde{\theta}_t^{(m)} \mid \Psi^{(m)})\exp(||\theta_t||^2)},
\end{eqnarray*}
}

where $\sum_{t_1, t_2, ..., t_q}$ is the shorthand for $\sum_{t_1=1}^{n_q} ... \sum_{t_q=1}^{n_q}$, $n_q$ is the number of quadrature points, $\theta_t = (\theta_{t_1}, \theta_{t_2}, ..., \theta_{t_q})^{\top}$ represents the abscissas with corresponding weights $\pi_t$, $\Tilde{\theta}_t^{(m)} = \sqrt{2} \Sigma_{\theta}^{(m)1/2}\theta_t$ are the re-scaled alternative abscissas, and $\Sigma_{\theta}^{(m)1/2}$ is the square root of $\Sigma_{\theta}^{(m)}$ \citep{elashoff2008joint}. It is well-known that the accuracy of the standard Gauss-Hermite approximation heavily depends on how closely the locations of quadrature points align with the main mass of the integrand. The above approximation method determines the location of quadrature points solely based on the prior distribution 
 $f(\theta_i|\Psi^{(m)})$, potentially resulting in poor approximation since $g(\theta_i) = h(\theta_i)f(\theta_i \mid Y_i, T_i, D_i, \Psi^{(m)})$ concentrates far from zero \citep{rizopoulos2012fast}. Therefore, to achieve satisfactory approximation, a sufficiently large number of quadrature points (at least 20-30 for 2-dimensional integration) is necessary. Furthermore, the required number of points typically increases with the dimension of random effects, making the evaluation of higher-dimensional integrals computationally expensive.

To address this computational bottleneck, \citet{li2022efficient} employed the pseudo-adaptive Gauss-Hermite quadrature rule proposed by \citet{rizopoulos2012fast} to fit a joint model with homogeneous WS variance. This rule centers and scales the integrand just once before the EM iterations begin, and does not relocate quadrature points thereafter. This approximation method not only requires a few quadrature points (3 to 6) but also avoids the computational demands of relocating quadrature points, thereby drastically reducing the computational burden of numerical integration.
However, applying the pseudo-adaptive Gauss-Hermite quadrature rule to our proposed joint model with heterogeneous WS variance requires efficiently fitting a mixed-effects multiple location and scale model in order to pre-calculate the locations of quadrature points. Despite recent methodological developments of this longitudinal model \citep{hedeker2008application,dzubur2020mixwild,williams2021bayesian}, to the best of our knowledge, computationally efficient implementations of these methods have not yet been developed. This makes it infeasible to apply the pseudo-adaptive approximation rule in the joint model proposed in this paper.
As an alternative to circumvent the computational burden of numerical integration, we employ an adaptive Gauss-Hermite quadrature approximation, proposed by \citet{naylor1982applications}, to appropriately center and scale the integrand at each iteration of the EM algorithm. Specifically, the posterior mode $\hat{\theta_i}^{(m)}$ and covariance $\hat{H}_i^{(m)}$ of $f(\theta_i \mid Y_i, T_i, D_i, \Psi^{(m)}), i = 1, \ldots, n,$ will be required for this approximation within each EM iteration. However, there is no explicit form of $\hat{\theta_i}^{(m)}$ and $\hat{H}_i^{(m)}$, so they have to be computed using a numerical optimizer \texttt{optim} with BFGS algorithm \citep{nocedal1999numerical}. The implementation of this approximation rule is detailed below:
\begin{enumerate}
    \item[Step 1.] At the beginning of each EM iteration, use the numerical optimizer \texttt{optim} to obtain the posterior mode $\hat{\theta_i}^{(m)}$ and variance $\hat{H}_i^{(m)}$.
    \item[Step 2.] Update the re-scaled alternative abscissas $\hat{\theta}_t^{(m)} = \hat{\theta}_i^{(m)} +  \sqrt{2} \hat{H}_i^{1/2(m)}\theta_t$, with $\hat{H}_i^{1/2(m)}$ the square root of $\hat{H}_i^{(m)}$.
    \item[Step 3.] Approximate integrals in equation (9) by
\end{enumerate}
{\footnotesize
\begin{eqnarray*}
E^{(m)}\{h(\theta_i)\} &\approx& \frac{\sum_{t_1, t_2, ..., t_q}\pi_t h(\hat{\theta}_t^{(m)})f(Y_i \mid \hat{\theta}_t^{(m)},\Psi^{(m)})f(T_i, D_i\mid \hat{\theta}_t^{(m)},\Psi^{(m)}) f(\hat{\theta}_t^{(m)} \mid \Psi^{(m)})\exp(||\theta_t||^2)}{\sum_{t_1, t_2, ..., t_q} \pi_t f(Y_i \mid \hat{\theta}_t^{(m)},\Psi^{(m)}) f(T_i, D_i\mid \hat{\theta}_t^{(m)},\Psi^{(m)})f(\hat{\theta}_t^{(m)} \mid \Psi^{(m)})\exp(||\theta_t||^2)}.
\end{eqnarray*}
}

Compared to the pseudo-adaptive Gauss-Hermite quadrature rule, the adaptive Gauss-Hermite quadrature approximation is a compromised version since it requires updates to quadrature points rather than pre-calculating the locations only once before an iteration starts. However, the estimation accuracy of integral approximation is still guaranteed with just a few quadrature points (3-6), substantially reducing the computational burden compared to the standard Gauss-Hermite quadrature rule. We adopt the adaptive Gauss-Hermite approximation rule throughout all simulation and application studies in this paper.  

\section{Score function formulas for standard error estimation}
\label{Sec:S2}
Note that $l^{(i)}(\Omega;Y, T, D)$ is the profile likelihood obtained by profiling out the baseline hazard $\lambda_0(.)$ from the observed data likelihood. Calculating its gradient, however, is difficult because there is no explicit expression for the profile likelihood. Here we approximate $\nabla_{\Omega}l^{(i)}(\hat{\Omega};Y, T, D)$ by the derivative of the profile expected complete-data log-likelihood, obtained in the last E-step when the EM algorithm has converged.

Using $\nabla_{\Omega}l^{(i)}$ to denote $\nabla_{\Omega}l^{(i)}(\hat{\Omega};Y, T, D)$ and $E$ to denote $E_{b_i, \omega_i|Y_i, T_i, D_i, \hat{\Psi}}$, the parametric components of the observed score vector in equation (10) is given by
{\small
\begin{eqnarray}
\label{SE:beta}
\nabla_{\beta}l^{(i)} &=& \sum_{j=1}^{n_i} \exp\left\{-W_{i}^{\top}(o_{ij})\tau\right\} \cr
&&\times \left(r_i(o_{ij}) E\left[\exp\left\{-V_{i}^{\top}(o_{ij})\omega_i\right\}\right] - Z_{i}^{\top}(o_{ij}) E\left[b_i \exp\left\{-V_{i}^{\top}(o_{ij})\omega_i\right\}\right]\right) X_{i}^{(1)}(o_{ij}), \\
\label{SE:tau}
\nabla_{\tau}l^{(i)} &=& \sum_{j=1}^{n_i}\frac{1}{2}\left[ \exp\left\{-W_{i}^{\top}(o_{ij}) \tau\right\} \right.\cr
&&\times \left. \left\{r_i^2(o_{ij}) E\left[\exp\left\{-V_{i}^{\top}(o_{ij})\omega_i\right\}\right] \right.\right.\cr
&& -\left.\left. 2r_i(o_{ij}) Z_{i}^{\top}(o_{ij}) E\left[b_i\exp\left\{-V_{i}^{\top}(o_{ij})\omega_i\right\}\right] \right.\right.\cr
&&\left.\left. + \text{tr}\left(Z_{ij}^{\bigotimes 2} E\left[b_i^{\bigotimes 2} \exp\left\{-V_{i}^{\top}(o_{ij})\omega_i\right\}\right]\right)\right\} - 1 \right]W_i(o_{ij}), 
\end{eqnarray}
\begin{eqnarray}
\label{SE:Sigma}
\nabla_{\Sigma}l^{(i)} &=& \frac{1}{2}\left[2\Sigma^{-1}E(\theta_i^{\bigotimes 2}) \Sigma^{-1} - \left\{\Sigma^{-1}E(\theta_i^{\bigotimes 2}) \Sigma^{-1} \circ I\right\} - 2\Sigma^{-1} + \Sigma^{-1} \circ I\right],
\end{eqnarray}
\begin{eqnarray}
\label{SE:gamma}
\nabla_{\gamma_k}l^{(i)} &=& I(D_i = k) \left( X_i^{(2)}(T_i) - \frac{\sum_{r \in R(T_i)} \exp\left\{X_r^{(2)\top}(T_i)\gamma_k\right\}  E\left[\exp\left\{M_{r}^{\top}(\theta_r, T_i)\alpha_k\right\}\right]X_r^{(2)}(T_i)}{\sum_{r \in R(T_i)}\exp\left\{X_r^{(2)\top}(T_i)\gamma_k\right\} E\left[\exp\left\{M_{r}^{\top}(\theta_r, T_i)\alpha_k\right\}\right]} \right) \cr
&& + \sum_{j = 1}^{t_{kj} \leq T_i} \left\{\frac{d_{kj}\sum_{r \in R(t_{kj})} \exp\left\{X_r^{(2)\top}(t_{kj})\gamma_k\right\} E\left[\exp\left\{M_{r}^{\top}(\theta_r, t_{kj})\alpha_k\right\}\right]X_r^{(2)}(t_{kj})}{\left(\sum_{r \in R(t_{kj})} \exp\left\{X_r^{(2)\top}(t_{kj})\gamma_k\right\}E\left[\exp\left\{M_{r}^{\top}(\theta_r, t_{kj})\alpha_k\right\}\right]\right)^2} \right.\cr
&& \times\left. \exp\left\{X_i^{(2)\top}(t_{kj})\gamma_k\right\}E\left[\exp\left\{M_{i}^{\top}(\theta_i, t_{kj})\alpha_k\right\}\right]\right. \cr
&& - \left. \frac{d_{kj}\exp\left\{X_i^{(2)\top}(t_{kj})\gamma_k\right\}E\left[\exp\left\{M_{i}^{\top}(\theta_i, t_{kj})\alpha_k\right\}\right]X_i^{(2)}(t_{kj})}{\sum_{r \in R(t_{kj})} \exp\left\{X_r^{(2)\top}(t_{kj})\gamma_k\right\}E\left[\exp\left\{M_{r}^{\top}(\theta_r, t_{kj})\alpha_k\right\}\right]}\right\},
\end{eqnarray}
\begin{eqnarray}
\label{SE:alpha}
\nabla_{\alpha_k}l^{(i)} &=& I(D_i = k)\left[ E(\theta_i) - \frac{\sum_{r \in R(T_i)} \exp\left\{X_r^{(2)\top}(T_i)\gamma_k\right\}E\left[M_{r}(\theta_r, T_i)\exp\left\{M_{r}^{\top}(\theta_r, T_i)\alpha_k\right\}\right]}{\sum_{r \in R(T_i)} \exp\left\{X_r^{(2)\top}(T_i)\gamma_k\right\}E\left[\exp\left\{M_{r}^{\top}(\theta_r, T_i)\alpha_k\right\}\right]}\right] \cr
&& + \sum_{j = 1}^{t_{kj} \leq T_i}\left\{ \frac{d_{kj}\sum_{r \in R(t_{kj})} \exp\left\{X_r^{(2)\top}(t_{kj})\gamma_k\right\}E\left[M_{r}(\theta_r,t_{kj})\exp\left\{M_{r}^{\top}(\theta_r, t_{kj})\alpha_k\right\}\right]}{\left(\sum_{r \in R(t_{kj})} \exp\left\{X_r^{(2)\top}(t_{kj})\gamma_k\right\}E\left[\exp\left\{M_{r}^{\top}(\theta_r,t_{kj})\alpha_k\right\}\right]\right)^2} \right.\cr
&& \times \left. E\left[\exp\left\{M_{i}^{\top}(\theta_i, t_{kj})\alpha_k\right\}\right]\exp\left\{X_i^{(2)\top}(t_{kj})\gamma_k\right\} \right.\cr
&&  \left. - \frac{d_{kj}\exp\left\{X_i^{(2)\top}(t_{kj})\gamma_k\right\}E\left[M_{i}(\theta_i,t_{kj})\exp\left\{M_{i}^{\top}(\theta_i, t_{kj})\alpha_k\right\}\right]}{\sum_{r \in R(t_{kj})} \exp\left\{X_r^{(2)\top}(t_{kj})\gamma_k\right\}E\left[\exp\left\{M_{r}^{\top}(\theta_r, T_i)\alpha_k\right\}\right]}\right\}.
\end{eqnarray}
}
\begin{remark}
We note a subtle yet important difference between our formulas for 
$\nabla_{\Omega}l^{(i)}(\hat{\Omega};Y, T, D)$
and those discussed by \citet{hsieh2006joint}. Specifically, our formulas for the score function are calculated by firstly profiling out $\lambda_0(.)$ from the expected complete-data log-likelihood  and then taking the first derivative 
of the resulting profile expected complete-data log-likelihood with respect to the parametric component $\Omega$. In contrast, the score function defined in equation (6) of \citet{hsieh2006joint} is calculated by firstly taking the first derivative of the expected complete-data log-likelihood with respect to the parametric component $\Omega$ and then replacing $\lambda_0(.)$ with the Breslow-type estimator $\hat{\lambda_0}(.)$ afterwards. The latter approach is intuitively invalid because, when taking the derivative, it does not take into account that the profile likelihood also depends on $\Omega$ through $\hat{\lambda_0}(.)$. \citet{hsieh2006joint} showed both theoretically and empirically that this approach suffers from information loss compared to the true Hessian, thus leading to underestimated standard errors. However, our method accounts for the fact that the profile likelihood also depends on  $\Omega$ through $\hat{\lambda_0}(.)$, and thus does not suffer from the same issue and has demonstrated satisfactory performance in our simulation studies.
\end{remark}

\section{Linear scan algorithms for efficient implementation for the shared random effects model}
\label{Appen:Comp}
When the latent association structure of the joint model (1) - (3) is assumed to be shared random effects and the survival covariates are assumed to be time-independent, i.e., $X_i^{(2)}(t) = X_i^{(2)}$, we apply the linear scan algorithms developed by \citet{li2022efficient} to reduce the computational burden appeared in our estimation algorithms under the setting of heterogeneous WS variability. Below we discuss some linear scan algorithms for implementation of the EM steps and standard error estimation.  
\subsection{Linear scan for the E-step}
As discussed in Supplementary Material \ref{Appen:Mstep}, the E-step involves evaluating expected values of multiple $h(\theta_i)$'s at each EM iteration, which requires calculating $f(T_i, D_i\mid \theta_i, \Psi^{(m)})$ across all subjects (See Supplementary Material \ref{Appen:Estep}). Note that $f(T_i, D_i\mid \theta_i, \Psi^{(m)})$ can be rewritten as 
\begin{eqnarray*}
f(T_i, D_i\mid \theta_i, \Psi) & = & \prod_{k=1}^K \left[\Delta\Lambda_{0k}^{(m)}(T_i) \exp^{(m)}\left\{X_i^{(2)\top}(T_i) \gamma_k + M_{i}^{\top}(\theta_i, T_i)\alpha_k\right\}\right]^{I(D_i=k)} \cr
     &&\times  \exp\left[-\sum_{k=1}^K \int_0^{T_i} \exp^{(m)}\left\{X_i^{(2)\top}(t) \gamma_k + M_{i}^{\top}(\theta_i, t)\alpha_k\right\}d\Lambda_{0k}^{(m)}(t)\right] \cr
& = & \prod_{k=1}^K \left\{\Delta\Lambda_{0k}^{(m)}(T_i) \exp^{(m)}(X_i^{(2)\top} \gamma_k + \alpha_{bk}^{\top} b_i + \alpha_{\omega k}^{\top} \omega_i)\right\}^{I(D_i=k)}  \cr
&& \quad \times \exp \left\{-\sum_{k=1}^K \Lambda_{0k}^{(m)}(T_i) \exp^{(m)}(X_i^{(2)\top} \gamma_k + \alpha_{bk}^{\top} b_i + \alpha_{\omega k}^{\top} \omega_i) \right\}.
\end{eqnarray*}
For each subject $i$, calculating $\Lambda_{0k}(T_i)$ would involve $O(n)$ operations if a global search is performed to find an interval of two adjacent uncensored event times that contains $T_i$. Consequently, calculating all $\Lambda_{0k}(T_i)$'s will require $O(n^2)$ operations. Taking advantage of the fact that $\Lambda_{0k}(t)$ is a right-continuous and non-decreasing step function, we define the following a linear scan map 
{\small
\begin{eqnarray}
\label{eqn:Lscan}
\{\Lambda_{0k}^{(m)}(t_{k1}), \Lambda_{0k}^{(m)}(t_{k2}), \ldots \Lambda_{0k}^{(m)}(t_{kq_k}) \} \mapsto \{\Lambda_{0k}^{(m)}(T_{(1)}), \Lambda_{0k}^{(m)}(T_{(2)}), \ldots, \Lambda_{0k}^{(m)}(T_{(n)})\},
\end{eqnarray}
}
where $t_{k1} > \cdots > t_{kq_k}$ are scanned forward from the largest to the smallest, and for each $t_{kj}$, only a subset of the ranked observation times $T_{(i)}$ are scanned forward to calculate  $\Lambda_{0k}^{(m)}(T_{(i)})$ as follows
\begin{equation*}
  \Lambda_{0k}^{(m)}(T_{(i)}) =
    \begin{cases}
     \Lambda_{0k}^{(m)}(t_{k1}) , & \text{if $T_{(i)} \geq t_{k1}$,}\\
      \Lambda_{0k}^{(m)}(t_{k(j+1)}), & \text{if $T_{(i)} \in [t_{k(j+1)}, t_{kj}), $ for some $j\in\{1,\ldots, q_k-1 \}$},\\
      0, & \text{$T_{(i)}<t_{kq_k}$}.
    \end{cases}       
\end{equation*}
Consequently, the entire algorithm for calculating all $\Lambda_{0k}(T_i)$'s costs only $O(n)$ operations since the scanned $T_{(i)}$'s for different $t_{kj}$'s do not overlap.
  
\subsection{Linear risk set scan for the M-step}
Multiple quantities in (\ref{Appen:gamma1})-(\ref{Appen:alpha2}) including the cumulative baseline hazard functions involve aggregating information over the risk set $R(t_{kj}) = \{r: T_r \geq t_{kj}\}$ at each uncensored event time $t_{kj}$, which are further aggregated across all $t_{kj}$'s. All subjects are scanned to determine the risk set $R(t_{kj})$ for all uncensored event times will require $O(n^2)$ operations. Specifically, to update $\Lambda_{0k}^{(m+1)}(t_{kj})$, $\gamma^{(m+1)}_k$ and $\alpha^{(m+1)}_k$, one needs to compute $\sum_{r \in R(t_{kj})} a_r(t_{kj})$, where $a_r(.)$ is any time-dependent quantity defined in equations (\ref{Appen:gamma1})-(\ref{Appen:alpha2}). Note that when $a_r(.)$ is assumed to be time-independent, i.e., $a_r(.) = a_r$, the risk set $R(t_{k(j+1)})$ can be decomposed into two disjoint sets: 
\begin{eqnarray}
\label{Appen:MstepScan}
\sum_{r \in R(t_{k(j+1)})} a_{r} = \sum_{r \in R(t_{kj})} a_{r} +\sum_{\{r: T_{(r)} \in [t_{k(j+1)}, t_{kj})\}}  a_{r},
\end{eqnarray}
where the distinct uncensored event times $t_{k1}> \cdots>t_{kq_k}$ are arranged in a decreasing order. 
it is easy to see that calculating $\sum_{r \in R(t_{kj})} a_{r}$, $j=1, \ldots q_k$, takes $O(n)$ operations when 
$T_{(r)}$'s are scanned backward in time, by following the recursive formula (\ref{Appen:MstepScan}) where the subjects in $R(t_{kj})$ do not need to be scanned to calculate the second term.

\subsection{Linear risk set scan for standard error estimation}
Standard error estimation formula in (10) relies on the observed score vectors from the profile likelihood where the baseline hazards are profiled out. It is seen from equations (\ref{SE:gamma})-(\ref{SE:alpha}) that obtaining the observed score vectors $\nabla_{\gamma_k}l^{(i)}(\hat{\Omega};Y, T, D)$ and $\nabla_{\alpha_k}l^{(i)}(\hat{\Omega};Y, T, D)$ involve aggregating information either over $\{ r\in R(T_i) \}$ or over both $\{r \in R(t_{kj})\}$ and $\{ j: t_{kj} \le T_i\}$,  which can takes either $O(n)$ or $O(n^2)$ operations, respectively, if not optimized. As a result, the empirical Fisher information matrix can take $O(n^3)$ operations as it requires summing up the information across all subjects. Specifically, to calculate the gradient $\nabla_{\gamma_k}l^{(i)}(\hat{\Omega};Y, T, D)$ and $\nabla_{\alpha_k}l^{(i)}(\hat{\Omega};Y, T, D)$, one needs to compute
\begin{eqnarray*}
    B(T_i) = \sum_{j: t_{kj} \le T_i}b_{kj}(t_{kj}), \quad\mbox{for $i=1,..., n$,}
\end{eqnarray*}
where $B(.)$ is a right-continuous non-decreasing step function and $b_{kj}(t_{kj}) = \sum_{r \in R(t_{kj})} a_{r}(t_{kj})$ is any time-dependent quantity defined in equations (\ref{SE:gamma}) - (\ref{SE:alpha}). When $a_r(.)$ is assumed to be time-independent, $b_{kj}(.) = b_{kj}$. Note that $B(t_{k1}), \ldots, B(t_{kq_k})$ can be computed in $O(n)$ operations as one scans through $t_{k1}, \ldots, t_{kq_k}$ backward in time, following the recursive formula (\ref{Appen:MstepScan}). Furthermore, analogous to (\ref{eqn:Lscan}), the following linear scan algorithm can be used to calculate $\{B(T_{(1)}), B(T_{(2)}), \cdots, B(T_{(n)})\}$ from $\{B(t_{k1}),\ldots, B(t_{kq_k})\}$:
\begin{eqnarray*}
\{B(t_{k1}),\ldots, B(t_{kq_k})\} \mapsto \{B(T_{(1)}), B(T_{(2)}), \cdots, B(T_{(n)})\},
\end{eqnarray*}
where for each $t_{kj}$, only a subset of the ranked observation times $T_{(i)}$'s are scanned forward to calculate  $B(T_{(i)})$'s as follows
\begin{equation*}
  B(T_{(i)}) =
    \begin{cases}
      B(t_{k1}), & \text{if $T_{(i)} \geq t_{k1}$,}\\
      B(t_{k(j+1)}), & \text{if $T_{(i)} \in [t_{k(j+1)}, t_{kj}), \; $ for some $j\in\{1,\ldots, q_k-1 \}$,}\\
      0, & \text{otherwise}.
    \end{cases}       
\end{equation*}
Consequently, calculating all $B(T_{(i)})$'s takes $O(n)$ operations.

\begin{remark}[Extension to Scalable Joint Modeling with Time-Dependent Covariates and General Association Structures via Landmarking]
Recall that our general EM algorithm, described in Section~\ref{sec:S1} of the Supplementary Material, accommodates time-dependent covariates and a wide range of latent association structures, but may become computationally intensive for large datasets. The efficient and scalable linear-scan algorithms for the shared random effects model with time-independent covariates described in this section can be extended to accommodate time-dependent covariates and more general latent association structures by integrating joint  models with the landmarking approach \citep{van2007dynamic,nicolaie2013dynamic}. The key idea of landmarking is to condition on a pre-specified landmark time $s$ and restrict model fitting to subjects who remain event-free at $s$. Time-dependent survival covariates and latent association components are then evaluated at $s$, rendering the survival submodel conditionally time-independent and thereby enabling the application of our linear-scan algorithms, as outlined below.

Specifically, for subjects at risk at time $s$, consider the following landmark survival submodel:
\begin{eqnarray}
\label{neweq2}
\lambda_{ik}^{(s)}(t\mid X_i^{(2)}(s), M_i(\theta_i, s))
& = & \lim_{h \to 0}\frac{P\left\{t \le \tilde{T}_i < t + h, \tilde{D}_i = k \mid T_i \ge t, X_i^{(2)}(s), M_i(\theta_i, s)\right\}}{h} \cr
& = &
\lambda_{0k}^{(s)}(t)
\exp\left\{X_i^{(2)\top}(s)\gamma_k + M_i^{\top}(\theta_i, s)\alpha_k\right\},
\\
&& \mbox{for}\quad t > s,\quad  k = 1,\ldots,K, \nonumber
\end{eqnarray}
where $\lambda_{0k}^{(s)}(t)$ denotes an unspecified baseline hazard defined on $(s,\infty)$. Here, $M_i(\theta_i,s)$ represents any functional of the longitudinal process based on data observed up to time $s$, such as the current value or slope, as well as latent process components. Conditional on the landmark time $s$, both $X_i^{(2)}(s)$ and $M_i(\theta_i,s)$ are fixed, enabling efficient fitting of the joint model on the landmark risk set by adapting the linear-scan algorithms developed earlier to the landmarking formulation.

By varying $s$ over a grid of landmark times, this modeling strategy provides a practical and scalable route to incorporate time-dependent survival covariates and flexible latent association structures while retaining the computational advantages of our linear scan algorithms.
\end{remark}

\section{Dynamic Prediction and Prediction performance metrics}
\label{appen:DP}
\subsection{Dynamic prediction}
\label{DynamicP}
Given the longitudinal biomarker history 
$Y_{i^*}^{(s)}=\{Y_{i^*}(o_{{i^*}j}), o_{{i^*}j}\leq s\}$ prior to a landmark time $s>0$ and that an event has yet to happen by time $s$,  the cumulative incidence probability 
$$P_{i^* k}(u, s|\Psi) = \text{Pr}(T_{i^*} \leq u, D_{i^*} = k |T_{i^*} > s, Y_{i^*}^{(s)}, \Psi)$$
for type $k$ failure at a horizon time $u>s$ is obtained by  replacing $\Psi$, $S_{i^*}(.)$, and $CIF_{i^* k}(.)$ 
in equation (11) with their corresponding sample estimates $\hat{\Psi}$, $\hat{S}_{i^*}(.)$, and $\widehat{CIF}_{i^* k}(.)$, respectively, as shown below:
\begin{eqnarray}
\label{competingdy-est}
{\hat P}_{i^* k}(u, s|\widehat\Psi) 
&=& \frac{\int CIF_{i^*k}(u,s|\theta_{i^*}, \widehat\Psi) f(Y_{i^*}^{(s)}|\theta_{i^*}, \widehat\Psi)f(\theta_{i^*} | \widehat\Psi) d\theta_{i^*}}{\int f(Y_{i^*}^{(s)}|\theta_{i^*}, \widehat\Psi)S_{i^*}(s|\theta_{i^*}, \widehat\Psi)f(\theta_{i^*} | \widehat\Psi) d\theta_{i^*}},
\end{eqnarray}
with
{\small
\begin{eqnarray}
\label{eq:surv}
\hat{S}_{i^*}(t) &=& \exp\left[-\sum_{k=1}^K \sum_{l: t_{kl} \leq t}
\exp\left\{X_{i^*}^{(2)\top}(t_{kl}) \hat{\gamma}_k + M_{i^*}^{\top}(\theta_{i^*}, t_{kl})\hat{\alpha}_k\right\}\Delta \hat{\Lambda}_{0k}(t_{kl}) \right],
\end{eqnarray}
\begin{eqnarray}
\label{eq:cif}
  \quad\;\;\;  \widehat{CIF}_{i^* k}(u,s) &=& \sum_{l: s \leq t_{kl} \leq u}\hat{S}_{i^*}(t_{kl})\Delta \hat{\Lambda}_{0k}(t_{kl})\exp\left\{X_{i^*}^{(2)\top}(t_{kl}) \hat{\gamma}_k + M_{i^*}^{\top}(\theta_{i^*}, t_{kl})\hat{\alpha}_k\right\},
\end{eqnarray}}
\begin{eqnarray}
\label{eq:Yi_s}
f(Y_{i^*}^{(s)}|\theta_{i^*}, \widehat \Psi) &=& \prod_{j=1}^{o_{{i^*}j}\leq s}\frac{\exp\left[-\frac{\left\{Y_{i^*}(o_{{i^*}j})-X_{i^*}^{(1)\top}(o_{{i^*}j})\hat{\beta} - Z_{i^*}^{\top}(o_{{i^*}j}) b_{i^*}\right\}^2}{2\exp\left\{W_{i^*}^{\top}(o_{{i^*}j}) \hat{\tau} + V_{i^*}^{\top}(o_{{i^*}j})\omega_{i^*}\right\}}\right]}{\sqrt{2\pi\exp\left\{W_{i^*}^{\top}(o_{{i^*}j}) \hat{\tau} + V_{i^*}^{\top}(o_{{i^*}j})\omega_{i^*}\right\}}},
\end{eqnarray}
and
\begin{eqnarray}
\label{eq:theta_i}
    f(\theta_{i^*} | \widehat\Psi) &=& \frac{1}{\sqrt{(2\pi)^{q}\mid \hat{\Sigma}_{\theta}\mid}}\exp\left(-\frac{1}{2}\theta_{i^*}^{\top} \hat{\Sigma}_{\theta}^{-1} \theta_{i^*}\right),
\end{eqnarray}
where $t_{kl}$ denotes the $l$-th uncensored $k$-th failure time ($l = 1, \ldots, q_k$), with $q_k$ representing the total number of unique uncensored $k$-th failure times, and $\Delta \hat{\Lambda}_{0k}(t_{kl})$ is the jump size of $\hat{\Lambda}_{0k}(t)$ at $t_{kl}$, with $\hat{\Lambda}_{0k}(t)$ defined as the limit of the Breslow-type estimator in Section \ref{Appen:Mstep}, equation \eqref{eq:Lambda0}, obtained upon convergence of the EM algorithm.

\subsection{Prediction performance metrics}
\label{appen:Predmetrics}
To evaluate the prediction performance of a joint model on a given dataset, we computed the 4-fold cross-validated mean absolute prediction error (MAPE4), the Brier score, and the C-index, as described below.
\begin{enumerate}
\item[Step 1.] Randomly partition the whole dataset into $4$ equal-sized disjoint subsets, $\mathcal{D}_n^{(1)}, \ldots,\mathcal{D}_n^{(4)}$. 

\item[Step 2.]
For each 
$l=1,\ldots,4$, designate
$\mathcal{D}_n^{(l)}$ as the validation set and the remaining $3$ subsets, denoted by $\mathcal{D}_n^{(-l)}$, as the training set. The training set $\mathcal{D}_n^{(-l)}$ is used to fit a joint model, and then for each subject $i^*$ in the validation set  $\mathcal{D}_n^{(l)}$, the fitted model is used to perform a dynamic prediction of its risk-$k$ cumulative incidence rate $\hat{P}_{i^*k}(u, s)$ at a horizon time $u$ from some landmark time $s$ using formula (11). 

\item[Step 3.]
Rank the subjects in the validation $\mathcal{D}_n^{(l)}$ according to the predicted risk-$k$ cumulative incidence rate $\hat{P}^{(l)}_{i^*k}(u, s)$, and denote by $Q_{kq}^{(l)}$ the $q^{th}$ quartile group, $q=1,\ldots,4$.  
Define the risk-$k$ mean absolute prediction error (MAPE) for the $q^{th}$ quartile group  $Q_{kq}^{(l)}$ by
\begin{eqnarray}
\label{MAEQ}
    {\text{MAPE4}^{(l)}_{k}}(u, s) = \frac{1}{4}\sum_{q=1}^4 \Bigg| \hat{F}_k(u | s, Q_{kq}^{(l)}) - \frac{\sum I(i^* \in Q_{kq}^{(l)})\hat{P}_{i^*k}(u, s)}{\sum I(i^* \in Q_{kq}^{(l)})}\Bigg|,
\end{eqnarray}
where 
$\hat{F}_k(u | s, Q_{kq}^{(l)}) = \int_s^u \hat{S}(t|s, Q_{kq}^{(l)})d\hat{H}_k(t|s, Q_{kq}^{(l)})$
is the empirical cumulative risk-$k$ incidence in the $q^{th}$ quartile group  $Q_{kq}^{(l)}$,
with $\hat{S}(t|s, Q_{kq}^{(l)})$ being the Kaplan-Meier estimator of the all-risk survival function and $\hat{H}_k(t|s, Q_{kq}^{(l)})$ the Nelson-Aalen estimator of the risk-$k$ cause-specific cumulative hazard function within the $q^{th}$ quartile group. 

\item[Step 4.]
Lastly, the 4-fold cross-validated mean absolute prediction error (MAPE4) for risk-$k$ is defined as
\begin{eqnarray}\label{MAEQ4}
    {\text{MAPE4}_{k}}(u, s) &=& \frac{1}{4}\sum_{l=1}^4 {\text{MAPE}^{(l)}_{k}}(u, s), \quad k=1,2.
\end{eqnarray}
\end{enumerate}
Analogously, we calculated the cross-validated Brier score by replacing equation (\ref{MAEQ}) with the corresponding formulas from \citet{wu2018quantifying}, and computed the cross-validated time-dependent C-index using the formulas from \citet{wolbers2014concordance}, with the time-dependent cumulative incidence probability $P_{ik}(u, s|\Psi)$—defined in Section 2.5—used as the risk score.

In our simulations, we set the landmark time $s=3$ so that all the longitudinal information of the subjects at risk in a validation set is available just prior to time $s$, and we predict the future event probabilities for each type of failure at pre-determined horizon times 
$u=(4, 5, 6, 7)$. 

To assess the discrimination power, we consider two types of risk scores representing the relative risk of experiencing a failure for assessing the discriminative capacity of a joint model. One is the time-dependent cumulative incidence probability $P_{ik}(u, s|\Psi)$ defined in Section 2.5, and the other one is the time-independent prognostic index (PI) \citep{royston2013external}, given by
\begin{eqnarray}
\label{eq:PI}
    \text{PI}_{i k} = \exp\left(X_{i}^{(2)\top} \hat{\gamma}_k + \hat{\theta}_{i}^{\top}\hat{\alpha}_k\right), \quad k=1,\ldots,K,
\end{eqnarray}
where $\hat{\gamma}_k$ and $\hat{\alpha}_k$ are the estimates based on a training dataset $\mathcal{D}_n^{(-l)}$; $\hat{\theta}_{i}$ is the mode of the posterior distribution with respect to $\theta_{i}$, $f(\theta_{i} | Y_{i}, T_{i}, D_{i}, \theta_{i},\hat{\Psi})$, with subject $i$ from a validation dataset $\mathcal{D}_n^{(l)}$. Notably, this risk score, the linear combination of time-independent linear predictors, was used to evaluate the overall prediction performance of our joint model, where the survival fixed effects covariates are all time-independent in our MESA analysis. Compared to $P_{ik}(u, s|\Psi)$, a distinctive feature of $\text{PI}_{i k}$ is that the overall prediction performance can be evaluated without relying on a specific time point, which is also used for our MESA analysis. Analogously, we then calculated the cross-validated C-index score for each of the two risk scores by substituting equation (\ref{MAEQ}) with formulas from \citet{wolbers2014concordance}, namely C-index and $\text{C-index}^*$, respectively. We then computed the 4-fold cross-validated C-index scores similar to equation (\ref{MAEQ4}).

\section{Additional simulations}
\label{sm:addsim}
\subsection{(Simulation 2: Impact of random effects correlation and sample size)}
\label{sm:addheter}
We considered extensive simulation scenarios where both the longitudinal and competing risks outcomes were generated from equations (16) - (19), which are linked together through the shared random effects $\theta_i = (b_i, \omega_i)^{\top}$. Furthermore, $\theta_i \sim N_{2}(0, \Sigma_{\theta})$ with $\sigma_{b}^2 = 0.5, \sigma_{\omega}^2 = 0.5, \text{ and } \sigma_{b\omega} = \text{cov}(b, \omega)=0.5\rho_{b\omega}$, where different levels of correlation $\rho_{b\omega} = (0, 0.25, 0.5, 0.75)$ were considered. $o_{ij}$, $X_{1i}$, $X_{2i}$, $X_{3i}$, $\lambda_{01}(t)$, $\lambda_{02}(t)$ are specified as the same as in the first generative joint model (12) - (15). We simulated non-informative censoring time $C_i \sim Uni(5,10)$ and let $T_i = min\{T_{i1}^*, T_{i2}^*, C_i\}$ be the observed survival time (possibly censored) for subject $i$, where $T_{i1}^*$ and $T_{i2}^*$ are independent conditional on the covariates $X_i$, $b_i$, and $\omega_i$, from models (18) and (19), respectively, $i=1,\ldots, n$. The longitudinal measurements for subject $i$ are assumed missing when $o_{ij}>T_i$.  The median censoring rate is 21\%, and the median rate of event 1 and event 2 is 47\% and 32\%, respectively. The average number of longitudinal measurements per subject is 13. 

We examined the finite sample estimation performance across various correlations and sample sizes $n=800, 2000, 10000$. The results are summarized in Table \ref{tab:simSE1} - \ref{tab:simSE12}. It is observed that our proposed joint model method (Model 1) demonstrates very small bias for all parameters and standard error estimates, and that CPs are close to the 95\% nominal level. On the other hand, ignoring the heterogeneous WS variability (Model 2) induces non-negligible bias in several parameters and standard error estimates, leading to significant under-coverage of the associated confidence intervals. For example, according to Table \ref{tab:simSE6}, the parameter estimates and standard error estimates for the association parameters $\alpha_{b1}$ and $\alpha_{b2}$, and for the random effects variance $\sigma_b^2$ in Model 2 are substantially biased, and their associated confidence interval coverages (20.8\%, 28.4\%, and 78.6\%) are unreasonably lower than the nominal 95\% level. These biases for Model 2 are consistently observed in our additional simulation studies for different combinations of correlations $\rho_{b\omega} = {0.75, 0.5, 0.25, 0}$ (high, medium, low, zero) and sample sizes $n={800, 2000, 10000}$, as reported in Tables \ref{tab:simSE1} through \ref{tab:simSE12}. 
 
It is worth pointing out that for Model 2, when the correlation $\rho_{b\omega}$ is reduced towards zero, the estimation bias in the association parameters $\alpha_{b1}$ and $\alpha_{b2}$ diminishes, resulting in CPs approaching the nominal 95\% level. However, estimates for some other fixed effects parameters, such as $\gamma_{11}$ and $\gamma_{13}$, and the random effects variance $\sigma_b^2$, remain biased (see, e.g., Table \ref{tab:simSE8}). 

\begin{table}[p]
\begin{center}
\def~{\hphantom{0}}
\caption{Comparison of the bias, standard error (SE),
estimated standard error (Est. SE), and coverage probability (CP) 
between the proposed joint model with heterogeneous WS variability (Model 1) and a classical joint model 
with homogeneous WS variability (Model 2) 
for the longitudinal outcome ($n=800, \rho_{b\omega} = 0.75$)}
\resizebox{1.0\textwidth}{!}{\begin{tabular}{lrrcccrccc}
\hline \hline \multicolumn{2}{c}{} &\multicolumn{4}{c}{Model 1 (heterogeneous WS variability)} &\multicolumn{4}{c}{Model 2 (homogeneous WS variability)} \\ \cline{3-10}
Parameter &True &Bias &SE &Est. SE &CP (\%) &Bias &SE &Est. SE &CP (\%) \\
\hline
$Longitudinal$ &  &    & &  &  &    & & & \\
\  Fixed effects &  &   & &   &  &    & & & \\
\ \ Mean &  &   & &   &  &    & & & \\
\ \ \ $\beta_{0}$ &5 &$<$0.001 &0.048 &0.050 &96.8 &0.001 &0.052 &0.066 &99.0 \\
\ \ \ $\beta_{1}$ &1.5 &0.001 &0.064 &0.069 &96.8 &-0.003 &0.069 &0.075 &96.2 \\
\ \ \ $\beta_{2}$ &2 &0.002 &0.060 &0.059 &95.8 &-0.002 &0.066 &0.066 &96.4 \\
\ \ \ $\beta_{3}$ &1 &-0.001 &0.018 &0.018 &94.8 &-0.003 &0.020 &0.020 &94.6 \\
\ \ \ $\beta_{4}$ &2 &$<$0.001 &0.002 &0.002 &94.8 &-0.001 &0.004 &0.003 &\textbf{88.2} \\
\ \ Variance &  &   & &   &  &    & & & \\
\ \ \ $\tau_{0}$ &0.5 &0.001 &0.048 &0.049 &95.2 &- &- &- & - \\
\ \ \ $\tau_{1}$ &0.5 &0.001 &0.063 &0.064 &95.4 &- &- &- & -  \\
\ \ \ $\tau_{2}$ &-0.2 &0.002 &0.055 &0.056 &94.4 &- &- &- & -  \\
\ \ \ $\tau_{3}$ &0.2 &$<$0.001 &0.017 &0.017 &96.4 &- &- &- & -  \\
\ \ \ $\tau_{4}$ &0.05 &$<$0.001 &0.002 &0.002 &96.2 &- &- &- & -  \\

$Competing \ risks $&  &  & &    &  &    & & & \\
\  Fixed effects &  &    & &  &  &    & & & \\
\ \ $\gamma_{11}$ &1 &0.008 &0.116 &0.120 &97.0 &-0.035 &0.113 &0.116 &93.8 \\
\ \ $\gamma_{12}$ &0.5 &0.005 &0.096 &0.100 &96.0 &-0.002 &0.094 &0.097 &95.8 \\
\ \ $\gamma_{13}$ &0.5 &0.004 &0.032 &0.035 &96.6 &-0.009 &0.031 &0.033 &96.2 \\
\ \ $\gamma_{21}$ &-0.5 &-0.009 &0.149 &0.139 &94.6 &0.010 &0.149 &0.138 &94.2 \\
\ \ $\gamma_{22}$ &0.5 &-0.003 &0.120 &0.132 &94.6 &$<$0.001 &0.120 &0.115 &94.8 \\
\ \ $\gamma_{23}$ &0.25 &0.004 &0.041 &0.041 &95.4 &0.003 &0.038 &0.037 &94.2 \\
\ Association &  &      &  &    & & & & &\\
\ \  $\alpha_{b1}$ &0.1 &-0.005 &0.296 &0.299 &96.2 &\textbf{0.308} &0.118 &0.112 &\textbf{30.0} \\
\ \  $\alpha_{b2}$ &-0.1 &-0.016 &0.273 &0.277 &97.0 &\textbf{-0.417} &0.158 &0.168 &\textbf{27.8} \\
\ \  $\alpha_{\omega 1}$ &0.5 &0.015 &0.241 &0.265 &98.0 &- &- &- & -  \\
\ \  $\alpha_{\omega 2}$ &-0.5 &-0.014 &0.249 &0.265 &95.8 &- &- &- & -   \\
\ Covariance matrix &  &      &  &    & & & & & \\
\ of random effects &  &      &  &    & & & & & \\
\ \  $\sigma_b^2$ &0.5  &-0.004 &0.050 &0.047 &92.0 &\textbf{-0.030} &0.051 &0.046 &\textbf{84.6} \\
\ \  $\sigma_{\omega}^2$ &0.5  &-0.003 &0.045 &0.045 &95.0 &- &- &- & - \\
\ \  $\sigma_{b\omega}$ &0.375 &-0.004 &0.038 &0.037 &94.2 &- &- &- & - \\

\hline \hline
\end{tabular}
\label{tab:simSE1}}
\end{center}
\vspace{2mm}
\noindent Note: Large error in confidence interval coverage probability (CP) compared to the 95\% nominal level are highlighted in boldface. Each entry is based on 500 Monte Carlo samples.
\end{table}
\clearpage

\begin{table}
\begin{center}
\def~{\hphantom{0}}
\caption{Comparison of the bias, standard error (SE),
estimated standard error (Est. SE), and coverage probability (CP) 
between the proposed joint model with heterogeneous WS variability (Model 1) and a classical joint model 
with homogeneous WS variability (Model 2)
for the longitudinal outcome ($n=800, \rho_{b\omega} = 0.5$) }
\label{tab:simSE2}
\resizebox{1.0\textwidth}{!}{\begin{tabular}{lrrcccrcccrccc}
\hline \hline \multicolumn{2}{c}{} &\multicolumn{4}{c}{Model 1 (heterogeneous WS variability)}
&\multicolumn{4}{c}{Model 2 (homogeneous WS variability)} \\ \cline{3-10}
Parameter &True &Bias &SE &Est. SE &CP (\%) &Bias &SE &Est. SE &CP (\%)\\
\hline
$Longitudinal$ &  &    & &  &  &    & & & \\
\  Fixed effects &  &   & &   &  &    & & & \\
\ \ Mean trajectory &  &   & &   &  &    & & & \\
\ \ \ $\beta_{0}$ &5 &0.003 &0.050 &0.049 &95.8 &0.004 &0.052 &0.065 &98.6\\
\ \ \ $\beta_{1}$ &1.5 &-0.001 &0.066 &0.070 &95.8 &-0.004 &0.071 &0.075 &95.4 \\
\ \ \ $\beta_{2}$ &2 &0.001 &0.061 &0.060 &95.8 &-0.003 &0.063 &0.066 &96.0\\
\ \ \ $\beta_{3}$ &1 &-0.001 &0.019 &0.018 &95.2 &-0.001 &0.019 &0.020 &95.8\\
\ \ \ $\beta_{4}$ &2 &$<$0.001 &0.002 &0.002 &97.2 &$<$0.001 &0.003 &0.003 &93.2\\
\ \ WS variability &  &   & &   &  &    & & & \\
\ \ \ $\tau_{0}$ &0.5 &0.002 &0.047 &0.050 &95.8 &- &- &- & - \\
\ \ \ $\tau_{1}$ &0.5 &-0.001 &0.063 &0.065 &95.2 &- &- &- & - \\
\ \ \ $\tau_{2}$ &-0.2 &0.005 &0.054 &0.057 &95.2 &- &- &- & - \\
\ \ \ $\tau_{3}$ &0.2 &$<$0.001 &0.017 &0.017 &94.8 &- &- &- & - \\
\ \ \ $\tau_{4}$ &0.05 &$<$0.001 &0.002 &0.002 &96.0 &- &- &- & - \\

$Competing \ risks $&  &  & &    &  &    & & & \\
\  Fixed effects &  &    & &  &  &    & & & \\
\ \ $\gamma_{11}$ &1 &0.008 &0.114 &0.119 &96.4 &-0.045 &0.111 &0.114 &94.4 \\
\ \ $\gamma_{12}$ &0.5 &0.004 &0.096 &0.099 &94.8 &-0.005 &0.094 &0.095 &95.0 \\
\ \ $\gamma_{13}$ &0.5 &0.004 &0.033 &0.034 &95.8 &-0.012 &0.032 &0.033 &93.8 \\
\ \ $\gamma_{21}$ &-0.5 &-0.010 &0.150 &0.139 &94.8 &0.028 &0.149 &0.136 &92.6 \\
\ \ $\gamma_{22}$ &0.5 &-0.002 &0.118 &0.116 &94.8 &-0.002 &0.116 &0.113 &95.0 \\
\ \ $\gamma_{23}$ &0.25 &$<$0.001 &0.039 &0.038 &94.2 &0.004 &0.038 &0.037 &93.8 \\
\ Association &  &      &  &    & & & & &\\
\ \  $\alpha_{b1}$ &0.1 &0.004 &0.143 &0.150 &95.6 &\textbf{0.211} &0.116 &0.124 &\textbf{62.2}\\
\ \  $\alpha_{b2}$ &-0.1 &-0.007 &0.144 &0.154 &96.2 &\textbf{-0.258} &0.145 &0.161 &\textbf{65.8}\\
\ \  $\alpha_{\omega 1}$ &0.5 &$<$0.001 &0.139 &0.141 &95.8 &- &- &- & - \\
\ \  $\alpha_{\omega 2}$ &-0.5 &-0.006 &0.142 &0.159 &96.8 &- &- &- & -  \\
\ Covariance matrix &  &      &  &    & & & & & \\
\ of random effects &  &      &  &    & & & & & \\
\ \  $\sigma_b^2$ &0.5 &-0.005 &0.045 &0.044 &93.8 &\textbf{-0.030} &0.053 &0.045 &\textbf{82.2} \\
\ \  $\sigma_{\omega}^2$ &0.5 &-0.003 &0.031 &0.032 &95.0 &- &- &- & - \\
\ \  $\sigma_{b\omega}$ &0.25 &-0.004 &0.038 &0.040 &96.0 &- &- &- & - \\
\hline \hline
\end{tabular}
}
\end{center}
\noindent Note: Large error in confidence interval coverage probability (CP) compared to the 95\% nominal level are highlighted in boldface. Each entry is based on 500 Monte Carlo samples.
\end{table}
\clearpage

\begin{table}[p]
\begin{center}
\def~{\hphantom{0}}
\caption{Comparison of the bias, standard error (SE),
estimated standard error (Est. SE), and coverage probability (CP) 
between the proposed joint model with heterogeneous WS variability (Model 1) and a classical joint model 
with homogeneous WS variability (Model 2) 
for the longitudinal outcome ($n=800, \rho_{b\omega} = 0.25$)}
\resizebox{1.0\textwidth}{!}{\begin{tabular}{lrrcccrccc}
\hline \hline \multicolumn{2}{c}{} &\multicolumn{4}{c}{Model 1 (heterogeneous WS variability)} &\multicolumn{4}{c}{Model 2 (homogeneous WS variability)} \\ \cline{3-10}
Parameter &True &Bias &SE &Est. SE &CP (\%) &Bias &SE &Est. SE &CP (\%) \\
\hline
$Longitudinal$ &  &    & &  &  &    & & & \\
\  Fixed effects &  &   & &   &  &    & & & \\
\ \ Mean &  &   & &   &  &    & & & \\
\ \ \ $\beta_{0}$ &5 &0.001 &0.051 &0.049 &93.6 &0.003 &0.054 &0.065 &99.2 \\
\ \ \ $\beta_{1}$ &1.5 &$<$0.001 &0.069 &0.070 &95.8 &-0.003 &0.073 &0.075 &96.0 \\
\ \ \ $\beta_{2}$ &2 &-0.003 &0.061 &0.060 &94.4 &-0.004 &0.063 &0.066 &96.0 \\
\ \ \ $\beta_{3}$ &1 &$<$0.001 &0.020 &0.018 &92.6 &-0.001 &0.020 &0.020 &95.0 \\
\ \ \ $\beta_{4}$ &2 &$<$0.001 &0.002 &0.002 &95.6 &$<$0.001 &0.003 &0.003 &90.4 \\
\ \ Variance &  &   & &   &  &    & & & \\
\ \ \ $\tau_{0}$ &0.5 &0.001 &0.049 &0.050 &95.4 &- &- &- & - \\
\ \ \ $\tau_{1}$ &0.5 &0.001 &0.065 &0.065 &94.6 &- &- &- & -  \\
\ \ \ $\tau_{2}$ &0.2 &0.002 &0.055 &0.057 &95.2 &- &- &- & -  \\
\ \ \ $\tau_{3}$ &0.2 &$<$0.001 &0.017 &0.017 &96.0 &- &- &- & -  \\
\ \ \ $\tau_{4}$ &0.05 &$<$0.001 &0.002 &0.002 &97.4 &- &- &- & -  \\

$Competing \ risks $&  &  & &    &  &    & & & \\
\  Fixed effects &  &    & &  &  &    & & & \\
\ \ $\gamma_{11}$ &1 &0.009 &0.116 &0.119 &96.8 &-0.053 &0.111 &0.113 &92.4 \\
\ \ $\gamma_{12}$ &0.5 &0.004 &0.098 &0.099 &94.8 &-0.008 &0.095 &0.095 &94.6 \\
\ \ $\gamma_{13}$ &0.5 &0.004 &0.033 &0.034 &94.8 &-0.014 &0.031 &0.032 &93.4 \\
\ \ $\gamma_{21}$ &-0.5 &-0.013 &0.150 &0.139 &94.4 &0.037 &0.146 &0.135 &92.2 \\
\ \ $\gamma_{22}$ &0.5 &-0.001 &0.116 &0.116 &95.2 &-0.004 &0.114 &0.112 &95.6 \\
\ \ $\gamma_{23}$ &0.25 &$<$0.001 &0.038 &0.038 &94.8 &0.006 &0.037 &0.037 &93.8 \\
\ Association &  &      &  &    & & & & &\\
\ \  $\alpha_{b1}$ &0.1 &-0.005 &0.121 &0.125 &96.0 &\textbf{0.094} &0.118 &0.122 &\textbf{89.0} \\
\ \  $\alpha_{b2}$ &-0.1 &-0.007 &0.134 &0.128 &94.4 &\textbf{-0.126} &0.141 &0.156 &90.6 \\
\ \  $\alpha_{\omega 1}$ &0.5 &0.007 &0.116 &0.116 &95.2 &- &- &- & -  \\
\ \  $\alpha_{\omega 2}$ &-0.5 &-0.010 &0.132 &0.134 &94.2 &- &- &- & -   \\
\ Covariance matrix &  &      &  &    & & & & & \\
\ of random effects &  &      &  &    & & & & & \\
\ \  $\sigma_b^2$ &0.5  &-0.004 &0.044 &0.045 &94.0 &\textbf{-0.028} &0.051 &0.045 &\textbf{87.4} \\
\ \  $\sigma_{\omega}^2$ &0.5  &-0.003 &0.038 &0.040 &96.2 &- &- &- & - \\
\ \  $\sigma_{b\omega}$ &0.125 &$<$0.001 &0.030 &0.031 &95.4 &- &- &- & - \\

\hline \hline
\end{tabular}
\label{tab:simSE3}}
\end{center}
\vspace{2mm}
\noindent Note: Large error in confidence interval coverage probability (CP) compared to the 95\% nominal level are highlighted in boldface. Each entry is based on 500 Monte Carlo samples.
\end{table}
\clearpage
\begin{table}[p]
\begin{center}
\def~{\hphantom{0}}
\caption{Comparison of the bias, standard error (SE),
estimated standard error (Est. SE), and coverage probability (CP) 
between the proposed joint model with heterogeneous WS variability (Model 1) and a classical joint model 
with homogeneous WS variability (Model 2) 
for the longitudinal outcome ($n=800, \rho_{b\omega} = 0$)}
\resizebox{1.0\textwidth}{!}{\begin{tabular}{lrrcccrccc}
\hline \hline \multicolumn{2}{c}{} &\multicolumn{4}{c}{Model 1 (heterogeneous WS variability)} &\multicolumn{4}{c}{Model 2 (homogeneous WS variability)} \\ \cline{3-10}
Parameter &True &Bias &SE &Est. SE &CP (\%) &Bias &SE &Est. SE &CP (\%) \\
\hline
$Longitudinal$ &  &    & &  &  &    & & & \\
\  Fixed effects &  &   & &   &  &    & & & \\
\ \ Mean &  &   & &   &  &    & & & \\
\ \ \ $\beta_{0}$ &5 &0.001 &0.052 &0.048 &94.0 &-0.001 &0.056 &0.065 &98.2 \\
\ \ \ $\beta_{1}$ &1.5 &$<$0.001 &0.073 &0.069 &92.8 &$<$0.001 &0.078 &0.075 &92.8 \\
\ \ \ $\beta_{2}$ &2 &0.001 &0.059 &0.059 &93.8 &-0.001 &0.060 &0.066 &97.4 \\
\ \ \ $\beta_{3}$ &1 &$<$0.001 &0.019 &0.018 &92.8 &$<$0.001 &0.021 &0.020 &93.2 \\
\ \ \ $\beta_{4}$ &2 &$<$0.001 &0.002 &0.002 &94.6 &$<$0.001 &0.003 &0.003 &92.0 \\
\ \ Variance &  &   & &   &  &    & & & \\
\ \ \ $\tau_{0}$ &0.5 &0.002 &0.048 &0.050 &95.8 &- &- &- & - \\
\ \ \ $\tau_{1}$ &0.5 &-0.001 &0.063 &0.065 &96.6 &- &- &- & -  \\
\ \ \ $\tau_{2}$ &-0.2 &0.001 &0.055 &0.057 &97.0 &- &- &- & -  \\
\ \ \ $\tau_{3}$ &0.2 &$<$0.001 &0.017 &0.017 &96.4 &- &- &- & -  \\
\ \ \ $\tau_{4}$ &0.05 &$<$0.001 &0.002 &0.002 &96.2 &- &- &- & -  \\

$Competing \ risks $&  &  & &    &  &    & & & \\
\  Fixed effects &  &    & &  &  &    & & & \\
\ \ $\gamma_{11}$ &1 &0.005 &0.114 &0.118 &97.0 &-0.058 &0.110 &0.112 &91.8 \\
\ \ $\gamma_{12}$ &0.5 &0.005 &0.094 &0.099 &95.4 &-0.007 &0.092 &0.094 &95.4 \\
\ \ $\gamma_{13}$ &0.5 &0.003 &0.033 &0.034 &96.2 &-0.014 &0.032 &0.032 &91.4 \\
\ \ $\gamma_{21}$ &-0.5 &-0.010 &0.149 &0.139 &93.8 &0.044 &0.144 &0.135 &91.0 \\
\ \ $\gamma_{22}$ &0.5 &-0.001 &0.122 &0.116 &94.4 &-0.003 &0.119 &0.112 &94.6 \\
\ \ $\gamma_{23}$ &0.25 &$<$0.001 &0.038 &0.038 &94.6 &0.007 &0.037 &0.037 &94.2 \\
\ Association &  &      &  &    & & & & &\\
\ \  $\alpha_{b1}$ &0.1 &-0.001 &0.113 &0.118 &96.0 &-0.008 &0.115 &0.122 &96.6 \\
\ \  $\alpha_{b2}$ &-0.1 &-0.008 &0.121 &0.122 &94.0 &-0.004 &0.130 &0.156 &98.8 \\
\ \  $\alpha_{\omega 1}$ &0.5 &0.001 &0.110 &0.111 &95.6 &- &- &- & -  \\
\ \  $\alpha_{\omega 2}$ &-0.5 &-0.016 &0.129 &0.127 &93.8 &- &- &- & -   \\
\ Covariance matrix &  &      &  &    & & & & & \\
\ of random effects &  &      &  &    & & & & & \\
\ \  $\sigma_b^2$ &0.5  &-0.006 &0.041 &0.044 &95.6 &-0.034 &0.049 &0.044 &83.2 \\
\ \  $\sigma_{\omega}^2$ &0.5  &-0.003 &0.037 &0.040 &96.2 &- &- &- & - \\
\ \  $\sigma_{b\omega}$ &0 &$<$0.001 &0.029 &0.031 &96.8 &- &- &- & - \\

\hline \hline
\end{tabular}
\label{tab:simSE4}}
\end{center}
\vspace{2mm}
\noindent Note: Large error in confidence interval coverage probability (CP) compared to the 95\% nominal level are highlighted in boldface. Each entry is based on 500 Monte Carlo samples.
\end{table}
\clearpage
\begin{table}[p]
\begin{center}
\def~{\hphantom{0}}
\caption{Comparison of the bias, standard error (SE),
estimated standard error (Est. SE), and coverage probability (CP) 
between the proposed joint model with heterogeneous WS variability (Model 1) and a classical joint model 
with homogeneous WS variability (Model 2) 
for the longitudinal outcome ($n=2000, \rho_{b\omega} = 0.75$)}
\resizebox{1.0\textwidth}{!}{\begin{tabular}{lrrcccrccc}
\hline \hline \multicolumn{2}{c}{} &\multicolumn{4}{c}{Model 1 (heterogeneous WS variability)} &\multicolumn{4}{c}{Model 2 (homogeneous WS variability)} \\ \cline{3-10}
Parameter &True &Bias &SE &Est. SE &CP (\%) &Bias &SE &Est. SE &CP (\%) \\
\hline
$Longitudinal$ &  &    & &  &  &    & & & \\
\  Fixed effects &  &   & &   &  &    & & & \\
\ \ Mean &  &   & &   &  &    & & & \\
\ \ \ $\beta_{0}$ &5 &$<$0.001 &0.028 &0.031 &96.4 &0.001 &0.032 &0.041 &99.0 \\
\ \ \ $\beta_{1}$ &1.5 &0.001 &0.040 &0.043 &96.0 &-0.004 &0.044 &0.047 &96.2 \\
\ \ \ $\beta_{2}$ &2 &-0.004 &0.037 &0.037 &94.6 &-0.006 &0.040 &0.041 &95.8 \\
\ \ \ $\beta_{3}$ &1 &$<$0.001 &0.012 &0.012 &94.2 &-0.002 &0.013 &0.013 &94.8 \\
\ \ \ $\beta_{4}$ &2 &$<$0.001 &0.001 &0.002 &95.8 &-0.001 &0.002 &0.002 &90.8 \\
\ \ Variance &  &   & &   &  &    & & & \\
\ \ \ $\tau_{0}$ &0.5 &0.001 &0.030 &0.031 &95.2 &- &- &- & - \\
\ \ \ $\tau_{1}$ &0.5 &-0.002 &0.039 &0.040 &95.0 &- &- &- & -  \\
\ \ \ $\tau_{2}$ &-0.2 &-0.002 &0.034 &0.035 &94.0 &- &- &- & -  \\
\ \ \ $\tau_{3}$ &0.2 &0.001 &0.010 &0.011 &96.2 &- &- &- & -  \\
\ \ \ $\tau_{4}$ &0.05 &$<$0.001 &0.001 &0.001 &96.0 &- &- &- & -  \\

$Competing \ risks $&  &  & &    &  &    & & & \\
\  Fixed effects &  &    & &  &  &    & & & \\
\ \ $\gamma_{11}$ &1 &$<$0.001 &0.073 &0.074 &94.8 &-0.040 &0.071 &0.072 &91.0 \\
\ \ $\gamma_{12}$ &0.5 &-0.001 &0.064 &0.062 &93.6 &-0.006 &0.063 &0.060 &93.8 \\
\ \ $\gamma_{13}$ &0.5 &0.003 &0.022 &0.021 &94.4 &-0.009 &0.022 &0.021 &90.2 \\
\ \ $\gamma_{21}$ &-0.5 &-0.007 &0.086 &0.087 &94.6 &0.013 &0.086 &0.086 &95.4 \\
\ \ $\gamma_{22}$ &0.5 &$<$0.001 &0.072 &0.072 &95.0 &0.002 &0.072 &0.071 &94.4 \\
\ \ $\gamma_{23}$ &0.25 &-0.001 &0.024 &0.023 &94.0 &0.001 &0.024 &0.023 &93.6 \\
\ Association &  &      &  &    & & & & &\\
\ \  $\alpha_{b1}$ &0.1 &0.003 &0.139 &0.148 &96.8 &\textbf{0.318} &0.072 &0.077 &\textbf{1.6} \\
\ \  $\alpha_{b2}$ &-0.1 &0.005 &0.142 &0.148 &96.0 &\textbf{-0.406} &0.094 &0.103 &\textbf{1.8} \\
\ \  $\alpha_{\omega 1}$ &0.5 &$<$0.001 &0.131 &0.142 &96.6 &- &- &- & -  \\
\ \  $\alpha_{\omega 2}$ &-0.5 &-0.014 &0.151 &0.151 &96.2 &- &- &- & -   \\
\ Covariance matrix &  &      &  &    & & & & & \\
\ of random effects &  &      &  &    & & & & & \\
\ \  $\sigma_b^2$ &0.5  &-0.001 &0.026 &0.027 &96.2 &\textbf{-0.030} &0.032 &0.028 &\textbf{76.2} \\
\ \  $\sigma_{\omega}^2$ &0.5  &-0.003 &0.024 &0.024 &95.2 &- &- &- & - \\
\ \  $\sigma_{b\omega}$ &0.375 &$<$0.001 &0.020 &0.021 &96.0 &- &- &- & - \\

\hline \hline
\end{tabular}
\label{tab:simSE5}}
\end{center}
\vspace{2mm}
\noindent Note: Large error in confidence interval coverage probability (CP) compared to the 95\% nominal level are highlighted in boldface. Each entry is based on 500 Monte Carlo samples.
\end{table}
\clearpage

\begin{table}[p]
\begin{center}
\def~{\hphantom{0}}
\caption{Comparison of the bias, standard error (SE),
estimated standard error (Est. SE), and coverage probability (CP) 
between the proposed joint model with heterogeneous WS variability (Model 1) and a classical joint model 
with homogeneous WS variability (Model 2)
 ($n=2000, \rho_{b\omega} = 0.5$) }
\resizebox{14cm}{!}{\begin{tabular}{lrrcccrccc}
\hline \hline \multicolumn{2}{c}{} &\multicolumn{4}{c}{Model 1 (heterogeneous WS variability)}
&\multicolumn{4}{c}{Model 2 (homogeneous WS variability)} \\ \cline{3-10}
Parameter &True &Bias &SE &Est. SE &CP (\%) &Bias &SE &Est. SE &CP (\%)\\
\hline
$Longitudinal$ &  &    & &  &  &    & & & \\
\  Fixed effects &  &   & &   &  &    & & & \\
\ \ Mean trajectory &  &   & &   &  &    & & & \\
\ \ \ $\beta_{0}$ &5 &$<$0.001 &0.031 &0.030 &94.8 &0.001 &0.035 &0.041 &97.4\\
\ \ \ $\beta_{1}$ &1.5 &0.001 &0.042 &0.044 &95.4 &-0.002 &0.047 &0.047 &94.8\\
\ \ \ $\beta_{2}$ &2 &-0.005 &0.037 &0.038 &95.0 &-0.006 &0.040 &0.041 &94.4\\
\ \ \ $\beta_{3}$ &1 &$<$0.001 &0.012 &0.012 &94.6 &-0.001 &0.013 &0.013 &94.8\\
\ \ \ $\beta_{4}$ &2 &$<$0.001 &0.002 &0.002 &94.6 &$<$0.001 &0.002 &0.002 &89.0\\
\ \ WS variability &  &   & &   &  &    & & & \\
\ \ \ $\tau_{0}$ &0.5 &0.001 &0.030 &0.031 &96.0 &- &- &- & -\\
\ \ \ $\tau_{1}$ &0.5 &-0.003 &0.041 &0.041 &94.6 &- &- &- & - \\
\ \ \ $\tau_{2}$ &-0.2 &-0.003 &0.035 &0.035 &96.0 &- &- &- & - \\
\ \ \ $\tau_{3}$ &0.2 &$<$0.001 &0.011 &0.011 &94.4 &- &- &- & - \\
\ \ \ $\tau_{4}$ &0.05 &$<$0.001 &0.001 &0.001 &95.8 &- &- &- & - \\

$Competing \ risks $&  &  & &    &  &    & & & \\
\  Fixed effects &  &    & &  &  &    & & & \\
\ \ $\gamma_{11}$ &1 &-0.001 &0.072 &0.074 &95.8 &\textbf{-0.053} &0.069 &0.071 &\textbf{87.8} \\
\ \ $\gamma_{12}$ &0.5 &-0.001 &0.064 &0.062 &93.6 &-0.009 &0.062 &0.060 &93.4\\
\ \ $\gamma_{13}$ &0.5 &0.002 &0.022 &0.021 &94.6 &\textbf{-0.012} &0.021 &0.020 &\textbf{88.4}\\
\ \ $\gamma_{21}$ &-0.5 &-0.006 &0.085 &0.087 &95.8 &0.032 &0.085 &0.085 &92.8\\
\ \ $\gamma_{22}$ &0.5 &$<$0.001 &0.073 &0.072 &93.8 &$<$0.001 &0.071 &0.070 &94.2\\
\ \ $\gamma_{23}$ &0.25 &-0.001 &0.024 &0.023 &93.8 &0.003 &0.023 &0.023 &93.8 \\
\ Association &  &      &  &    & & & & &\\
\ \  $\alpha_{b1}$ &0.1 &0.003 &0.088 &0.092 &97.2 &\textbf{0.207} &0.069 &0.076 &\textbf{20.8}\\
\ \  $\alpha_{b2}$ &-0.1 &0.003 &0.092 &0.094 &96.6 &\textbf{-0.247} &0.089 &0.100 &\textbf{28.4}\\
\ \  $\alpha_{\omega 1}$ &0.5 &-0.004 &0.085 &0.087 &95.0 &- &- &- & - \\
\ \  $\alpha_{\omega 2}$ &-0.5 &-0.007 &0.100 &0.097 &95.2 &- &- &- & -\\
\ Covariance matrix &  &      &  &    & & & & & \\
\ of random effects &  &      &  &    & & & & & \\
\ \  $\sigma_b^2$ &0.5 &-0.001 &0.029 &0.027 &95.2 &\textbf{-0.028} &0.032 &0.028 &\textbf{78.6}\\
\ \  $\sigma_{\omega}^2$ &0.5 &-0.002 &0.025 &0.025 &94.6 &- &- &- & -\\
\ \  $\sigma_{b\omega}$ &0.25 &$<$0.001 &0.021 &0.020 &93.8 &- &- &- & -\\
\hline \hline
\end{tabular}
}
\label{tab:simSE6}
\end{center}
\noindent Note: Large error in confidence interval coverage probability (CP) compared to the 95\% nominal level are highlighted in boldface. Each entry is based on 500 Monte Carlo samples.
\end{table}
\clearpage

\begin{table}[p]
\begin{center}
\def~{\hphantom{0}}
\caption{Comparison of the bias, standard error (SE),
estimated standard error (Est. SE), and coverage probability (CP) 
between the proposed joint model with heterogeneous WS variability (Model 1) and a classical joint model 
with homogeneous WS variability (Model 2) 
for the longitudinal outcome ($n=2000, \rho_{b\omega} = 0.25$)}
\resizebox{1.0\textwidth}{!}{\begin{tabular}{lrrcccrccc}
\hline \hline \multicolumn{2}{c}{} &\multicolumn{4}{c}{Model 1 (heterogeneous WS variability)} &\multicolumn{4}{c}{Model 2 (homogeneous WS variability)} \\ \cline{3-10}
Parameter &True &Bias &SE &Est. SE &CP (\%) &Bias &SE &Est. SE &CP (\%) \\
\hline
$Longitudinal$ &  &    & &  &  &    & & & \\
\  Fixed effects &  &   & &   &  &    & & & \\
\ \ Mean &  &   & &   &  &    & & & \\
\ \ \ $\beta_{0}$ &5 &$<$0.001 &0.030 &0.031 &95.0 &0.001 &0.034 &0.041 &98.6 \\
\ \ \ $\beta_{1}$ &1.5 &0.002 &0.044 &0.044 &95.4 &-0.001 &0.046 &0.047 &95.6 \\
\ \ \ $\beta_{2}$ &2 &-0.003 &0.038 &0.037 &95.8 &-0.002 &0.039 &0.041 &96.6 \\
\ \ \ $\beta_{3}$ &1 &$<$0.001 &0.013 &0.011 &90.8 &-0.001 &0.013 &0.013 &93.4 \\
\ \ \ $\beta_{4}$ &2 &$<$0.001 &0.001 &0.002 &95.0 &$<$0.001 &0.002 &0.002 &91.0 \\
\ \ Variance &  &   & &   &  &    & & & \\
\ \ \ $\tau_{0}$ &0.5 &$<$0.001 &0.030 &0.031 &95.8 &- &- &- & - \\
\ \ \ $\tau_{1}$ &0.5 &-0.003 &0.041 &0.041 &96.2 &- &- &- & -  \\
\ \ \ $\tau_{2}$ &0.2 &-0.003 &0.035 &0.036 &95.0 &- &- &- & -  \\
\ \ \ $\tau_{3}$ &0.2 &$<$0.001 &0.011 &0.011 &94.6 &- &- &- & -  \\
\ \ \ $\tau_{4}$ &0.05 &$<$0.001 &0.001 &0.001 &94.4 &- &- &- & -  \\

$Competing \ risks $&  &  & &    &  &    & & & \\
\  Fixed effects &  &    & &  &  &    & & & \\
\ \ $\gamma_{11}$ &1 &$<$0.001 &0.072 &0.074 &95.8 &\textbf{-0.060} &0.069 &0.070 &\textbf{86.4} \\
\ \ $\gamma_{12}$ &0.5 &-0.001 &0.063 &0.061 &93.6 &-0.011 &0.061 &0.059 &93.0 \\
\ \ $\gamma_{13}$ &0.5 &0.003 &0.021 &0.021 &93.6 &\textbf{-0.014} &0.021 &0.020 &\textbf{86.8} \\
\ \ $\gamma_{21}$ &-0.5 &-0.007 &0.086 &0.087 &95.0 &0.041 &0.084 &0.084 &91.6 \\
\ \ $\gamma_{22}$ &0.5 &$<$0.001 &0.072 &0.072 &94.4 &-0.001 &0.070 &0.070 &94.6 \\
\ \ $\gamma_{23}$ &0.25 &-0.001 &0.024 &0.023 &94.4 &0.005 &0.023 &0.023 &93.4 \\
\ Association &  &      &  &    & & & & &\\
\ \  $\alpha_{b1}$ &0.1 &0.004 &0.073 &0.077 &96.0 &\textbf{0.100} &0.068 &0.075 &\textbf{78.4} \\
\ \  $\alpha_{b2}$ &-0.1 &-0.002 &0.077 &0.078 &95.2 &\textbf{-0.118} &0.083 &0.097 &\textbf{81.8} \\
\ \  $\alpha_{\omega 1}$ &0.5 &-0.001 &0.071 &0.072 &95.4 &- &- &- & -  \\
\ \  $\alpha_{\omega 2}$ &-0.5 &-0.005 &0.085 &0.082 &94.6 &- &- &- & -   \\
\ Covariance matrix &  &      &  &    & & & & & \\
\ of random effects &  &      &  &    & & & & & \\
\ \  $\sigma_b^2$ &0.5  &-0.001 &0.028 &0.027 &94.4 &\textbf{-0.028} &0.031 &0.028 &\textbf{78.2} \\
\ \  $\sigma_{\omega}^2$ &0.5  &-0.002 &0.025 &0.025 &94.6 &- &- &- & - \\
\ \  $\sigma_{b\omega}$ &0.125 &$<$0.001 &0.020 &0.019 &95.0 &- &- &- & - \\

\hline \hline
\end{tabular}
\label{tab:simSE7}}
\end{center}
\vspace{2mm}
\noindent Note: Large error in confidence interval coverage probability (CP) compared to the 95\% nominal level are highlighted in boldface. Each entry is based on 500 Monte Carlo samples.
\end{table}
\clearpage
\begin{table}[p]
\begin{center}
\def~{\hphantom{0}}
\caption{Comparison of the bias, standard error (SE),
estimated standard error (Est. SE), and coverage probability (CP) 
between the proposed joint model with heterogeneous WS variability (Model 1) and a classical joint model 
with homogeneous WS variability (Model 2) 
for the longitudinal outcome ($n=2000, \rho_{b\omega} = 0$)}
\resizebox{1.0\textwidth}{!}{\begin{tabular}{lrrcccrccc}
\hline \hline \multicolumn{2}{c}{} &\multicolumn{4}{c}{Model 1 (heterogeneous WS variability)} &\multicolumn{4}{c}{Model 2 (homogeneous WS variability)} \\ \cline{3-10}
Parameter &True &Bias &SE &Est. SE &CP (\%) &Bias &SE &Est. SE &CP (\%) \\
\hline
$Longitudinal$ &  &    & &  &  &    & & & \\
\  Fixed effects &  &   & &   &  &    & & & \\
\ \ Mean &  &   & &   &  &    & & & \\
\ \ \ $\beta_{0}$ &5 &0.001 &0.032 &0.031 &94.4 &0.002 &0.034 &0.041 &97.8 \\
\ \ \ $\beta_{1}$ &1.5 &0.001 &0.045 &0.043 &93.4 &-0.002 &0.046 &0.047 &94.6 \\
\ \ \ $\beta_{2}$ &2 &-0.001 &0.037 &0.037 &94.8 &$<$0.001 &0.039 &0.041 &96.2 \\
\ \ \ $\beta_{3}$ &1 &$<$0.001 &0.013 &0.011 &91.6 &$<$0.001 &0.013 &0.013 &94.0 \\
\ \ \ $\beta_{4}$ &2 &$<$0.001 &0.001 &0.002 &95.8 &$<$0.001 &0.002 &0.002 &93.0 \\
\ \ Variance &  &   & &   &  &    & & & \\
\ \ \ $\tau_{0}$ &0.5 &0.002 &0.031 &0.031 &95.0 &- &- &- & - \\
\ \ \ $\tau_{1}$ &0.5 &-0.003 &0.039 &0.041 &96.2 &- &- &- & -  \\
\ \ \ $\tau_{2}$ &-0.2 &-0.004 &0.036 &0.036 &94.6 &- &- &- & -  \\
\ \ \ $\tau_{3}$ &0.2 &-0.001 &0.011 &0.011 &94.8 &- &- &- & -  \\
\ \ \ $\tau_{4}$ &0.05 &$<$0.001 &0.001 &0.001 &94.0 &- &- &- & -  \\

$Competing \ risks $&  &  & &    &  &    & & & \\
\  Fixed effects &  &    & &  &  &    & & & \\
\ \ $\gamma_{11}$ &1 &0.001 &0.072 &0.074 &95.8 &\textbf{-0.062} &0.069 &0.070 &\textbf{85.0} \\
\ \ $\gamma_{12}$ &0.5 &$<$0.001 &0.065 &0.061 &92.6 &-0.010 &0.063 &0.059 &92.8 \\
\ \ $\gamma_{13}$ &0.5 &0.002 &0.022 &0.021 &94.4 &\textbf{-0.015} &0.021 &0.020 &\textbf{84.8} \\
\ \ $\gamma_{21}$ &-0.5 &-0.006 &0.086 &0.086 &95.4 &0.046 &0.085 &0.084 &91.2 \\
\ \ $\gamma_{22}$ &0.5 &-0.002 &0.072 &0.071 &94.8 &-0.004 &0.070 &0.070 &94.2 \\
\ \ $\gamma_{23}$ &0.25 &-0.001 &0.024 &0.023 &94.4 &0.006 &0.024 &0.023 &94.2 \\
\ Association &  &      &  &    & & & & &\\
\ \  $\alpha_{b1}$ &0.1 &0.004 &0.070 &0.073 &95.2 &-0.008 &0.069 &0.075 &96.4 \\
\ \  $\alpha_{b2}$ &-0.1 &0.005 &0.076 &0.075 &95.8 &0.011 &0.077 &0.096 &98.0 \\
\ \  $\alpha_{\omega 1}$ &0.5 &0.001 &0.066 &0.068 &95.4 &- &- &- & -  \\
\ \  $\alpha_{\omega 2}$ &-0.5 &-0.007 &0.075 &0.078 &95.4 &- &- &- & -   \\
\ Covariance matrix &  &      &  &    & & & & & \\
\ of random effects &  &      &  &    & & & & & \\
\ \  $\sigma_b^2$ &0.5  &-0.004 &0.028 &0.027 &94.4 &\textbf{-0.031} &0.033 &0.028 &\textbf{73.2} \\
\ \  $\sigma_{\omega}^2$ &0.5  &-0.002 &0.025 &0.025 &95.0 &- &- &- & - \\
\ \  $\sigma_{b\omega}$ &0 &0.001 &0.019 &0.019 &96.6 &- &- &- & - \\

\hline \hline
\end{tabular}
\label{tab:simSE8}}
\end{center}
\vspace{2mm}
\noindent Note: Large error in confidence interval coverage probability (CP) compared to the 95\% nominal level are highlighted in boldface. Each entry is based on 500 Monte Carlo samples.
\end{table}
\clearpage
\begin{table}[p]
\begin{center}
\def~{\hphantom{0}}
\caption{Comparison of the bias, standard error (SE),
estimated standard error (Est. SE), and coverage probability (CP) 
between the proposed joint model with heterogeneous WS variability (Model 1) and a classical joint model 
with homogeneous WS variability (Model 2) 
for the longitudinal outcome ($n=10000, \rho_{b\omega} = 0.75$)}
\resizebox{1.0\textwidth}{!}{\begin{tabular}{lrrcccrccc}
\hline \hline \multicolumn{2}{c}{} &\multicolumn{4}{c}{Model 1 (heterogeneous WS variability)} &\multicolumn{4}{c}{Model 2 (homogeneous WS variability)} \\ \cline{3-10}
Parameter &True &Bias &SE &Est. SE &CP (\%) &Bias &SE &Est. SE &CP (\%) \\
\hline
$Longitudinal$ &  &    & &  &  &    & & & \\
\  Fixed effects &  &   & &   &  &    & & & \\
\ \ Mean &  &   & &   &  &    & & & \\
\ \ \ $\beta_{0}$ &5 &0.001 &0.014 &0.014 &95.4 &0.001 &0.015 &0.018 &98.8 \\
\ \ \ $\beta_{1}$ &1.5 &-0.001 &0.019 &0.019 &96.0 &-0.005 &0.021 &0.021 &94.2 \\
\ \ \ $\beta_{2}$ &2 &-0.001 &0.016 &0.017 &95.2 &-0.003 &0.018 &0.018 &94.8 \\
\ \ \ $\beta_{3}$ &1 &$<$0.001 &0.005 &0.005 &95.4 &-0.002 &0.005 &0.006 &93.6 \\
\ \ \ $\beta_{4}$ &2 &$<$0.001 &0.001 &0.001 &94.4 &$<$0.001 &0.001 &0.001 &\textbf{83.4} \\
\ \ Variance &  &   & &   &  &    & & & \\
\ \ \ $\tau_{0}$ &0.5 &$<$0.001 &0.014 &0.014 &93.6 &- &- &- & - \\
\ \ \ $\tau_{1}$ &0.5 &-0.001 &0.018 &0.018 &95.8 &- &- &- & -  \\
\ \ \ $\tau_{2}$ &-0.2 &-0.002 &0.015 &0.016 &96.2 &- &- &- & -  \\
\ \ \ $\tau_{3}$ &0.2 &$<$0.001 &0.005 &0.005 &96.4 &- &- &- & -  \\
\ \ \ $\tau_{4}$ &0.05 &$<$0.001 &0.001 &0.001 &93.8 &- &- &- & -  \\

$Competing \ risks $&  &  & &    &  &    & & & \\
\  Fixed effects &  &    & &  &  &    & & & \\
\ \ $\gamma_{11}$ &1 &0.001 &0.033 &0.033 &93.4 &\textbf{-0.038} &0.033 &0.032 &\textbf{77.2} \\
\ \ $\gamma_{12}$ &0.5 &0.001 &0.028 &0.027 &93.0 &-0.004 &0.028 &0.027 &93.0 \\
\ \ $\gamma_{13}$ &0.5 &$<$0.001 &0.009 &0.009 &95.2 &\textbf{-0.012} &0.009 &0.009 &\textbf{72.5} \\
\ \ $\gamma_{21}$ &-0.5 &-0.002 &0.039 &0.038 &95.0 &0.017 &0.039 &0.038 &91.2 \\
\ \ $\gamma_{22}$ &0.5 &$<$0.001 &0.031 &0.032 &95.6 &0.003 &0.031 &0.032 &95.2 \\
\ \ $\gamma_{23}$ &0.25 &-0.001 &0.010 &0.010 &95.2 &0.002 &0.010 &0.010 &95.4 \\
\ Association &  &      &  &    & & & & &\\
\ \  $\alpha_{b1}$ &0.1 &-0.007 &0.060 &0.065 &95.6 &\textbf{0.307} &0.031 &0.033 &\textbf{0} \\
\ \  $\alpha_{b2}$ &-0.1 &0.006 &0.065 &0.065 &95.0 &\textbf{-0.396} &0.043 &0.045 &\textbf{0} \\
\ \  $\alpha_{\omega 1}$ &0.5 &0.004 &0.060 &0.062 &95.4 &- &- &- & -  \\
\ \  $\alpha_{\omega 2}$ &-0.5 &-0.008 &0.069 &0.066 &95.2 &- &- &- & -   \\
\ Covariance matrix &  &      &  &    & & & & & \\
\ of random effects &  &      &  &    & & & & & \\
\ \  $\sigma_b^2$ &0.5  &$<$0.001 &0.012 &0.012 &93.8 &\textbf{-0.026} &0.016 &0.013 &\textbf{45.7} \\
\ \  $\sigma_{\omega}^2$ &0.5  &-0.002 &0.011 &0.011 &95.4 &- &- &- & - \\
\ \  $\sigma_{b\omega}$ &0 &$<$0.001 &0.010 &0.009 &94.6 &- &- &- & - \\

\hline \hline
\end{tabular}
\label{tab:simSE9}}
\end{center}
\vspace{2mm}
\noindent Note: Large error in confidence interval coverage probability (CP) compared to the 95\% nominal level are highlighted in boldface. Each entry is based on 500 Monte Carlo samples.
\end{table}
\clearpage

\begin{table}
\begin{center}
\def~{\hphantom{0}}
\caption{Comparison of the bias, standard error (SE),
estimated standard error (Est. SE), and coverage probability (CP) 
between the proposed joint model with heterogeneous WS variability (Model 1) and a classical joint model 
with homogeneous WS variability (Model 2)
for the longitudinal outcome ($n=10000, \rho_{b\omega} = 0.5$)}
\resizebox{1.0\textwidth}{!}{\begin{tabular}{lrrcccrccc}
\hline \hline \multicolumn{2}{c}{} &\multicolumn{4}{c}{Model 1 (heterogeneous WS variability)}
&\multicolumn{4}{c}{Model 2 (homogeneous WS variability)}\\ \cline{3-10}
Parameter &True &Bias &SE &Est. SE &CP (\%) &Bias &SE &Est. SE &CP (\%)\\
\hline
$Longitudinal$ &  &    & &  &  &    & & & \\
\  Fixed effects &  &   & &   &  &    & & & \\
\ \ Mean &  &   & &   &  &    & & & \\
\ \ \ $\beta_{0}$ &5 &$<$0.001 &0.014 &0.014 &96.6 &0.001 &0.015 &0.018 &98.2 \\
\ \ \ $\beta_{1}$ &1.5 &$<$0.001 &0.020 &0.020 &95.4 &-0.004 &0.021 &0.021 &95.6\\
\ \ \ $\beta_{2}$ &2 &$<$0.001 &0.017 &0.017 &94.8 &-0.001 &0.018 &0.018 &95.6\\
\ \ \ $\beta_{3}$ &1 &$<$0.001 &0.005 &0.005 &93.8 &-0.001 &0.006 &0.006 &94.2\\
\ \ \ $\beta_{4}$ &2 &$<$0.001 &0.001 &0.001 &94.8 &$<$0.001 &0.001 &0.001 &\textbf{88.0}\\
\ \ Variance &  &   & &   &  &    & & & \\
\ \ \ $\tau_{0}$ &0.5 &$<$0.001 &0.014 &0.014 &94.8 &- &- &- & - \\
\ \ \ $\tau_{1}$ &0.5 &-0.001 &0.018 &0.018 &95.4 &- &- &- & - \\
\ \ \ $\tau_{2}$ &-0.2 &-0.001 &0.015 &0.016 &97.8 &- &- &- & - \\
\ \ \ $\tau_{3}$ &0.2 &$<$0.001 &0.005 &0.005 &96.6 &- &- &- & - \\
\ \ \ $\tau_{4}$ &0.05 &$<$0.001 &0.001 &0.001 &95.0 &- &- &- & - \\

$Competing \ risks $&  &  & &    &  &    & & & \\
\  Fixed effects &  &    & &  &  &    & & & \\
\ \ $\gamma_{11}$ &1 &$<$0.001 &0.033 &0.033 &94.8 &\textbf{-0.051} &0.033 &0.032 &\textbf{63.0} \\
\ \ $\gamma_{12}$ &0.5 &0.001 &0.028 &0.027 &93.8 &-0.006 &0.027 &0.026 &94.2\\
\ \ $\gamma_{13}$ &0.5 &$<$0.001 &0.009 &0.009 &94.4 &\textbf{-0.015} &0.009 &0.009 &\textbf{61.6} \\
\ \ $\gamma_{21}$ &-0.5 &-0.002 &0.039 &0.038 &95.0 &\textbf{0.036} &0.038 &0.038 &\textbf{81.8}\\
\ \ $\gamma_{22}$ &0.5 &-0.001 &0.030 &0.032 &95.8 &$<$0.001 &0.030 &0.031 &96.0\\
\ \ $\gamma_{23}$ &0.25 &-0.001 &0.010 &0.010 &95.8 &0.004 &0.010 &0.010 &94.2 \\
\ Association &  &      &  &    & & & & &\\
\ \  $\alpha_{b1}$ &0.1 &0.001 &0.040 &0.041 &95.6 &\textbf{0.202} &0.031 &0.033 &\textbf{0}\\
\ \  $\alpha_{b2}$ &-0.1 &-0.002 &0.040 &0.041 &94.8 &\textbf{-0.246} &0.038 &0.044 &\textbf{0}\\
\ \  $\alpha_{\omega 1}$ &0.5 &-0.003 &0.038 &0.038 &95.6 &- &- &- & - \\
\ \  $\alpha_{\omega 2}$ &-0.5 &-0.002 &0.042 &0.043 &95.2 &- &- &- & - \\
\ Covariance matrix &  &      &  &    & & & & & \\
\ of random effects &  &      &  &    & & & & & \\
\ \  $\sigma_b^2$ &0.5  &-0.001 &0.013 &0.012 &93.0 &\textbf{-0.027} &0.016 &0.013 &\textbf{46.2} \\
\ \  $\sigma_{\omega}^2$ &0.5  &-0.001 &0.012 &0.011 &94.4 &- &- &- & -  \\
\ \  $\sigma_{b\omega}$ &0 &$<$0.001 &0.009 &0.009 &94.2 &- &- &- & - \\

\hline \hline
\end{tabular}
\label{tab:simSE10}}
\end{center}
\vspace{2mm}
\noindent Note: Large error in confidence interval coverage probability (CP) compared to the 95\% nominal level are highlighted in boldface. Each entry is based on 500 Monte Carlo samples.
\end{table}
\clearpage

\begin{table}[p]
\begin{center}
\def~{\hphantom{0}}
\caption{Comparison of the bias, standard error (SE),
estimated standard error (Est. SE), and coverage probability (CP) 
between the proposed joint model with heterogeneous WS variability (Model 1) and a classical joint model 
with homogeneous WS variability (Model 2) 
for the longitudinal outcome ($n=10000, \rho_{b\omega} = 0.25$)}
\resizebox{1.0\textwidth}{!}{\begin{tabular}{lrrcccrccc}
\hline \hline \multicolumn{2}{c}{} &\multicolumn{4}{c}{Model 1 (heterogeneous WS variability)} &\multicolumn{4}{c}{Model 2 (homogeneous WS variability)} \\ \cline{3-10}
Parameter &True &Bias &SE &Est. SE &CP (\%) &Bias &SE &Est. SE &CP (\%) \\
\hline
$Longitudinal$ &  &    & &  &  &    & & & \\
\  Fixed effects &  &   & &   &  &    & & & \\
\ \ Mean &  &   & &   &  &    & & & \\
\ \ \ $\beta_{0}$ &5 &$<$0.001 &0.013 &0.014 &96.4 &$<$0.001 &0.015 &0.018 &98.4 \\
\ \ \ $\beta_{1}$ &1.5 &$<$0.001 &0.019 &0.020 &96.0 &-0.002 &0.020 &0.021 &96.2 \\
\ \ \ $\beta_{2}$ &2 &$<$0.001 &0.018 &0.017 &94.4 &$<$0.001 &0.019 &0.018 &94.4 \\
\ \ \ $\beta_{3}$ &1 &$<$0.001 &0.005 &0.005 &95.2 &-0.001 &0.006 &0.006 &95.2 \\
\ \ \ $\beta_{4}$ &2 &$<$0.001 &0.001 &0.001 &95.8 &$<$0.001 &0.001 &0.001 &\textbf{89.8} \\
\ \ Variance &  &   & &   &  &    & & & \\
\ \ \ $\tau_{0}$ &0.5 &$<$0.001 &0.014 &0.014 &93.8 &- &- &- & - \\
\ \ \ $\tau_{1}$ &0.5 &$<$0.001 &0.018 &0.018 &95.4 &- &- &- & -  \\
\ \ \ $\tau_{2}$ &-0.2 &-0.002 &0.016 &0.016 &94.2 &- &- &- & -  \\
\ \ \ $\tau_{3}$ &0.2 &$<$0.001 &0.005 &0.005 &95.4 &- &- &- & -  \\
\ \ \ $\tau_{4}$ &0.05 &$<$0.001 &0.001 &0.001 &93.2 &- &- &- & -  \\

$Competing \ risks $&  &  & &    &  &    & & & \\
\  Fixed effects &  &    & &  &  &    & & & \\
\ \ $\gamma_{11}$ &1 &0.001 &0.033 &0.033 &95.0 &\textbf{-0.059} &0.032 &0.031 &\textbf{52.2} \\
\ \ $\gamma_{12}$ &0.5 &0.002 &0.028 &0.027 &93.2 &-0.008 &0.027 &0.026 &93.6 \\
\ \ $\gamma_{13}$ &0.5 &$<$0.001 &0.009 &0.009 &94.8 &\textbf{-0.017} &0.009 &0.009 &\textbf{50.6} \\
\ \ $\gamma_{21}$ &-0.5 &-0.001 &0.038 &0.038 &95.6 &\textbf{0.046} &0.038 &0.038 &\textbf{77.2} \\
\ \ $\gamma_{22}$ &0.5 &-0.001 &0.031 &0.032 &95.6 &-0.002 &0.030 &0.031 &95.0 \\
\ \ $\gamma_{23}$ &0.25 &-0.001 &0.010 &0.010 &95.6 &0.005 &0.010 &0.010 &93.2 \\
\ Association &  &      &  &    & & & & &\\
\ \  $\alpha_{b1}$ &0.1 &-0.001 &0.032 &0.034 &96.4 &\textbf{0.095} &0.031 &0.033 &\textbf{16.8} \\
\ \  $\alpha_{b2}$ &-0.1 &$<$0.001 &0.034 &0.035 &95.4 &\textbf{-0.113} &0.038 &0.043 &\textbf{22.2} \\
\ \  $\alpha_{\omega 1}$ &0.5 &0.001 &0.032 &0.031 &94.4 &- &- &- & -  \\
\ \  $\alpha_{\omega 2}$ &-0.5 &-0.001 &0.036 &0.036 &95.0 &- &- &- & -   \\
\ Covariance matrix &  &      &  &    & & & & & \\
\ of random effects &  &      &  &    & & & & & \\
\ \  $\sigma_b^2$ &0.5  &-0.002 &0.013 &0.012 &93.2 &\textbf{-0.027} &0.015 &0.013 &\textbf{40.2} \\
\ \  $\sigma_{\omega}^2$ &0.5  &-0.002 &0.012 &0.011 &93.8 &- &- &- & - \\
\ \  $\sigma_{b\omega}$ &0 &$<$0.001 &0.009 &0.009 &93.8 &- &- &- & - \\

\hline \hline
\end{tabular}
\label{tab:simSE11}}
\end{center}
\vspace{2mm}
\noindent Note: Large error in confidence interval coverage probability (CP) compared to the 95\% nominal level are highlighted in boldface. Each entry is based on 500 Monte Carlo samples.
\end{table}
\clearpage
\begin{table}[p]
\begin{center}
\def~{\hphantom{0}}
\caption{Comparison of the bias, standard error (SE),
estimated standard error (Est. SE), and coverage probability (CP) 
between the proposed joint model with heterogeneous WS variability (Model 1) and a classical joint model 
with homogeneous WS variability (Model 2) 
for the longitudinal outcome ($n=10000, \rho_{b\omega} = 0$)}
\resizebox{1.0\textwidth}{!}{\begin{tabular}{lrrcccrccc}
\hline \hline \multicolumn{2}{c}{} &\multicolumn{4}{c}{Model 1 (heterogeneous WS variability)} &\multicolumn{4}{c}{Model 2 (homogeneous WS variability)} \\ \cline{3-10}
Parameter &True &Bias &SE &Est. SE &CP (\%) &Bias &SE &Est. SE &CP (\%) \\
\hline
$Longitudinal$ &  &    & &  &  &    & & & \\
\  Fixed effects &  &   & &   &  &    & & & \\
\ \ Mean &  &   & &   &  &    & & & \\
\ \ \ $\beta_{0}$ &5 &-0.001 &0.014 &0.014 &94.2 &$<$0.001 &0.016 &0.018 &97.6 \\
\ \ \ $\beta_{1}$ &1.5 &0.001 &0.019 &0.020 &95.6 &$<$0.001 &0.020 &0.021 &96.8 \\
\ \ \ $\beta_{2}$ &2 &0.001 &0.017 &0.017 &94.2 &0.001 &0.017 &0.018 &96.2 \\
\ \ \ $\beta_{3}$ &1 &$<$0.001 &0.005 &0.005 &95.6 &$<$0.001 &0.006 &0.006 &94.8 \\
\ \ \ $\beta_{4}$ &2 &$<$0.001 &0.001 &0.001 &94.4 &$<$0.001 &0.001 &0.001 &\textbf{88.8} \\
\ \ Variance &  &   & &   &  &    & & & \\
\ \ \ $\tau_{0}$ &0.5 &$<$0.001 &0.014 &0.014 &94.0 &- &- &- & - \\
\ \ \ $\tau_{1}$ &0.5 &$<$0.001 &0.019 &0.018 &94.4 &- &- &- & -  \\
\ \ \ $\tau_{2}$ &-0.2 &$<$0.001 &0.015 &0.016 &95.2 &- &- &- & -  \\
\ \ \ $\tau_{3}$ &0.2 &$<$0.001 &0.005 &0.005 &94.6 &- &- &- & -  \\
\ \ \ $\tau_{4}$ &0.05 &$<$0.001 &0.001 &0.001 &94.8 &- &- &- & -  \\

$Competing \ risks $&  &  & &    &  &    & & & \\
\  Fixed effects &  &    & &  &  &    & & & \\
\ \ $\gamma_{11}$ &1 &0.001 &0.033 &0.033 &95.0 &\textbf{-0.061} &0.031 &0.031 &\textbf{51.0} \\
\ \ $\gamma_{12}$ &0.5 &0.002 &0.028 &0.027 &93.6 &-0.009 &0.027 &0.026 &92.0 \\
\ \ $\gamma_{13}$ &0.5 &$<$0.001 &0.009 &0.009 &94.8 &\textbf{-0.017} &0.009 &0.009 &\textbf{51.8} \\
\ \ $\gamma_{21}$ &-0.5 &-0.001 &0.038 &0.038 &96.4 &\textbf{0.050} &0.036 &0.037 &\textbf{73.2} \\
\ \ $\gamma_{22}$ &0.5 &$<$0.001 &0.031 &0.032 &97.0 &-0.001 &0.030 &0.031 &96.6 \\
\ \ $\gamma_{23}$ &0.25 &-0.001 &0.010 &0.010 &95.0 &0.006 &0.010 &0.010 &92.0 \\
\ Association &  &      &  &    & & & & &\\
\ \  $\alpha_{b1}$ &0.1 &0.001 &0.031 &0.032 &96.6 &-0.012 &0.031 &0.033 &95.6 \\
\ \  $\alpha_{b2}$ &-0.1 &-0.001 &0.034 &0.033 &95.8 &0.006 &0.036 &0.042 &97.6 \\
\ \  $\alpha_{\omega 1}$ &0.5 &0.001 &0.028 &0.030 &95.8 &- &- &- & -  \\
\ \  $\alpha_{\omega 2}$ &-0.5 &$<$0.001 &0.035 &0.035 &94.4 &- &- &- & -   \\
\ Covariance matrix &  &      &  &    & & & & & \\
\ of random effects &  &      &  &    & & & & & \\
\ \  $\sigma_b^2$ &0.5  &$<$0.001 &0.013 &0.012 &94.0 &\textbf{-0.026} &0.015 &0.012 &\textbf{44.0} \\
\ \  $\sigma_{\omega}^2$ &0.5  &-0.001 &0.012 &0.011 &91.2 &- &- &- & - \\
\ \  $\sigma_{b\omega}$ &0 &0.001 &0.008 &0.009 &94.8 &- &- &- & - \\

\hline \hline
\end{tabular}
\label{tab:simSE12}}
\end{center}
\vspace{2mm}
\noindent Note: Large error in confidence interval coverage probability (CP) compared to the 95\% nominal level are highlighted in boldface. Each entry is based on 500 Monte Carlo samples.
\end{table}
\clearpage

\subsection{(Simulation 3: Impact of non-ignorable monotone missing data on the mixed-effects multiple location-scale model)}
\label{sm:addsim1} 
When WS variability itself is of scientific interest, the mixed-effects multiple location-scale model (16)-(17) has commonly been used for longitudinal data analysis \citep[among others]{german2022wiser, hedeker2008application}. In this simulation, we illustrate that non-ignorable missing data due to informative dropout can induce substantial bias and invalid inference when fitting the mixed-effects multiple location-scale model alone, and that the proposed joint model can be used as a method to address this issue. Using the same simulation setting for Table \ref{tab:simSE6} as described in Section 3.1, we have fitted the mixed-effects location-scale model (16)-(17) for the longitudinal outcome only using the WiSER method proposed by \citet{german2022wiser}, with various sample sizes of $n=800, 2000, 10000$, and the results are summarized in Table \ref{tab:simSEWiSER}. It is observed that
WiSER produced substantial biases and severe undercoverage in estimating $ (\tau_0, \tau_1, \tau_3, \tau_4)$ and severe underestimation of the standard error of $\sigma_b^2$ across all sample sizes and in $ (\beta_1, \beta_3) $ even when $ n=10000$. These biases are largely due to the non-ignorable missing data from terminal events, which the WiSER method does not account for. In contrast, estimation biases for the same parameters have been corrected by our proposed joint model for the same simulation scenarios, as shown in Table \ref{tab:simSE2}, \ref{tab:simSE6}, and \ref{tab:simSE10}.

\begin{table}
\begin{center}
\def~{\hphantom{0}}
\caption{Simulated bias, standard error (SE),
estimated standard error (Est. SE), and coverage probability (CP) 
for a mixed-effects multiple location-scale model for the longitudinal outcome only (Model 3) using WiSER \citep{german2022wiser} ($n=800, 2000, 10000, \rho_{b\omega} = 0.5$)} 
\label{tab:simSEWiSER}
\resizebox{1.0\textwidth}{!}{\begin{tabular}{lrrcccrcccrccc}
\hline \hline \multicolumn{2}{c}{} &\multicolumn{4}{c}{$n=800$}
&\multicolumn{4}{c}{$n=2000$} & \multicolumn{4}{c}{$n=10000$}\\ \cline{3-14}
Parameter &True &Bias &SE &Est. SE &CP (\%) &Bias &SE &Est. SE &CP (\%) &Bias &SE &Est. SE &CP (\%)\\
\hline
$Longitudinal$ &  &    & &  &  &    & & & \\
\  Fixed effects &  &   & &   &  &    & & & \\
\ \ Mean trajectory &  &   & &   &  &    & & & \\
\ \ \ $\beta_{0}$ &5 &0.009 & 0.053 & 0.052 & 94.8& 0.009 &  0.033 &  0.033 &  94.0 & 0.008 & 0.015 & 0.015 & 90.4\\
\ \ \ $\beta_{1}$ &1.5 & -0.025 & 0.071 & 0.071 & 93.0&  -0.024 &  0.043 &  0.045 &  91.6& \textbf{-0.025} & 0.020 & 0.020 & \textbf{76.2}\\
\ \ \ $\beta_{2}$ &2 & -0.002 & 0.062 & 0.061 & 94.2&  0.001 &  0.036 &  0.039 &  96.8&$<$0.001 & 0.017 & 0.017 & 95.0\\
\ \ \ $\beta_{3}$ &1 & -0.004 & 0.019 & 0.019 & 93.8&  -0.005 &  0.012 &  0.012 &  93.4& \textbf{-0.005} & 0.005 & 0.005 & \textbf{86.6}\\
\ \ \ $\beta_{4}$ &2 &$<$0.001 & 0.003 & 0.003 & 94.4& $<$0.001 & 0.002 & 0.002 & 95.0&$<$0.001 & 0.001 & 0.001 & 95.2\\
\ \ WS variability &  &   & &   &  &    & & & \\
\ \ \ $\tau_{0}$ &0.5 & \textbf{0.284} & 0.058 & 0.058 & \textbf{0}& \textbf{0.291} & 0.039 & 0.037 & \textbf{0}& \textbf{0.289} & 0.016 & 0.017 & \textbf{0}\\
\ \ \ $\tau_{1}$ &0.5  & \textbf{-0.090} & 0.08 & 0.076 & \textbf{76.6}& \textbf{-0.100} & 0.049 & 0.049 & \textbf{46.8}& \textbf{-0.094} & 0.022 & 0.022 & \textbf{0.6}\\
\ \ \ $\tau_{2}$ &-0.2 & 0 & 0.066 & 0.066 & 95.8& -0.011 & 0.042 & 0.043 & 95.2& -0.005 & 0.019 & 0.019 & 94.6\\
\ \ \ $\tau_{3}$ &0.2 & \textbf{-0.018} & 0.021 & 0.020 & \textbf{82.8}& \textbf{-0.019} & 0.012 & 0.013 & \textbf{69.2} & \textbf{-0.02} & 0.006 & 0.006 & \textbf{8}\\
\ \ \ $\tau_{4}$ &0.05 & \textbf{-0.002} & 0.003 & 0.003 & \textbf{87.6}& \textbf{-0.002} & 0.002 & 0.002 & \textbf{77.2}& \textbf{-0.002} & 0.001 & 0.001 & \textbf{35.4}\\
Covariance matrix of random effects &\\
\ \  $\sigma_b^2$ &0.5& -0.004 & 0.048 & \textbf{0.024} & \textbf{66.6} &  <$0.001$ &0.030  & \textbf{0.015} & \textbf{65.8}&-0.001 & 0.013 & \textbf{0.007} & \textbf{69.0}\\
\hline \hline
\end{tabular}
}
\end{center}
\noindent Note: Large error in confidence interval coverage probability (CP) compared to the 95\% nominal level are highlighted in boldface. Each entry is based on 500 Monte Carlo samples.
\end{table}

\subsection{(Simulation 4: Generative joint model with homogeneous WS variance)}
\label{sm:homo}
In this simulation scenario, the longitudinal measurements $Y_i(o_{ij})$ were generated from the linear mixed-effects model
(\ref{simplerlongmean}) - (\ref{simplerlongvar})
\begin{eqnarray}
\label{simplerlongmean}
    m_i(o_{ij}) &=& \beta_0 + \beta_1 X_{1i} + \beta_2 X_{2i} + \beta_3 X_{3i} + \beta_4 o_{ij} + b_{i},\\
    \label{simplerlongvar}
    \sigma_{i}^2(o_{ij}) &\equiv& exp(\tau_0) = \sigma^2,
\end{eqnarray}
and the competing risks event data were generated from the following proportional cause-specific hazards models:
\begin{eqnarray}
\label{simplerh01}
\lambda_1(t) &=& \lambda_{01}(t) \exp\{\gamma_{11}X_{1i} + \gamma_{12}X_{2i} + \gamma_{13}X_{3i} + \alpha_{b1} b_i\}, \\
\label{simplerh02}
\lambda_2(t) &=& \lambda_{02}(t) \exp\{\gamma_{21}X_{1i} + \gamma_{22}X_{2i} + \gamma_{23}X_{3i} + \alpha_{b2} b_i\},
\end{eqnarray}
where the four submodels (\ref{simplerlongmean})-(\ref{simplerh02}) are linked together through the shared random effects $b_i$. Here $o_{ij}$'s represent the scheduled visiting times for subject $i$ with an increment of 0.25,  $X_{1i} \sim Bernoulli(0.5)$, $X_{2i} \sim Uni(-1, 1)$, and $X_{3i} \sim N(1,4)$, $b_i \sim N(0, 0.5)$. The baseline hazards $\lambda_{01}(t), \lambda_{02}(t)$ are set to constants 0.05 and 0.1, respectively. We simulated non-informative censoring time $C_i \sim Uni(5,10)$ and let $T_i = min\{T_{i1}^*, T_{i2}^*, C_i\}$ be the observed survival time (possibly censored) for subject $i$, where $T_{i1}^*$ and $T_{i2}^*$ are independent event times from models (\ref{simplerh01}) and (\ref{simplerh02}), respectively, $i=1,\ldots, n$. The longitudinal measurements for subject $i$ are assumed missing when $o_{ij}>T_i$. The average number of longitudinal measurements per subject is around 10. Results are summarized in Table \ref{tab:simplersimSE}. It can be seen that under the sample size of $n = 2000$, the Model 1 can still yield low bias and satisfying CPs that are close to 95\%, similar to Model 2. This result shows that 
Model 1 has almost the same efficiency as Model 2 in terms of $\beta$ and $\gamma$ and marginally lower efficiency in terms of $\alpha$, but the bias is comparable.
It is concluded that Model 1 remains robust in terms of estimation and inference when the homogeneity assumption of WS variability of the longitudinal outcome holds.     

\begin{table}[ht]
\begin{center}
\def~{\hphantom{0}}
\caption{Comparison of the bias, standard error (SE),
estimated standard error (Est. SE), and coverage probability (CP) 
between the proposed joint model with heterogeneous WS variability (Model 1) and a classical joint model 
with homogeneous WS variability (Model 2) 
for the longitudinal outcome when the simulated data are generated from the simpler model (\ref{simplerlongmean})-(\ref{simplerh02}) with homogeneous WS variability, $n$=2000}
\label{tab:simplersimSE}
\resizebox{1.0\textwidth}{!}{\begin{tabular}{lrrcccrccc}
\hline \hline \multicolumn{2}{c}{} &\multicolumn{4}{c}{Model 1 (heterogeneous WS variability)}
&\multicolumn{4}{c}{Model 2 (homogeneous WS variability)} \\ \cline{3-10}
Parameter &True &Bias &SE &Est. SE &CP (\%) &Bias &SE &Est. SE &CP (\%) \\
\hline
$Longitudinal$ &  &    & &  &  &    & & & \\
\  Fixed effects &  &   & &   &  &    & & & \\
\ \ Mean trajectory &  &   & &   &  &    & & & \\
\ \ \ $\beta_{0}$ &5 &0.001 &0.029 &0.029 &95.5 &0.001 &0.029 &0.029 &95.4 \\
\ \ \ $\beta_{1}$ &1.5 &$<$0.001 &0.038 &0.038 &94.9 &-0.001 &0.038 &0.038 &95.0 \\
\ \ \ $\beta_{2}$ &2 &0.001 &0.034 &0.033 &95.3 &0.001 &0.034 &0.033 &95.0 \\
\ \ \ $\beta_{3}$ &1 &$<$0.001 &0.010 &0.010 &94.1 &$<$0.001 &0.010 &0.010 &94.4 \\
\ \ \ $\beta_{4}$ &2 &$<$0.001 &0.001 &0.001 &96.6 &$<$0.001 &0.001 &0.001 &96.6 \\
\ \ WS variability &  &   & &   &  &    & & & \\
\ \ \ $\tau_{0}$ &0.5 &-0.004 &0.017 &0.019 &96.6 &$<$0.001 &0.014 &0.015 &95.2 \\
$Competing \ risks $&  &  & &    &  &    & & & \\
\  Fixed effects &  &    & &  &  &    & & & \\
\ \ $\gamma_{11}$ &1 &0.011 &0.074 &0.073 &94.5 &$<$0.001 &0.071 &0.071 &94.4 \\
\ \ $\gamma_{12}$ &0.5 &0.007 &0.057 &0.060 &96.0 &0.002 &0.057 &0.059 &95.2 \\
\ \ $\gamma_{13}$ &0.5 &0.006 &0.020 &0.021 &96.2 &0.001 &0.019 &0.020 &95.4 \\
\ \ $\gamma_{21}$ &-0.5 &-0.001 &0.090 &0.088 &95.3 &0.005 &0.088 &0.086 &94.4 \\
\ \ $\gamma_{22}$ &0.5 &0.010 &0.069 &0.072 &95.1 &0.004 &0.067 &0.071 &94.8 \\
\ \ $\gamma_{23}$ &0.25 &0.002 &0.025 &0.024 &94.5 &$<$0.001 &0.023 &0.023 &95.6 \\
\ Association &  &      &  &    & & & & &\\
\ \  $\alpha_{b1}$ &0.1 &$<$0.001 &0.060 &0.066 &98.3 &$<$0.001 &0.055 &0.058 &96.6 \\
\ \  $\alpha_{b2}$ &-0.1 &-0.003 &0.077 &0.077 &97.2 &$<$0.001 &0.070 &0.070 &96.0 \\
\ Covariance matrix &  &      &  &    & & & & & \\
\ of random effects &  &      &  &    & & & & & \\
\ \  $\sigma_b^2$ &0.5 &-0.001 &0.022 &0.022 &95.3 &-0.001 &0.022 &0.022 &95.0 \\
\hline \hline
\end{tabular}
}
\end{center}
\noindent Note: Large error in confidence interval coverage probability (CP) compared to the 95\% nominal level are highlighted in boldface. Each entry is based on 500 Monte Carlo samples.
\end{table}

\subsection{(Simulation 6: Generative joint model with non-linear mean evolution and homogeneous WS variability for the longitudinal outcome)}
\label{appen:NLhomo}
In this simulation, we consider a scenario where non-linear individual mean trajectories drives the biomarker variability over time instead of heterogeneous WS variability. The details of this data generating mechanism is described below.

For the longitudinal process, we used a B-splines of time with 2 degrees of freedom and generated simulated data based on the following form:
\begin{eqnarray}
\label{sim:homoeq1.1}
m_i(o_{ij}) &=& \beta_0 + \sum_{\kappa=1}^2 \beta_{\kappa} B_{\kappa}(o_{ij}) + \beta_3 X_{1i} + b_{i0} + \sum_{\kappa=1}^2 b_{i\kappa} B_{\kappa}(o_{ij}),\\
    \sigma_{i}^2(o_{ij}) &=& \sigma^2,
\end{eqnarray}
where $\beta = (5, 2, 1, 1.5)^{\top}$, $\sigma^2 = 1.65$, $B_{\kappa}(o_{ij}), \kappa =1, 2, $ denotes the 2-degree B-spline basis at  the scheduled visiting times $o_{ij}$'s for subject $i$ with an increment of 0.25, $X_{1i} \sim Bernulli(0.5)$, random effects $b_i = (b_{i0}, b_{i1}, b_{i2})^{\top} \sim N(0, \Sigma_{b})$ with
\begin{eqnarray*}
\Sigma_{b} =  \left( {\begin{array}{cccc}
   10 & 3 & 2 \\
   3 & 4 & 2 \\
   2 & 2 & 4 \\
  \end{array} } \right).
\end{eqnarray*}
For the competing risks event process, we generated two competing events, each modeled by a separate Cox proportional hazards submodel: 
\begin{eqnarray}
\lambda_{i1}(t) &=& \lambda_{01}(t) \exp\{\gamma_{11}X_{1i} + \gamma_{12}X_{2i} + \gamma_{13}X_{3i} + \sum_{\kappa=0}^2 \alpha_{b\kappa1} b_{i\kappa}\}, \\
\label{sim:homoeq2}
\lambda_{i2}(t) &=& \lambda_{02}(t) \exp\{\gamma_{21}X_{1i} + \gamma_{22}X_{2i} + \gamma_{23}X_{3i} + \sum_{\kappa=0}^2 \alpha_{b\kappa2} b_{i\kappa}\},
\end{eqnarray}
where $X_{2i} \sim Uni(-1, 1)$, $X_{3i} \sim N(1, 4)$, $(\gamma_{11}, \gamma_{12}, \gamma_{13}) = (1, 0.5, 0.5), (\gamma_{21}, \gamma_{22}, \gamma_{23}) = (-0.5, 0.5, 0.25)$. The association between the longitudinal and competing risks event process was modeled through the association parameters $(\alpha_{b01}, \alpha_{b11}, \alpha_{b21}) = (0.05, 0.01, 0.02)$, $(\alpha_{b02}, \alpha_{b12}, \alpha_{b22}) = (-0.05, 0.02, 0.03)$. The two competing risks event times were simulated with the exponential baseline hazard functions $\lambda_{01}(t) = 0.05$ and $\lambda_{02}(t) =1$. We simulated non-informative censoring time $C_i \sim Uni(4,8)$ and let $T_i = min\{T_{i1}^*, T_{i2}^*, C_i\}$ be the observed survival time (possibly censored) for subject $i$, where $T_{i1}^*$ and $T_{i2}^*$ are independent event times, which results in the median event rate of 44.5\% and 28.1\% for both failures, respectively. we generated 10 random datasets with a sample size of $n = 3000$, where the homogeneous WS variability and non-linear individual mean trajectory of the longitudinal outcome is present. 

We compared three joint models:
\begin{enumerate}[label=Model \arabic*:]
    \item (Mis-specified model: linear individual mean trajectory + heterogeneous WS variability) 
    \begin{eqnarray*}
        m_i(o_{ij}) &=& \beta_0 + \beta_1 o_{ij} + \beta_{2} X_{1i} + b_{i0} +  b_{i1}o_{ij}, \quad b_i = (b_{i0}, b_{i1})^{\top},\\
    \sigma_{i}^2(o_{ij}) &=& \exp\left\{\tau_0 + \sum_{\kappa=1}^2 \tau_{\kappa} B_{\kappa}(o_{ij}) + \tau_{3} X_{1i} + \omega_i\right\}, \cr
       \lambda_{i1}(t) &=& \lambda_{01}(t)\exp(\gamma_{11} X_{1i} + \gamma_{12}X_{2i} + \gamma_{13}X_{3i} + \alpha_{b1}^{\top} b_{i} + \alpha_{\omega 1}\omega_i), \\
    \lambda_{i2}(t) &=& \lambda_{02}(t)\exp(\gamma_{21} X_{1i} + \gamma_{22}X_{2i} + \gamma_{23}X_{3i} + \alpha_{b2}^{\top} b_{i} + \alpha_{\omega 2}\omega_i),
\end{eqnarray*}

\item (Correctly specified model: non-linear individual mean trajectory + homogeneous WS variability)
\begin{eqnarray*}
m_i(o_{ij}) &=& \beta_0 + \sum_{\kappa=1}^2 \beta_{\kappa} B_{\kappa}(o_{ij}) + \beta_3 X_{1i} + b_{i0} + \sum_{\kappa=1}^2 b_{i\kappa} B_{\kappa}(o_{ij}), \quad b_i = (b_{i0}, b_{i1}, b_{i2})^{\top}\\
    \sigma_{i}^2(o_{ij}) &=& \sigma^2, \cr
    \lambda_{i1}(t) &=& \lambda_{01}(t)\exp(\gamma_{11} X_{1i} + \gamma_{12}X_{2i} + \gamma_{13}X_{3i} + \alpha_{b1}^{\top} b_{i}), \\
    \lambda_{i2}(t) &=& \lambda_{02}(t)\exp(\gamma_{21} X_{1i} + \gamma_{22}X_{2i} + \gamma_{23}X_{3i} + \alpha_{b2}^{\top} b_{i}),
\end{eqnarray*}

\item (Nest Model 2 as a special case: non-linear individual mean trajectory + heterogeneous WS variability)
\begin{eqnarray*}
        m_i(o_{ij}) &=& \beta_0 + \sum_{\kappa=1}^2 \beta_{\kappa} B_{\kappa}(o_{ij}) + \beta_3 X_{1i} + b_{i0} + \sum_{\kappa=1}^2 b_{i\kappa} B_{\kappa}(o_{ij}), \quad b_i = (b_{i0}, b_{i1}, b_{i2})^{\top},\\
    \sigma_{i}^2(o_{ij}) &=& \exp\left\{\tau_0 + \sum_{\kappa=1}^2 \tau_{\kappa} B_{\kappa}(o_{ij}) + \tau_{3} X_{1i} + \omega_i\right\}, \cr
       \lambda_{i1}(t) &=& \lambda_{01}(t)\exp(\gamma_{11} X_{1i} + \gamma_{12}X_{2i} + \gamma_{13}X_{3i} + \alpha_{b1}^{\top} b_{i} + \alpha_{\omega 1}\omega_i), \\
    \lambda_{i2}(t) &=& \lambda_{02}(t)\exp(\gamma_{21} X_{1i} + \gamma_{22}X_{2i} + \gamma_{23}X_{3i} + \alpha_{b2}^{\top} b_{i} + \alpha_{\omega 2}\omega_i),
\end{eqnarray*}
\end{enumerate}

Because the fitted Model 1 and the generative joint model (Model 2) do not share the same covariate structure, a direct comparison of estimation behavior—such as bias, standard error, estimated standard error, and coverage probability—is not meaningful, as true parameter values are not defined for some Model 1 parameters. Accordingly, we report findings on estimation behavior only for comparisons between Models 2 and 3, while evaluating predictive performance across all three models. The estimation performance for Model 2 and Model 3 was summarized in Table \ref{sim:tabSim6Est}. Across all parameters, Model 2 and Model 3 demonstrated comparable estimation accuracy, with small biases and coverage probabilities near 95\%. This similarity is expected, since Model 3 nests Model 2 as a special case. In conclusion, when the data were generated under the assumption of homogeneous WS variability, the additional flexibility afforded by Model 3 had little impact on estimation accuracy.
\begin{table}[htbp]
\centering
\caption{Comparison of bias, empirical standard error (SE), estimated standard error (Est. SE), and coverage probability (CP) between Model 2 (correctly specified model: non-linear individual mean trajectory + homogeneous WS variability) and Model 3 (non-linear individual mean trajectory + heterogeneous WS variability), which nest Model 2 as a special case, when the simulated data are generated from the generative joint model (\ref{sim:homoeq1.1})-(\ref{sim:homoeq2}) (Model 2), $n$=3000}
\label{sim:tabSim6Est}
\resizebox{\textwidth}{!}{
\begin{tabular}{lccccccccc}
\hline \hline
& & \multicolumn{4}{c}{\textbf{Model 2}} 
& \multicolumn{4}{c}{\textbf{Model 3}} \\
 \cline{3-10}
Parameter & True
& Bias & SE & Est.SE & CP (\%) 
& Bias & SE & Est.SE & CP (\%) \\
\hline
$Longitudinal$ &  &    & &  &  &    & & & \\
\  Fixed effects &  &   & &   &  &    & & & \\
\ \ Mean trajectory &  &   & &   &  &    & & & \\
\ \ \ $\beta_{0}$ & 5.00
& -0.008 & 0.082 & 0.083 & 93.3 
& -0.006 & 0.082 & 0.083 & 93.7 \\
\ \ \ $\beta_{1}$ & 2.00
& -0.005 & 0.055 & 0.058 & 95.7 
&  0.005 & 0.056 & 0.058 & 94.7 \\
\ \ \ $\beta_{2}$ & 1.00
& -0.023 & 0.050 & 0.050 & 92.7 
& -0.004 & 0.050 & 0.050 & 95.0 \\
\ \ \ $\beta_{3}$ & 1.50
& -0.011 & 0.115 & 0.116 & 94.3 
& -0.002 & 0.115 & 0.116 & 94.7 \\
\ \ WS variability &  &   & &   &  &    & & & \\
\ \ \ $\sigma^2 \,/\, \tau_{0}\;(\log \sigma^2)$ & 1.65 / 0.50
& -0.001 & 0.014 & 0.014 & 94.0 
& -0.007 & 0.023 & 0.023 & 93.0 \\
\ \ \ $\tau_{1}$ & 0.00
& --- & --- & --- & --- 
& -0.002 & 0.018 & 0.020 & 96.0 \\
\ \ \ $\tau_{2}$ & 0.00
& --- & --- & --- & --- 
&  0.001 & 0.047 & 0.046 & 95.0 \\
\ \ \ $\tau_{3}$ & 0.00
& --- & --- & --- & --- 
& -0.001 & 0.030 & 0.032 & 97.0 \\
$Competing \ risks $&  &  & &    &  &    & & & \\
\  Fixed effects &  &    & &  &  &    & & & \\
\ \ $\gamma_{11}$ & 1.00
& -0.002 & 0.062 & 0.064 & 95.0 
&  0.012 & 0.065 & 0.065 & 95.7 \\
\ \ $\gamma_{12}$ & 0.50
&  0.003 & 0.056 & 0.052 & 93.0 
&  0.007 & 0.057 & 0.053 & 93.3 \\
\ \ $\gamma_{13}$ & 0.50
& -0.001 & 0.017 & 0.018 & 96.3 
&  0.004 & 0.018 & 0.018 & 96.7 \\
\ \ $\gamma_{21}$ & -0.50
&  0.006 & 0.056 & 0.063 & 97.0 
&  0.001 & 0.058 & 0.064 & 97.3 \\
\ \ $\gamma_{22}$ & 0.50
&  0.000 & 0.051 & 0.052 & 93.3 
&  0.004 & 0.051 & 0.053 & 94.3 \\
\ \ $\gamma_{23}$ & 0.25
&  0.001 & 0.017 & 0.017 & 95.3 
&  0.002 & 0.018 & 0.018 & 95.0 \\
\ Association &  &  & &    &  &    & & & \\
\ \ $\alpha_{b01}$ & 0.05
&  0.001 & 0.013 & 0.014 & 97.3 
&  0.001 & 0.015 & 0.015 & 95.3 \\
\ \ $\alpha_{b11}$ & 0.01
&  0.003 & 0.028 & 0.034 & 98.3 
&  0.002 & 0.033 & 0.037 & 98.0 \\
\ \ $\alpha_{b21}$ & 0.02
& -0.002 & 0.023 & 0.026 & 96.3 
&  0.000 & 0.026 & 0.028 & 97.0 \\
\ \ $\alpha_{b02}$ & -0.05
&  0.001 & 0.013 & 0.014 & 96.0 
&  0.000 & 0.014 & 0.015 & 96.3 \\
\ \ $\alpha_{b12}$ & 0.02
& -0.003 & 0.029 & 0.034 & 97.7 
& -0.002 & 0.034 & 0.036 & 96.7 \\
\ \ $\alpha_{b22}$ & 0.10
& -0.002 & 0.024 & 0.026 & 95.0 
&  0.002 & 0.027 & 0.028 & 96.3 \\
\ \ $\alpha_{\omega_{1}}$ & 0.00
& --- & --- & --- & --- 
& -0.018 & 1.398 & 1.247 & 95.0 \\
\ \ $\alpha_{\omega_{2}}$ & 0.00
& --- & --- & --- & --- 
&  0.027 & 1.353 & 1.270 & 95.0 \\
\ Covariance matrix &  &      &  &    & & & & & \\
\ of random effects &  &      &  &    & & & & & \\
\ \ $\Sigma_{b_{11}}$ & 10.00
& -0.019 & 0.308 & 0.278 & 92.3 
& -0.019 & 0.307 & 0.279 & 92.7 \\
\ \ $\Sigma_{b_{21}}$ & 3.00
& -0.010 & 0.173 & 0.182 & 95.3 
& -0.010 & 0.174 & 0.184 & 95.7 \\
\ \ $\Sigma_{b_{31}}$ & 2.00
& -0.019 & 0.155 & 0.159 & 95.7 
& -0.018 & 0.155 & 0.159 & 96.0 \\
\ \ $\Sigma_{b_{22}}$ & 4.00
&  0.010 & 0.223 & 0.215 & 92.3 
&  0.013 & 0.223 & 0.217 & 94.7 \\
\ \ $\Sigma_{b_{32}}$ & 2.00
& -0.011 & 0.126 & 0.136 & 96.7 
& -0.008 & 0.127 & 0.137 & 95.7 \\
\ \ $\Sigma_{b_{33}}$ & 4.00
& -0.024 & 0.162 & 0.159 & 93.3 
& -0.019 & 0.166 & 0.161 & 93.3 \\
\ \ $\Sigma_{b_{41}}$ & 0
& --- & --- & --- & --- 
& -0.002 & 0.004 & 0.032 & 98.3 \\
\ \ $\Sigma_{b_{42}}$ & 0
& --- & --- & --- & --- 
& -0.001 & 0.004 & 0.025 & 97.3 \\
\ \ $\Sigma_{b_{43}}$ & 0
& --- & --- & --- & --- 
& $<0.001$ & 0.004 & 0.022 & 98.0 \\
\hline \hline
\end{tabular}}
\end{table}

\begin{figure}[ht]
    \includegraphics[width=14cm]{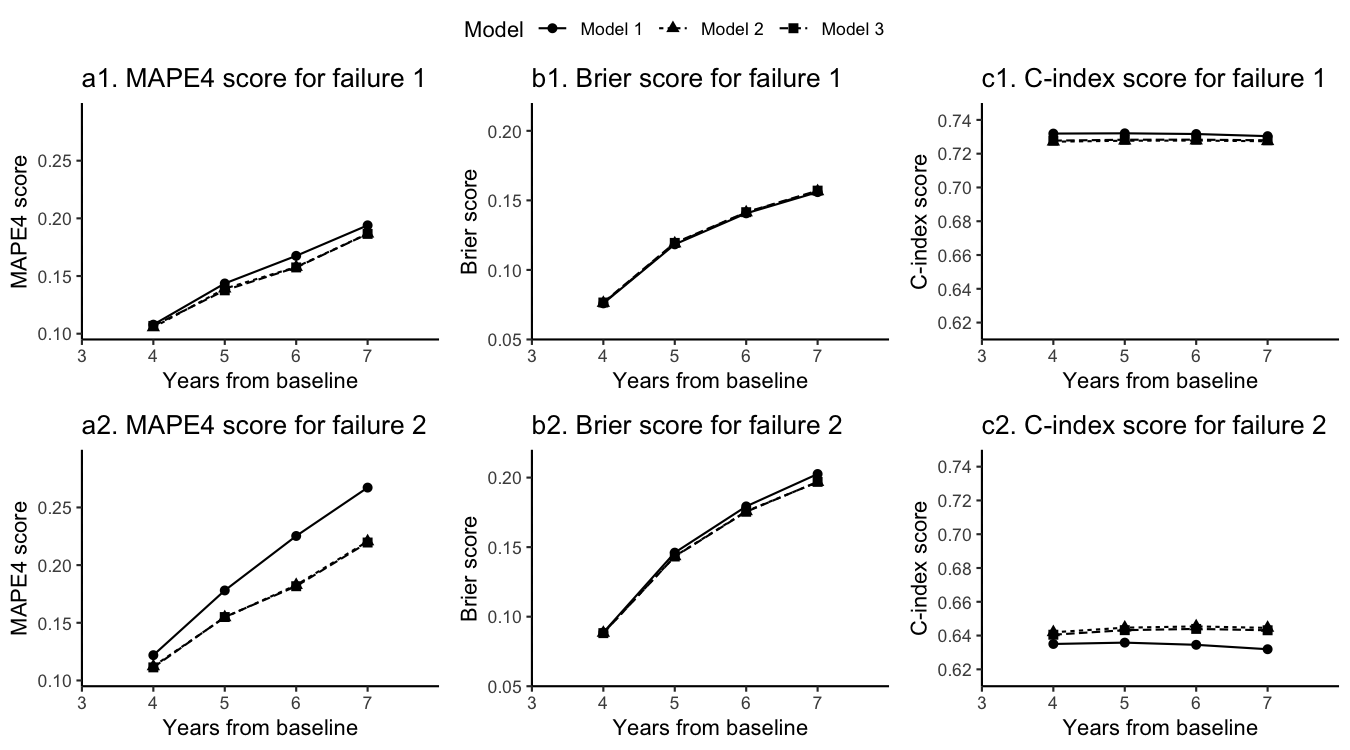}
    \caption{ Prediction performance of Model 1, Model 2, and Model 3 evaluated by MAPE4 score (Panels a1 and a2), Brier score (Panels b1 and b2), and C-index (Panels c1 and c2), based on 10 random splits of cross-validation across 10 simulated datasets at the horizon times $u = (4, 5, 6, 7)$ and the landmark time $s = 3$ under the generative joint model specified by equations (\ref{sim:homoeq1.1}) - (\ref{sim:homoeq2}).}
    \label{sim:fig:NLhomo}
\end{figure}

To evaluate predictive performance, for each simulated dataset, we computed three metrics: the 4-fold cross-validated mean absolute prediction error (MAPE4), the Brier score, and the C-index, as defined in Section \ref{appen:DP}. The procedure was repeated with 10 random partitions for each dataset, and the scores for each metric were averaged across the partitions. We then calculated the overall average scores across all 10 datasets. The results are depicted in Figure \ref{sim:fig:NLhomo}.

As expected, Figure~\ref{sim:fig:NLhomo} shows that, although the magnitudes varied across metrics, horizon times, and outcomes, Model~1 consistently exhibited inferior predictive performance, with the highest MAPE4 and Brier score and the lowest C-index for failure type~2; these differences were more pronounced for MAPE4 and the C-index for failure type~2 at later horizon times. This indicates that simply incorporating heterogeneous WS variance is insufficient when a linear mean trajectory is used to model a non-linear time evolution. Furthermore, Model 3, which includes Model 2 as a special case, exhibited nearly identical predictive performance to Model 2 across all three metrics, confirming that explicitly modeling WS variability does not materially degrade predictive performance even under the homogeneous WS variability scenario.

\clearpage
\section{Time Plots of Mean SBP and Logarithm of Residual Variance for the MESA Data}
\label{sm:addfig}
\begin{figure}[ht]
    \includegraphics[width=11cm]{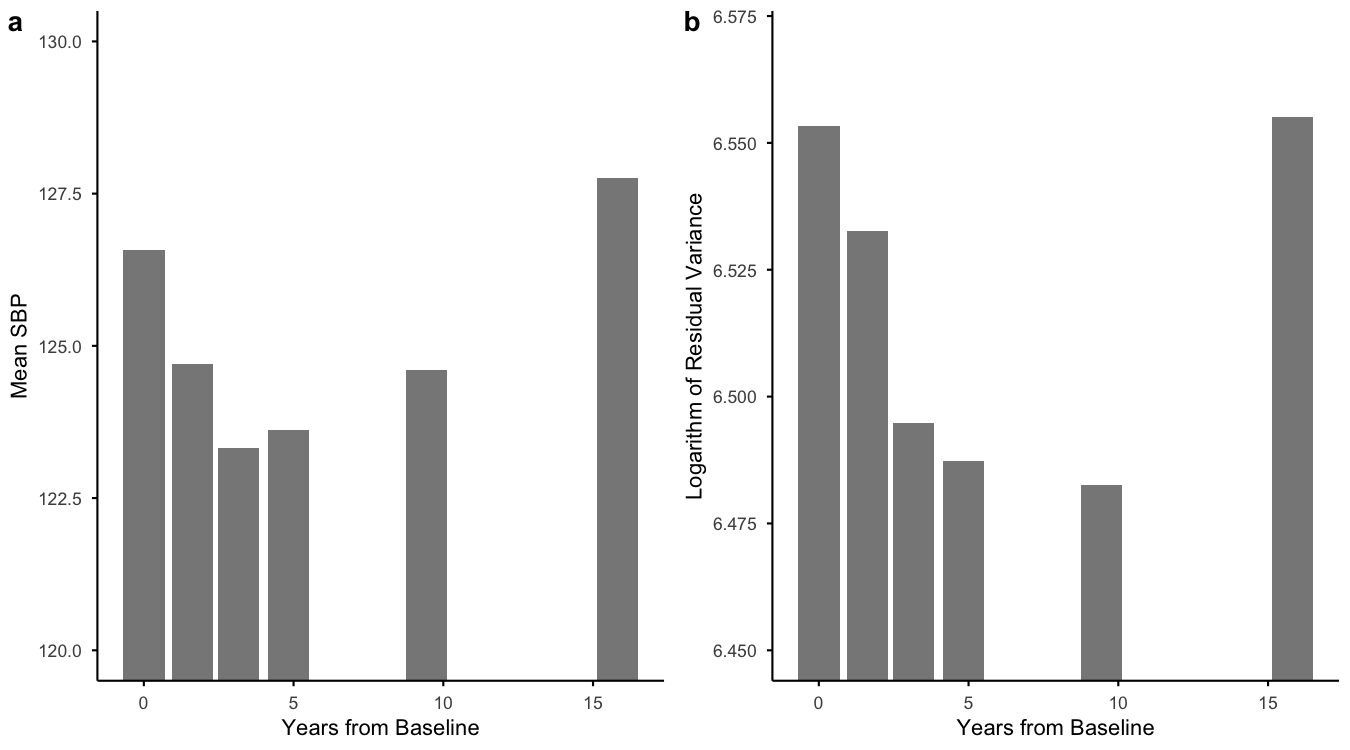}
    \caption{(MESA data) (a) Time plot of mean SBP at baseline and five subsequent visit times; (b) Time plot of the logarithm of the residual variance, obtained after fitting the preliminary submodel (21) under the assumption of homogeneous WS variance, at baseline and five subsequent visit times.
}
    \label{fig:MESA_rawdata}
\end{figure}

Figure~\ref{fig:MESA_rawdata}(a) shows the time plot of local mean SBP for the MESA data at baseline and five subsequent follow-up time points, where the local mean SBP at each time point is obtained by averaging the observed SBP values observed within a $\pm 0.2$-year window centered at the mean clinical visit time for that visit. This plot suggests a non-linear temporal trend that can be adequately captured by a quadratic spline with a single internal knot. A linear spline would likely be insufficient unless additional internal knots were introduced; however, this could lead to overfitting and the capture of spurious variation, particularly given the sparsity of SBP measurements in the MESA data (median = 5, IQR = [4, 6]). While a cubic spline with one or two internal knots could also be used, it introduces additional parameters and complexity, offering limited practical benefit in this setting, as the more parsimonious quadratic spline appears sufficient to capture the observed non-linearity in the mean response.

Figure~\ref{fig:MESA_rawdata}(b) displays the time trajectory of the logarithm of the residual variance, computed as the variance of the residuals (after centering by their visit-specific mean), obtained from fitting the preliminary submodel~(20) to the MESA data under the assumption of homogeneous within-subject (WS) variance. Although this assumption may introduce some bias in the residuals, the plot provides a useful exploratory tool for assessing temporal patterns of WS variability. As shown in Figure~\ref{fig:MESA_rawdata}(b), it suggests that a quadratic spline with a single internal knot may adequately capture the time-varying WS variability in submodel (21).

\bibliographystyle{imsart-nameyear} 
\bibliography{reference}